\documentclass[aps,pre,longbibliography,notitlepage,floatfix,nofootinbib,rmp,superscriptaddress]{revtex4-1}

\usepackage{bm,amsmath,amssymb,graphicx,stmaryrd,float,subcaption,upgreek,sidecap,MnSymbol,mathtools,lipsum}
\usepackage{epstopdf}
\usepackage{physics}
\usepackage[abs]{overpic}
\usepackage[justification=raggedright]{caption}
\usepackage{enumitem}
\usepackage{tikz}
\usepackage{hyperref}
\usepackage{wasysym}
\usepackage{relsize}

\usepackage{xcolor}

\newcommand{\lightdiamond}{\color[rgb]{.98,.5,.45} \text{\smaller[1.0] $\!\!\blacklozenge$}}
\newcommand{\reddiamond}{\color{red} \text{\smaller[1.0] $\!\!\blacklozenge$}}
\newcommand{\blackdiamond}{\color{black} \text{\smaller[1.0] $\!\!\blacklozenge$}}

\newcommand{\redstar}{\color{red} \text{\smaller[3]$\bigstar$}}
\newcommand{\blackstar}{\color{black} \text{\smaller[3]$\bigstar$}}

\newcommand{\redtriangle}{\color{red} \blacktriangle}
\newcommand{\lighttriangleleft}{\color[rgb]{.98,.5,.45} \text{\larger[1.0] $\!\!\blacktriangleleft\!$}}
\newcommand{\lighttriangleright}{\color[rgb]{.98,.5,.45} \text{\larger[1.0] $\!\!\!\blacktriangleright$}}
\newcommand{\redtriangleleft}{\color{red} \text{\larger[1.0] $\!\!\blacktriangleleft\!$}}
\newcommand{\redtriangleright}{\color{red} \text{\larger[1.0] $\!\!\!\blacktriangleright$}}

\newcommand{\lightsquare}{\color[rgb]{.98,.5,.45} \text{\smaller[1.0] $\!\!\blacksquare$}}
\newcommand{\redsquare}{\color{red} \text{\smaller[1.0] $\!\!\blacksquare$}}

\newcommand{\blasquare}{\color{black} \text{\smaller[1.0] $\!\!\blacksquare$}}

\graphicspath{{BigonFigures/BifurandOther/}}

\begin{document}
	\title{Numerical modeling of static equilibria and bifurcations in bigons and bigon rings}
	\author{Tian Yu}
	\email{tiany@princeton.edu}
	\affiliation{Department of Civil and Environmental Engineering, Princeton University, Princeton, NJ 08544}
	\author{Lauren Dreier}
	\email{ldreier@princeton.edu}
	\affiliation{School of Architecture, Princeton University, Princeton, NJ 08544}
	\author{Francesco Marmo}
	\email{f.marmo@unina.it}
	\affiliation{Department of Structures for Engineering and Architecture, University of Naples Federico II, Naples, Italy}
	\author{Stefano Gabriele}
	\email{stefano.gabriele@uniroma3.it}
	\affiliation{Department of Architecture, Roma Tre University, Italy}
	\author{Stefana Parascho}
	\email{parascho@princeton.edu}
	\affiliation{School of Architecture, Princeton University, Princeton, NJ 08544}
	\author{Sigrid Adriaenssens}
	\email{sadriaen@princeton.edu}
	\affiliation{Department of Civil and Environmental Engineering, Princeton University, Princeton, NJ 08544}
	\date{\today}

	\begin{abstract}
		In this study, we explore the mechanics of a \emph{bigon} and a \emph{bigon ring} from a combination of experiments and numerical simulations. A bigon is a simple elastic network consisting of two initially straight strips that are deformed to intersect with each other through a fixed intersection angle at each end. A bigon ring is a novel multistable structure composed of a series of bigons arranged to form a loop. We find that a bigon ring usually contains several families of stable states and one of them is a multiply-covered loop, which is similar to the folding behavior of a bandsaw blade.
			To model bigons and bigon rings, we propose a numerical framework combining several existing techniques to study mechanics of elastic networks consisting of thin strips. Each strip is modeled as a Kirchhoff rod, and the entire strip network is formulated as a two-point boundary value problem (BVP) that can be solved by a general-purpose BVP solver.
			Together with numerical continuation, we apply the numerical framework to study static equilibria and bifurcations of the bigons and bigon rings. Both numerical and experimental results show that the intersection angle and the aspect ratio of the strip's cross section contribute to the bistability of a bigon and the multistability of a bigon ring; the latter also depends on the number of bigon cells in the ring. The numerical results further reveal interesting connections among various stable states in a bigon ring. Our numerical framework can be applied to general elastic rod networks that may contain flexible joints, naturally curved strips of different lengths, etc. The folding and multistable behaviors of a bigon ring may inspire the design of novel deployable and morphable structures.
	\end{abstract}
	
	\maketitle
	
	\section{Introduction}	
	The manipulation of slender rods and strips into structural networks is a technique borrowed from traditional textile craft that can yield not only decorative but also highly functional, self-supporting objects, such as a hat brim to shade the sun or a basket to hold items \cite{perez2015design,martin2015basketmaker,vekhter2019weaving}. Elastic rod and strip networks also have important applications in engineered structures, such as deployable structures \cite{olson2013deployable,mchale2020morphing,panetta2019x,bouleau2019chi,pillwein2020elastic}, flexible robotics \cite{till2017elastic,black2017parallel}, grid shells \cite{baek2018form,baek2019rigidity}, lattice metamaterials \cite{chen2017lattice,leimer2020reduced}, and 3D mesostructures \cite{yan2016mechanical}. Elastic networks normally exhibit rich and unconventional mechanical behaviors due to their potentially complex geometries, connections, and topologies.
	The final form of an elastic network embodies the balancing of forces and moments of each member of the network, which is often in a largely deformed state. Rods and strips interact with each other through coupled nodes, which can be rigid or flexible. Understanding the mechanics of elastic networks is important for designing novel flexible structures with targeted shapes and functions \cite{yang2018multistable,celli2018shape,giorgio2016buckling,baek2020smooth,liu146mechanics,celli2020compliant}.

	In this work, we introduce a structure called a \emph{bigon}, which consists of two thin, initially straight strips that are bent to join with each other through a fixed intersection angle at their two ends, where the two strips share a surface normal (see Figure \ref{fig:bigon0}). A bigon with strips that have a flat cross section (i.e., highly anisotropic in terms of the two bending rigidities) is found to be bistable, which also depends on the intersection angle at the two ends. The bigon proposed in this work could be used as a building block to construct complex elastic networks. 
	
	We then propose a novel multistable structure called a \emph{bigon ring}, where several bigons are connected in series to form a closed loop.  The geometry of a bigon ring can be tuned by varying the intersection angle of each bigon and the number of bigon cells, resulting in interesting folding and multistable behaviors. Examples of manipulating a 6-bigon ring into various states can be found in the \emph{supplementary videos}. The different stable configurations of a bigon ring depend on the shape assumed by the constituent bigons. Some of these configurations are similar to the folded loops in overcurved rods and strips \cite{audoly2015buckling,manning2001stability,mouthuy2012overcurvature}, inextensible bistrips \cite{dias2014non}, and annular ribbons \cite{heijdenannular}. In fact, it is not uncommon for an elastic network to exhibit multiple stable equilibria, separated by unstable energy barriers that normally cannot be obtained in experiments \cite{guan2018structural,baek2018form}.
	
	Various modeling frameworks have been proposed to simulate mechanical behaviors of elastic rod and strip networks, such as analytical analysis based on an approximate 3D beam theory \cite{liu2019postbuckling}, finite element modeling \cite{guan2018structural}, constrained nonlinear optimization \cite{nabaei2015form}, and discrete elastic rods \cite{huang2020numerical,lestringant2020modeling,baek2018form}. The key idea is to simulate elastic networks as coupled rods, which are commonly modeled as \emph{elastica} for planar branched structures \cite{o2011static}, and as Kirchhoff rods or a more general Cosserat rods for spatial elastic networks \cite{baek2018form,baek2019rigidity,spillmann2008cosserat}. These models are robust enough to capture equilibrium shapes, but seem to be inefficient for conducting parametric studies and identifying potential bifurcations in elastic networks. Continuum theories have also been developed to study mechanical behaviors of elastic networks \cite{wang1986inextensible,steigmann2018continuum,eremeyev2019two}.

	Common to many of the aforementioned numerical frameworks is to treat an elastic network as a multi-point boundary value problem (MPBVP), in which various rods and strips are coupled together through nodes. However, general-purpose BVP solvers normally do not accept an MPBVP and require the form of a standard two-point BVP (TPBVP). Mathematically, a MPBVP can be converted to a standard TPBVP by imposing a correct number of boundary conditions that are consistent with the number of unknowns \cite{ascher1981reformulation}. To model the mechanics of a bigon and a bigon ring, we propose a numerical framework that formulates elastic rod networks as a standard TPBVP with mixed boundary conditions (BCs), i.e., the ``0" and ``1" ends of the TPBVP could be coupled together. 
	
	Our numerical framework is based on inextensible Kirchhoff rod theory, which models an elastic strip as a spatial curve (the centerline of the strip) with finite bending and torsional rigidities. The orientations of each strip's cross section are described with quaternions to avoid the potential polar singularity caused by Euler angles. However, using four quaternions to describe spatial rotations may lead to inconsistent prescription of boundary conditions, since a 3D rotation can be described by three independent parameters (e.g., three Euler angles). Usually a unit length quaternion field needs to be imposed pointwise by including an algebraic constraint, which could be difficult to implement. Healey and Mehta introduced a dummy parameter to the differential equations of quaternions and achieved a consistent prescription of boundary conditions for a single Cosserat rod \cite{healey2006straightforward}. The advantage of introducing the dummy parameter is that the unit length constraint of the quaternion field is only required at the two ends, which can be implemented with general-purpose BVP solvers. This technique has also been applied to an inextensible strip \cite{MooreHealey18}. In this study, we add a dummy parameter to each strip of an elastic network, which facilitates the consistent prescription of boundary conditions for the whole elastic network.

	By applying the numerical framework to study the mechanics of bigons and bigon rings, we prove both numerically and experimentally that a bigon with large intersection angle and high anisotropy of the constituent strip cross section tends to buckle out of plane to release the high energy cost of planar bending. Our numerical framework also captures the folding and multistable behaviors of a bigon ring well, which further reveal interesting connections between various states that have quite different shapes.

	This paper is organized as follows. In Section \ref{se:BigonGeometry}, we introduce the geometry of a bigon and a bigon ring. Section \ref{se:rodmodel} briefly introduces the Kirchhoff rod model, and shows how to formulate an elastic strip network as a standard TPBVP. In Section \ref{se: well-posed joints}, we discuss the formulation of elastic networks as a well-posed TPBVP, in which the number of boundary conditions equal the number of unknowns. Section \ref{se:bigonresults} presents the numerical results, analytical predictions, and experimental measurements of a bigon. Experimental configurations and numerical predictions of bigon rings are included in Section \ref{se:bigonringresults}, where we further study the bifurcations of various states as the intersection angle varies. The influence of the aspect ratio of the strip's cross section on the stable states are studied in Section \ref{se:anisotropy}. We summarize our work and give a further discussion in Section \ref{se:conanddiscuss}. In the Appendices, we apply our numerical framework to a bigon arm with hinged joints (Appendix \ref{appse:hingedarm}), briefly discuss the experimental method that measures the tangent angle of a bigon (Appendix \ref{app:anglemeasurement}), address the inside-out flip of a planar bigon (Appendix \ref{app:bilobatebigon}), and document the 2D projections of the solution curves of a 6-bigon ring for the interest of the reader (Appendix \ref{appse:2Dprojection}).
	
	\section{Geometry of a bigon and a bigon ring}\label{se:BigonGeometry}
	
\begin{figure}[h!]
		\centering
		\includegraphics[width=0.6\textwidth]{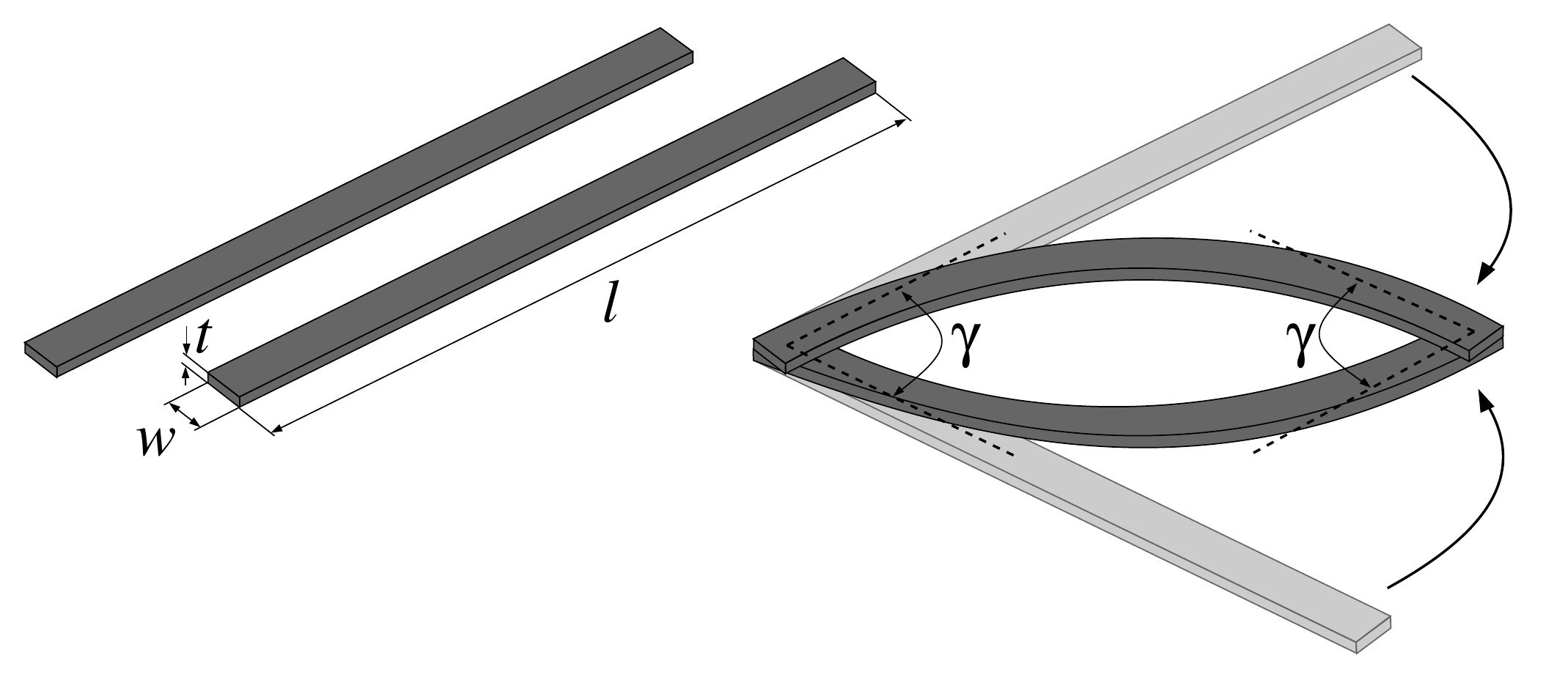}
		\caption{Bigons consist of two identical strips with a width $w$, thickness $t$, and length $l$, fixed at a prescribed intersection angle $\gamma$ at the two ends. Shown is the planar mode of a bigon, where the two strips are deformed into circular arcs.} \label{fig:bigon0}
	\end{figure}

	A bigon consists of two flexible strips that are initially straight and successively bent in such a way to mutually superimpose with each other at the two ends. After the two strips are bent, their extremities are joined together to form a prescribed intersection angle $\gamma$ at both ends (see Figure \ref{fig:bigon0}). The geometry of a bigon depends on the value of the intersection angle $\gamma$ and on the dimensions of the strips, which have length $l$, width $w$ and thickness $t$.

	Due to symmetry, the planar configuration of a bigon contains two circular arcs.
	Varying the intersection angle $\gamma$ generates a family of bigon shapes that have two extreme configurations associated to $\gamma = 180^{\circ}$ and $\gamma = 0^{\circ}$, corresponding to a circular and straight configuration, respectively. With large $\gamma$ and $w/t$, a bigon tends to buckle out of plane due to expensive planar bending, with a snap behavior similar to that of ``click-clack'' hair barrettes. On the other hand, a bigon tends to stay in plane with small $w/t$ and $\gamma$. 
	
\begin{figure}[h!]
		\centering
		\includegraphics[width=0.75\textwidth]{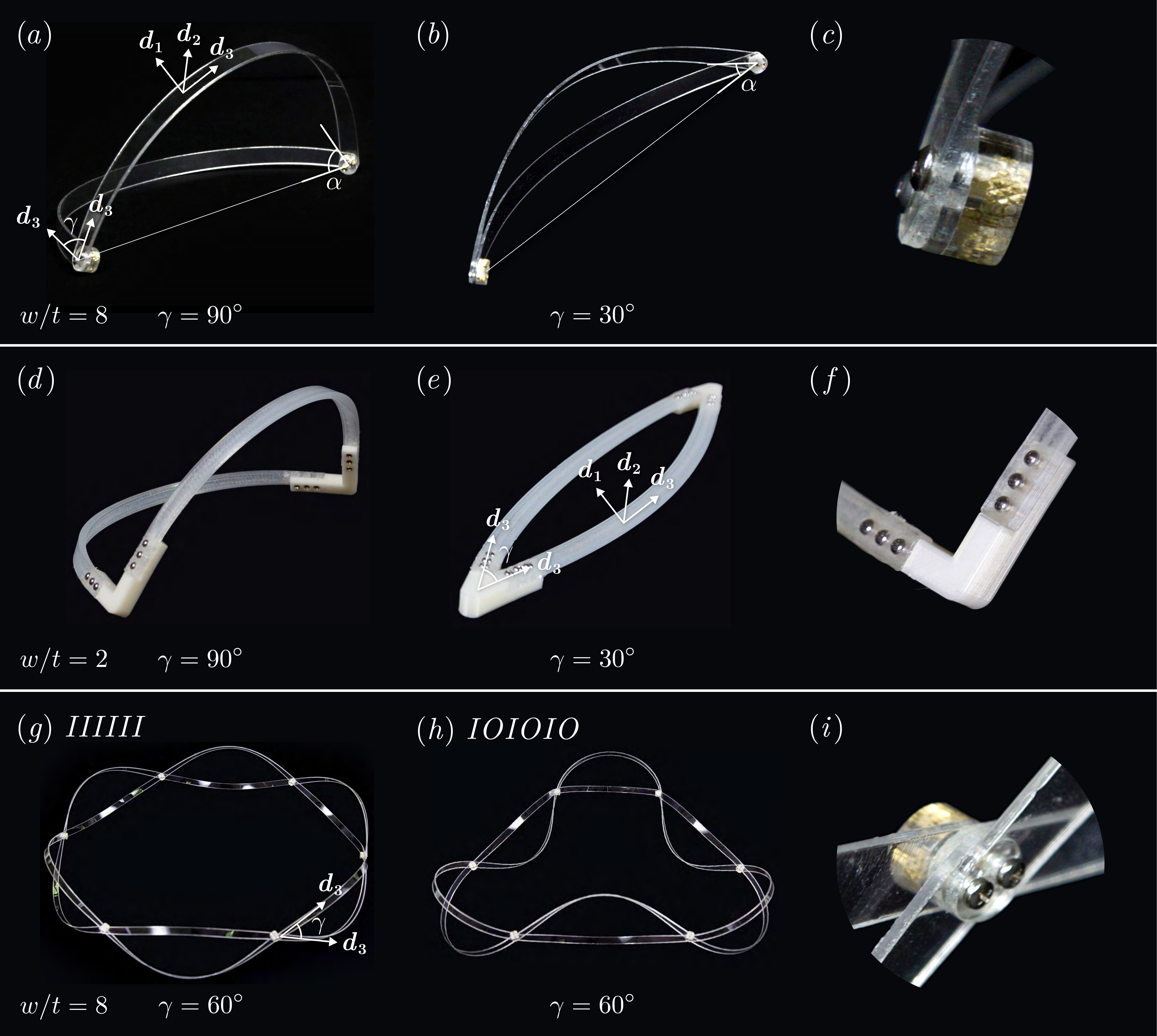}
		\caption{Bigon and bigon ring models. With $w = 8$ mm, $t = 1$ mm, $l = 200$ mm, and (a) $\gamma = 90^{\circ}$ and (b) $\gamma=30^{\circ}$, both bigons buckle out of plane. Shown also is the tangent angle $\alpha$. (c) A close-up of a node in (a)-(b). For 3D-printed bigons with $w/t \approx 2$: (d) $\gamma = 90^{\circ}$, the bigon buckles out of plane; (e) $\gamma = 30^{\circ}$, the bigon stays in plane. (f) Details of a node in (d)-(e). Different states of a 6-bigon ring with $\gamma = 60^{\circ}$: (g) $IIIIII$ (h) $IOIOIO$. (i) A close-up of a node in (g)-(h).} \label{fig:bigonmodel}
	\end{figure}

	Some bigons are assembled in order to experimentally study their behavior (Figure \ref{fig:bigonmodel}). To this end, we laser cut thin strips from PETG sheets with desired dimensions of length $l = 208$ mm, width $w = 8$ mm, and thickness $t = 1$ mm. The laser cutting imparts a small rest curvature on the strip, forming a shallow arch. The actual dimensions are
	$l = 208$ mm, $w = 7.78 \pm 0.10$ mm, and $t = 1.03 \pm 0.03$ mm, with a rising height of approximately 1 mm at the center of the arch. In order to connect strips together, two holes with 1.36 mm diameter are made at each end of the strip for accepting screws. The corners of strips are rounded with a diameter of 8 mm to match with the connecting nodes, which are made from a cast acrylic (2.86 mm thickness) that has been cut to a disk of 8 mm diameter with two through-holes of 1.36 mm diameter, accepting two screws that hold the strips together and prevent rotation. The length of all the strips, measured from node's center, has the same dimension $l=200$ mm. Hereafter, the length of a strip will be referred to as the arc length measured from a node's center.
	
	Two examples with $l = 200$ mm, $w = 8$ mm, and $t = 1$ mm, and $\gamma = 90^{\circ}$ and $\gamma = 30^{\circ}$ are shown in Figures \ref{fig:bigonmodel}(a) and \ref{fig:bigonmodel}(b), respectively. A node that fixes the two ends at a prescribed angle $\gamma$ is shown in Figure \ref{fig:bigonmodel}(c). Figures \ref{fig:bigonmodel}(a-b) define $\alpha$ as the tangent angle between the bisector of the two tangents at one end and the chord connecting the two ends. $\alpha$ is used to characterize the out-of-plane deformation of a bigon and a vanishing $\alpha$ corresponds to the planar state of a bigon.
	Figures \ref{fig:bigonmodel}(d)-\ref{fig:bigonmodel}(e) 
	show that with $w/t \! \approx \! 2$, a bigon with $\gamma = 90^{\circ}$ buckles out of plane, and a bigon with $\gamma = 30^{\circ}$ stays in plane, respectively. We have 3D-printed the straight and thick strips of these two models with NinjaTek Cheetah Flexible Filament (NinjaTek, Manheim, PA). The nodes at the two ends are printed with a more rigid PLA material. The strips are connected to the nodes through three screws at the end of each strip.
	The designed geometry is $w = 6$ mm and $t = 3$ mm. The final printed models have the following dimensions: $w = 6.34 \pm 0.07$ mm, $t = 2.92 \pm 0.06$ mm, and $l = 200$ mm. Also shown in Figures \ref{fig:bigonmodel}(a) and \ref{fig:bigonmodel}(e) is a set of right-handed orthonormal director frames $(\bm{d}_1, \bm{d}_2, \bm{d}_3)$, with $\bm{d}_1$ aligned with the width, $\bm{d}_2$ aligned with the thickness, and $\bm{d}_3$ aligned with the tangents of the strip's centerline. 
	
	A bigon ring is made by connecting several bigons with the same intersection angle to form a closed loop.
	Figure \ref{fig:bigonmodel}(g) shows a 6-bigon ring with $\gamma = 60^{\circ}$; several more 6-bigon rings with $\gamma=10^{\circ}, 110^{\circ}$, and $150^{\circ}$ are constructed and displayed in Section \ref{se:bigonringresults}. All strips have the same dimensions $l = 200$ mm, $w = 8$ mm, and $t = 1$ mm. 
	
	In experiments, we test the stability of various states by switching the bending direction of each bigon cell between \emph{inward} and \emph{outward}. Figures \ref{fig:bigonmodel}(g-h) show the $IIIIII$ and $IOIOIO$ mode of a 6-bigon ring, respectively, where $I$ represents a cell that bends inward and $O$ represents a cell that bends outward. Most of the stable shapes can be obtained consistently, however we observe that a few states change their stability after keeping the ring in a given position for a few days. As such, we perform our experiments soon after the ring is built and neglect the effects of long-term deformations in this work.
	
	To verify the accuracy of the intersection angles in a bigon ring, we measure the six intersection angles in the $IIIIII$ configuration and take the average. The angle error is found to be less than $3^{\circ}$, resulting in deviations less than $5\%$ for bigon rings with an intersection angle $\gamma \ge 60^{\circ}$ and as large as $30\%$ for a $10^{\circ}$ bigon ring. We note that the measured angle is consistently less than the target angle, resulting mainly from the clearances between the holes and the screws. In this small amount of slop, the strips tend to straighten, forcing the intersection angle closed slightly. Despite the relatively large percentage of error in the $10^{\circ}$ bigon ring, the slight closing of the intersection angle does not affect our qualitative comparison between experimental observations and numerical results, which is that a bigon ring with a small intersection angle will lose stability in most states. It is expected that construction errors become smaller in larger models, where a stronger joint design can be employed and construction tolerances are easier to control.

	In order to model the behavior of bigons and bigon rings we employ Kirchhoff rod theory to model single strips of elastic networks. Kirchhoff rod theory is appropriate for a strip that has a length much larger than its width, which is comparable to the thicknesses, i.e., $l \!>>\! w$, and $w \! \sim \! t$. For a strip with a high anisotropic cross section (i.e. $l>>w>>t$), Kirchhoff rod theory fails to capture the bending of the cross section and other plate-like behaviors \cite{audoly2015buckling}. Wide strips tend to deform into developable surfaces, which could be better described by an inextensible strip model \cite{starostin2015equilibrium}. However, an inextensible strip model is known to have singular issues that could be problematic for numerical simulations \cite{yu2019bifurcations,freddi2016corrected}. On the other hand, anisotropic rod models have been successfully applied to study the shape of a narrow {M}{\"o}bius strip \cite{mahadevan1993shape}, the cascade unlooping of helical strips \cite{starostin2009cascade}, and bifurcations of buckled narrow strips \cite{yu2019bifurcations}.
	In this work, we use the anisotropic rod theory to model elastic strip networks. In the 6-bigon ring models, we always keep $l = 200$ mm, $w = 8$ mm, and $t = 1$ mm. In experimental models, the midsurfaces of the two strips in the bigon with $w/t=8$ are shifted by a thickness at the two ends (Figure \ref{fig:bigonmodel}(c)), and the midsurfaces of the four strips in a 6-bigon ring are shifted up to three thicknesses at a node (Figure \ref{fig:bigonmodel}(i)). An exception is the bigon model with $w/t=2$, which has its midsurfaces fixed to the same plane at the two ends via a 3D-printed node (Figure \ref{fig:bigonmodel}(f)). All the numerical results presented in this paper did not include this ``shifting" caused by the thickness of the strip. At a node of our numerical model, all the strips are coupled together through a single point (corresponding to the node center), such that all the strips share a surface normal there.

	\section{Anisotropic rod model}\label{se:rodmodel}
	We use anisotropic rod theory to study the mechanical behaviors of elastic strip networks. Kirchhoff rod theory assumes balances of forces and moments on the centerline of the strip $\bm{X}(s)$, where $s$ is the arc length. An orthonormal right-handed material frame $(\bm{d}_1,\bm{d}_2,\bm{d}_3)$ is attached to the centerline of the strip (see Figure \ref{fig:bigonmodel}). The tangent can be identified with one of the directors, $\bm{X}'=\bm{d_3}$. Here, a prime denotes an $s$ derivative, where $s \in [0 \,, l]$, and $l$ is the length of the strip measured from node's center to center. The kinematics of the material frame $(\bm{d}_1,\bm{d}_2,\bm{d}_3)$ are given by $\bm{d}_i'=\bm{\omega} \times \bm{d}_i$, where $\bm{\omega}=\kappa_{1} \bm{d}_1 + \kappa_{2} \bm{d}_2 +\tau \bm{d}_3$ is the Darboux vector, and $\kappa_{1}, \kappa_{2}$ and $\tau$ represent the two bending curvatures and twist, respectively. The Kirchhoff equations describe the force and moment balances as following, 
	
	\begin{equation}\label{eq:F&Mbalance}
		\begin{aligned}
		\bm{N}'  &=\bm{0} \,,  \\
		\bm{M}' +\bm{d_3} \times \bm{N} &=\bm{0} \,, \\
		\end{aligned}
		\end{equation}

	where $\bm{N}$ and $\bm{M}$ are contact forces and moments, which can be resolved on the material frame as $\bm{N}=N_1 \bm{d_1} +N_2 \bm{d_2} +N_3 \bm{d_3}$ and $\bm{M}=M_1 \bm{d_1} +M_2 \bm{d_2} +M_3 \bm{d_3}$. We assume linear constitutive relations $M_1=EI_1\kappa_1$, $M_2=EI_2\kappa_2$, and $M_3=GJ\tau$, where $E$ and $G$ are the Young's modulus and shear modulus, respectively; $EI_1$, $EI_2$, and $GJ$ are the two bending rigidities and the torsional rigidity, respectively. For an anisotropic rod with a rectangular cross section, the two bending rigidities $EI_1$ and $EI_2$ are different. Pushing the ratio $I_2/I_1$ away from unity increases the flatness of the cross section. A perfectly anisotropic rod can be obtained by pushing $I_2/I_1$ to $\infty$ or $0$, which would penalize one of the bending curvatures $\kappa_{2}$ or $\kappa_{1}$ to zero. Equation \eqref{eq:F&Mbalance} leads to six equilibrium equations that can be written as
	
	\begin{equation}\label{F&Mequilibrium1} 
	\begin{aligned}
	N_1'-N_2 \tau+N_3 \kappa_2=0 \, , N_2'+N_1 \tau-N_3 \kappa_1=0 \, , N_3'+N_2 \kappa_1-N_1 \kappa_2 =0 \, , \\
	M_1'-M_2 \tau-N_2+M_3 \kappa_2 =0 \, , M_2'+M_1 \tau-M_3 \kappa_1+N_1 =0 \, , M_3'+M_2 \kappa_1-M_1 \kappa_2 =0 \, .
	\end{aligned}
	\end{equation}
	
	Normalizing the forces and moments by $GJ$, and defining the rigidity ratios $a=EI_1/(GJ)$ and $b=EI_2/(GJ)$, these equations become  
	
	\begin{equation}\label{eq:normF&M} 
	\begin{aligned}
	N_1' = N_2 \tau -N_3 \kappa_2 \, ,N_2' = -N_1 \tau + N_3 \kappa_1 \, , N_3' =-N_2 \kappa_1 +N_1 \kappa_2  \, , \\
	a\kappa_1 ' =(b-1) \kappa_2 \tau+N_2 \, , b\kappa_2 ' =(1-a) \kappa_1 \tau -N_1 \, , \tau' =(a-b) \kappa_1 \kappa_2 \, .\\
	\end{aligned}
	\end{equation}
	
	Assuming that a strip is composed of an elastically isotropic material, the bending and torsional rigidities are \cite{timoshenko1951theory},
	
	\begin{equation}\label{eq:inertiaofmoment} 
	\begin{aligned} 
	EI_1=\tfrac{1}{12} E w t^3\,,\;  EI_2=\tfrac{1}{12} E w^3 t \,,\; 
	GJ=\lambda G w t^3=\lambda w \frac{E}{2(1+\nu)} t^3 \, ,
	\end{aligned}
	\end{equation}
	in which $\lambda$ is a function of $w/t$, and $\nu$ is Poisson's ratio. In this study, $\nu$ is set to 0.33. The ratios of bending to torsional rigidity $a$ and $b$ are thus
	\begin{equation}\label{eq:coefficients} 
	a=\frac{(1+\nu)}{6\lambda} \,,\;  b=\frac{(1+\nu)}{6\lambda}\left(\frac{w}{t} \right)^2 \, .
	\end{equation}
	
	For a strip with a rectangular cross section, $\lambda=\tfrac{1}{3} \left( 1-\tfrac{192}{\pi^5} \tfrac{t}{w} \sum _{k=1} ^{\infty} \tfrac{1}{(2k-1)^5} \tanh \left( \tfrac{\pi (2k-1) w}{2t} \right)  \right)$ \cite{timoshenko1951theory}.  In numerical simulations, we keep the first ten terms (i.e., $k \in [1,10]$), which leads to an accurate $\lambda$ for $w/t$ up to 20 (error is less than $10^{-6}$). 
	We model each strip of an elastic network as a Kirchhoff rod, assign to it a full set of governing equations, and formulate an elastic network as a TPBVP.
	The governing equations of the $i$th strip in an elastic network can be summarized as
	\begin{equation}\label{eq:fullsetgovern} 
	\begin{aligned}
	N_{1i}' &=( N_{2i} \tau_{i} -N_{3i} \kappa_{2i} ) l_i \,, \; N_{2i}' =( -N_{1i} \tau_i + N_{3i} \kappa_{1i} ) l_i  \,, \; N_{3i}' =(-N_{2i} \kappa_{1i} +N_{1i} \kappa_{2i}) l_i \,,  \\
	a_{i} \kappa_{1i} ' &=( (b_{i}-1) \kappa_{2i} \tau_{i}+N_{2i} )l_i \,, \; b_{i} \kappa_{2i} ' =( (1-a_{i}) \kappa_{1i} \tau_{i} -N_{1i} ) l_i \,, \; \tau_{i}' =(a_{i}-b_{i}) \kappa_{1i} \kappa_{2i} l_i \,,\\
	q'_{1i}&=\left( \tfrac{1}{2}(-q_{2i} \tau_{i} +q_{3i} \kappa_{2i} -q_{4i} \kappa_{1i})+ \mu_{i} q_{1i} \right) l_i \,, \; q'_{2i}=\left( \tfrac{1}{2}(q_{1i} \tau_{i} + q_{4i} \kappa_{2i} +q_{3i} \kappa_{1i}) + \mu_{i} q_{2i} \right) l_i  \,, \\
	q'_{3i}&=\left( \tfrac{1}{2}(q_{4i} \tau_{i} - q_{1i} \kappa_{2i} - q_{2i} \kappa_{1i}) + \mu_{i} q_{3i} \right) l_i \,, \; q'_{4i}=\left( \tfrac{1}{2}(-q_{3i} \tau_{i} -q_{2i} \kappa_{2i} +q_{1i} \kappa_{1i}) + \mu_{i} q_{4i} \right) l_i  \,, \\
	x_{i}' &=2(q_{1i}^2 + q_{2i}^2-\tfrac{1}{2}) l_i \,, \; y_{i}' =2(q_{2i} q_{3i} + q_{1i} q_{4i}) l_i \,,\; z_{i}' =2(q_{2i} q_{4i} - q_{1i} q_{3i}) l_i \, ,
	\end{aligned}
	\end{equation}
	where, $i \in [1,K]$, and $K$ represents the total number of strips. We use quaternions ($q_{1i},q_{2i},q_{3i},q_{4i}$) to describe the orientations of the $i_{\mathrm{th}}$ strip's cross section, and Cartesian coordinates ($x_i, y_i, z_i$) to describe the position of the $i_{\mathrm{th}}$ strip's centerline (see Figures \ref{fig:bigonKinematics} and \ref{fig:bigonRingKinematics} for the Cartesian coordinate system). Relationships between quaternions and a $3-1-2$ rotational sequence of Euler angles are presented in Section \ref{se:bigonresults}. 
	$l_i$ represents the length of the $i_{\mathrm{th}}$ strip, and a prime denotes an $s_i$ derivative. By introducing the scaling factor $l_i$, the length of the integral interval is normalized to unity, i.e., $s_i \in [0\,,1]$ for all strips. In doing so, we have formulated an elastic network as having only two ends, namely ``0"s and ``1"s. Together with appropriate boundary conditions that couple the strips together, we obtain a TPBVP. We have introduced a ``dummy" parameter $\mu_{i}$ for each strip, which enables a consistent prescription of boundary conditions for quaternions \cite{healey2006straightforward}. These dummy parameters could be converted into state variables by introducing additional differential equations $\mu_i'=0$. In this study, we treat them as parameters that vary freely in numerical continuation. The computed values of these dummy parameters should be numerically zero \cite{healey2006straightforward}. In numerical continuation we keep monitoring them, and find their values are normally in the order of $10^{-15} - 10^{-13}$. 
	
	Adding a nondimensional gravity $-g \bm{\hat{z}}$ ($\bm{\hat{z}}$ is the basis vector along $z$ direction of the Cartesian coordinate) changes the first three equations of \eqref{eq:fullsetgovern} to
	
	\begin{equation}\label{Gravity}
	\begin{aligned}
	N_{1i}' &= (N_{2i} \tau_{i} -N_{3i} \kappa_{2i} +2 g (q_{1i}^2 + q_{4i}^2-\tfrac{1}{2}) ) l_i\, ,\\
	N_{2i}' &= (-N_{1i} \tau_i + N_{3i} \kappa_{1i} -2 g (q_{3i} q_{4i} + q_{1i} q_{2i})) l_i \, , \\
	N_{3i}' &=(-N_{2i} \kappa_{1i} +N_{1i} \kappa_{2i} + 2 g (q_{2i} q_{4i} - q_{1i} q_{3i})) l_i \, , \\
	\end{aligned}
	\end{equation}

	\section{Formulation of an elastic network as a well-posed  two-point boundary value problem}\label{se: well-posed joints}
	
	It has been shown in \cite{healey2006straightforward} that with the introduction of a dummy parameter to a single Cosserat rod, one can always obtain seven boundary conditions at each end of the strip, which could be a mixture of forces and moments, and positions and orientations. This agrees with the fourteen unknowns in Equation \eqref{eq:fullsetgovern} (i.e., three forces, two curvatures and one twist, three Cartesian positions, four quaternions, and one dummy parameter), which we can think of as being evenly distributed to the two ends of a strip, such that the seven boundary conditions at each end \emph{balance} the seven unknowns, resulting in a well-posed boundary value problem for a single strip. Similarly we can move the unknowns of elastic networks to the nodes: if the number of boundary conditions equals the number of unknowns at every node, we obtain a well-posed boundary value problem for the entire elastic network. We call such a node, \emph{well-posed}.  
	
	The unknowns at a general node may include the orientations of various cross sections that are part of the solution. We impose the quaternion boundary conditions at nodes through Euler angles that describe the orientations of strips there (see Section \ref{se:bigonresults}). Thus, the unit length constraint of quaternion field is automatically satisfied. If the orientations of a strip at a node is part of the solution, then the corresponding Euler angles become scalar unknowns. In numerical continuation, these scalar unknowns are treated as parameters that vary freely \cite{doedel2007auto}.

	Nodes in an elastic network can be rigid (such that angles between various strips at a node are fixed), flexible (such that angle variations between various strips at a node follow certain constitutive laws),  purely hinged (such that moments cannot be transferred at the node), etc. In addition, an elastic network can be fixed in space (i.e., global translations and rotations are forbidden) by constraining one or several of its nodes. Here, we give two examples to show that we can always find the exact number of boundary conditions that equals the number of unknowns at different types of nodes.


	For a rigid node that is clamped in space and has $n$ strips connected at the node, we have $7n$ unknowns from equation \eqref{eq:fullsetgovern}; each end of the strip provides seven boundary conditions that include the fixing of three Cartesian coordinates and the prescription of four quaternions, resulting in $7n$ boundary conditions in total. Thus, a rigid node clamped in space is well-posed.

\begin{figure}[h!]
		\centering
		\includegraphics[width=0.60\textwidth]{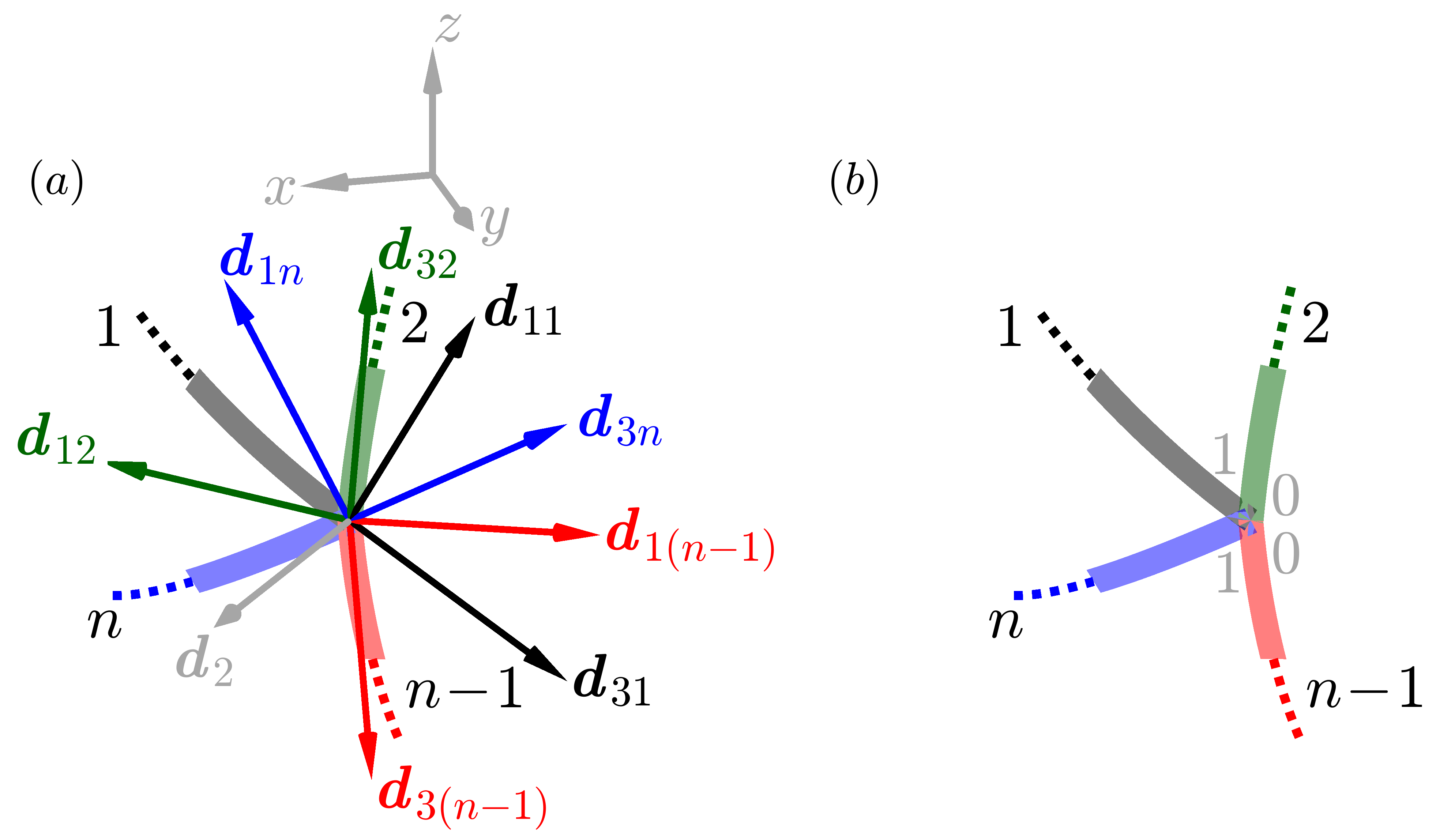}
		\caption{A rigid node of an elastic network that joins $n$ strips with a shared normal \textcolor{gray}{$\bm{d}_2$}. The rigid node is free to move and rotate in space. 
			(a) Only renderings and director frames of the first, second, $(n-1)_{\textrm{th}}$, and $n_{\textrm{th}}$ strips are shown. The director frames $(\bm{d}_{1i},\textcolor{gray}{\bm{d}_2}, \bm{d}_{3i})$ refer to the $i_{th}$ surface with the same color (shown lighter).
			(b) Based on (a), meeting at the node is the ``1" end of the first and $n_{\textrm{th}}$ strips, and the ``0" end of the second and $(n-1)_{\textrm{th}}$ strips.}
		\label{fig:generalrigidnode}
	\end{figure}

	If we relax certain degrees of freedom of the clamped node through rotations/translations, then corresponding moment/force boundary conditions should be added in. 
	For example, for a rigid node that is free to move and rotate in space, and has $n$ strips connected such that the strips share the surface normal at the node (Figure \ref{fig:generalrigidnode}(a)), the orientations of all the strips at the node only differ through a rotation about the normal, which is known \emph{a priori} as intersection angles. Once the orientation of any strip at the node is known, the orientations of all the other strips at the node are known. 
	At a free rigid node, aside from the $7n$ unknowns from equation \ref{eq:fullsetgovern}, we introduce three Euler angles as scalar unknowns to describe the orientations of an arbitrary strip at the node; the orientations of other strips can be obtained by the prescribed intersection angles. In total, we have $(7n+3)$ unknowns at a free rigid node, where we impose the continuity of forces, moments, and positions. Each of the former two gives three boundary conditions that can be imposed through Cartesian components, and the position continuities give $3(n-1)$ boundary conditions. In addition, the orientations of each strip at the node give four boundary conditions through quaternions. In total, we have $(7n+3)$ boundary conditions at a free rigid node, which balance the $(7n+3)$ unknowns. Figure \ref{fig:generalrigidnode}(b) shows a specific combination of ``0" and ``1" ends at the node: A strip with its tangent going out of the node has a ``0" end, while a strip with its tangent coming in the node has a ``1" end. Other choices of ``0" and ``1" ends are possible. For the entire elastic network, we ensure that every strip is assigned with a ``0" and ``1" end. While it does not matter which end is ``0" or ``1", we should avoid two ``0"s or two ``1"s for a single strip.

	Our formulation can be adapted to study elastic networks with external forces/moments applied at the nodes by adding the external forces/moments to the boundary conditions representing the continuity of forces/moments. It is also straightforward to extend our formulation to account for other types of nodes, such as a flexible node, where the variation of the angle between different strips follows certain constitutive law, e.g., a nonlinear hinge. Relaxing rigid nodes to be flexible, generally creates more unknowns since the angles between strips can vary. The additional unknowns can be balanced by adding boundary conditions representing the constitutive law of the flexible nodes.
	
	We solve the resulting two-point boundary value problem by conducting numerical continuation with AUTO 07P, which uses orthogonal collocation and pseudo-arclength continuation to solve ordinary differential equations as a bifurcation parameter varies. AUTO 07P is also able to detect various kinds of bifurcations and folds, switch to and compute the bifurcated branches, and allows us to track the loci of the bifurcations and folds through two-parameter continuation.  
	More information on AUTO 07P can be found in \cite{doedel2007auto}.
	
	Throughout the rest of the paper, numerical results of bigons and bigon rings are presented as solution curves with several renderings that correspond to the arrowed or marked locations. The renderings are not necessarily displayed in the same scale, as some are enlarged for clarity. In the renderings, each strip is normalized to a unit length, and the strip surface is reconstructed by sweeping the width (aligned with the director $\bm{d}_1$) along the centerline of the strip. The thickness of the strip is not shown on the numerical renderings. Black and gray circles represent bifurcation and fold points. Markers of the renderings are colored red for states that are stable in experimental models and light red for states that are physically unrealistic, i.e., unstable. No other stability information is shown on the solution curves.
	
	\section{Numerical results of a bigon}\label{se:bigonresults}
	
	\subsection{Modeling of a bigon}\label{sse:numericalbigonmodel}

	A bigon has two mirror symmetries: it is symmetric about the plane spanned by a bisector at a node and the chord connecting the two nodes, and symmetric about a second plane that is perpendicular to and bisects the chord. Due to these two mirror symmetries, the contact forces inside the strips vanish identically, and the two strips interact with each other through \emph{pure moments} at the nodes.

\begin{figure}[h!]
		\centering
		\includegraphics[width=0.75\textwidth]{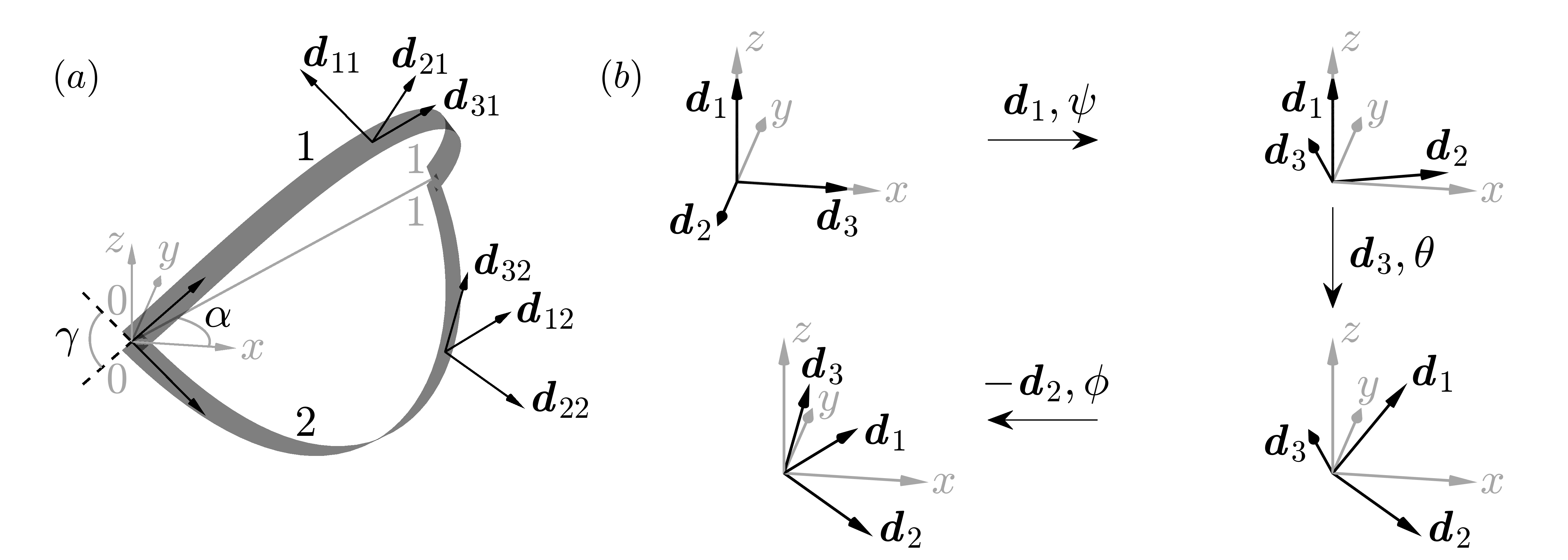}
		\caption{Relations between the Euler angles, material frame, and the Cartesian coordinate system $(x,y,z)$. (a) One end of the bigon is clamped at the origin, with the shared normal $\bm{d}_2$ aligned with $-y$, and the bisector of the intersection angle aligned with $x$. The other end is free to move and rotate. Also shown is an assignment of ``0" and ``1" ends (gray numbers), and the numbering of the strips (black).
			(b) With $\bm{d_1}$, $\bm{d_2}$, and $\bm{d_3}$ aligned with $z$, $-y$, and $x$ axis at the beginning, respectively, we sequentially rotate the director frame around $\bm{d}_1$ by $\psi$, around $\bm{d}_3$ by $\theta$, and around $-\bm{d}_2$ by $\phi$.}\label{fig:bigonKinematics}
	\end{figure}

	In the simulation, one end of the bigon is clamped at the origin of a Cartesian coordinate system $({x,y,z})$, such that the shared normal is aligned with $-y$ and the bisector of the intersection angle is aligned with $x$ (Figure \ref{fig:bigonKinematics}(a)). The tangent angle $\alpha$ is measured from the bisector to the chord connecting the two ends, with a counter clockwise angle being positive. The other end of the bigon is free to move and rotate. In numerical continuation, we vary the intersection angle $\gamma$ from $0^{\circ}$ to $180^{\circ}$.
	Euler angles enter our formulation as scalar unknowns to describe orientations of the free node. We follow a $3-1-2$ (i.e., $z-x-y$) rotation sequence, where at the beginning, the directors $\bm{d}_1$, $\bm{d}_2$, and $\bm{d}_3$ are aligned with $z$, $-y$, and $x$, respectively.
	We first rotate the director frame $(\bm{d}_1, \bm{d}_2, \bm{d}_3)$ about $\bm{d}_1$ axis by $\psi$ (corresponding to a ``3" rotation), then rotate the updated director frame about the new $\bm{d}_3$ axis by $\theta$ (corresponding to a ``1" rotation), and finally rotate the current director frame about the new $\bm{-d}_2$ axis by $\phi$ (corresponding to a ``2" rotation). Figure \ref{fig:bigonKinematics} shows the application of such a rotation sequence to the description of the rotation of the material frame attached to the centerline of a bigon. It also shows an assignment of ``0" and ``1" ends (gray numbers), and the numbering of the strips (black numbers). Other choices of ``0" and ``1" ends are possible.

	For a $3-1-2$ rotation, Euler angles and quaternions are related by \cite{henderson1977euler}
	\begin{equation}\label{eq:eulerparameter} 
	\begin{aligned} 
	q_1&=-\sin \tfrac{\psi}{2} \sin \tfrac{\theta}{2} \sin \tfrac{\phi}{2} + \cos \tfrac{\psi}{2} \cos \tfrac{\theta}{2} \cos \tfrac{\phi}{2}\,, q_2=-\sin \tfrac{\psi}{2} \sin \tfrac{\phi}{2} \cos \tfrac{\theta}{2} + \sin \tfrac{\theta}{2} \cos \tfrac{\psi}{2} \cos \tfrac{\phi}{2}\,, \\
	q_3&=\sin \tfrac{\psi}{2} \sin \tfrac{\theta}{2} \cos \tfrac{\phi}{2} + \sin \tfrac{\phi}{2} \cos \tfrac{\psi}{2} \cos \tfrac{\theta}{2}\,, q_4=\sin \tfrac{\psi}{2} \cos \tfrac{\theta}{2} \cos \tfrac{\phi}{2} + \sin \tfrac{\theta}{2} \sin \tfrac{\phi}{2} \cos \tfrac{\psi}{2}\,.
	\end{aligned} 
	\end{equation}
	
	The orientations of all the strips at a node are imposed as boundary conditions through Equation \ref{eq:eulerparameter}. At the fixed end where two strips are connected to each other, we have 14 boundary conditions and 14 unknowns (i.e., $7n$ with $n=2$).
	At the free end, we have 17 unknowns, namely 14 from the two strips and three Euler angles describing the orientations of the first strip's cross section. The Euler angles at the free node enter the problem through Equation \ref{eq:eulerparameter} as boundary conditions. On the other hand, we have 17 boundary conditions at the free end (i.e., $7n+3$ with $n=2$), including imposing the orientations of the two strips through quaternions (8 BCs), imposing the continuity of positions (3 BCs), the continuity of forces (3 BCs), and the continuity of moments (3 BCs). In total, we obtain a well-posed boundary value problem that has 31 unknowns. We choose the planar shape of a bigon consisting of two circular arcs as the start solution for numerical continuation.

	\subsection{Out-of-plane behavior of a bigon}\label{sse:oopbigon}
	
\begin{figure}[h!]
		\centering
		\includegraphics[width=0.8\textwidth]{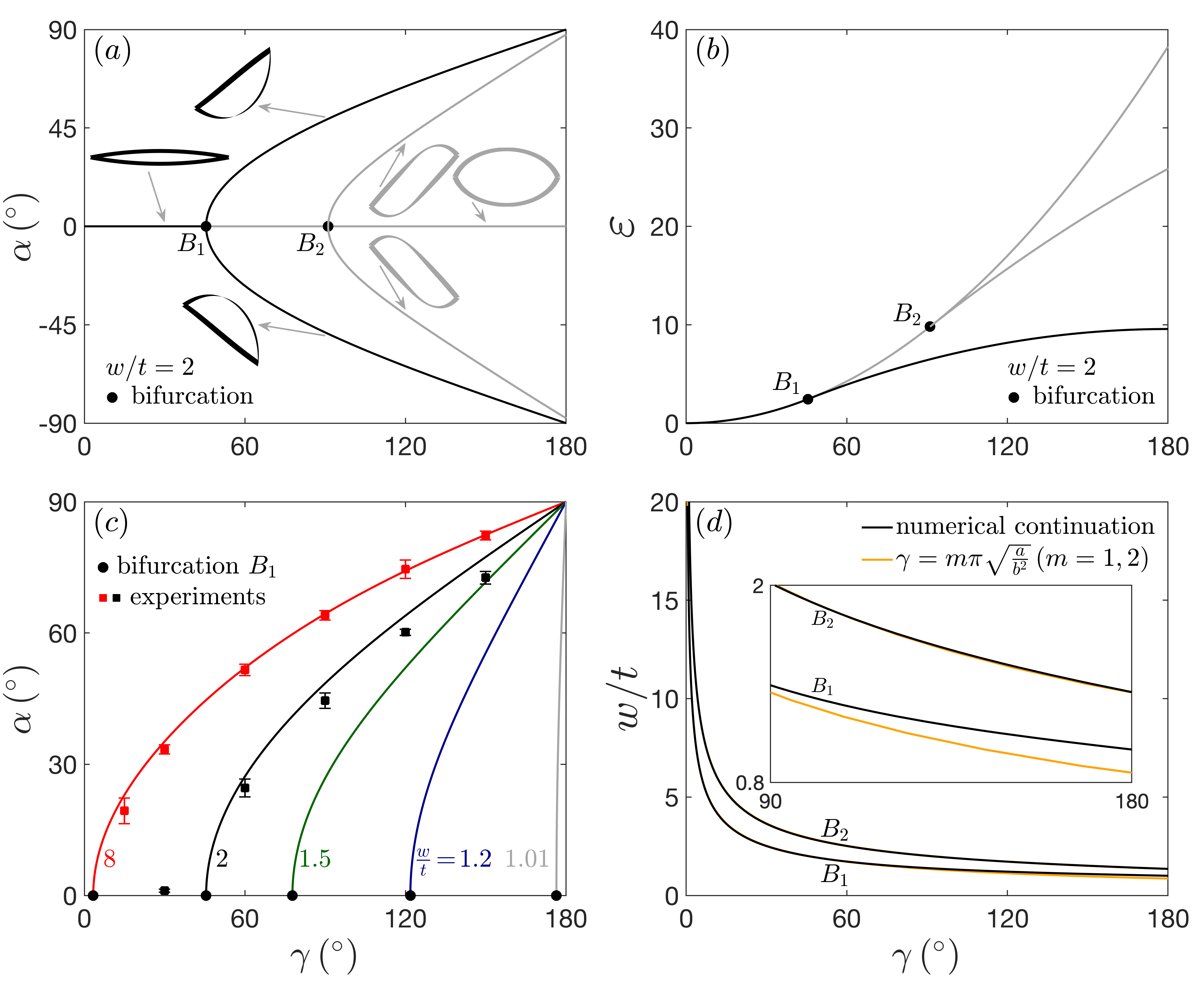}
		\caption{Numerical solutions and bifurcations (black circles) of a bigon. (a) The relations between the intersection angle $\gamma$ and the tangent angle $\alpha$ with $w/t=2$. The planar branch loses stability through a supercritical pitchfork bifurcation $B_1$. An $S$-like mode connects to the planar branch through another bifurcation $B_2$. (b) The results of (a) presented in the $\gamma - \varepsilon$ plane, where $\varepsilon$ represents the total elastic energy. (c) Solution curves of the stable buckled branch ($\alpha>0$) with $w/t=1.01, 1.2, 1.5, 2, $ and 8. 
			Experimental measurements for $w/t=8$ (red) and 2 (black) are presented as mean (squares) and standard deviation. 
			(d) Loci of the bifurcations $B_1$ and $B_2$ in $\gamma$ versus $w/t$ plane, from the Kirchhoff rod model (black) and an approximate analytical prediction  \cite{timoshenko2009theory} (orange)}.
		\label{fig:BigonAnisotropy}
	\end{figure}

	Figure \ref{fig:BigonAnisotropy} shows the numerical solutions of a bigon. The intersection angle $\gamma \in [0, \pi]$ is chosen as the bifurcation parameter, and the tangent angle $\alpha$ and the normalized elastic energy $\varepsilon=0.5 \sum_{i=1}^{2} \int _0 ^{1} (a_{i} \kappa_{1i} ^2 +b_{i} \kappa_{2i} ^2 +\tau_i^2) \, d s_i  $ are chosen as the solution measures. The two normalized bending rigidities $a_i$ and $b_i$ are calculated through Equation \eqref{eq:coefficients}. 
	Figure \ref{fig:BigonAnisotropy}(a) shows that with $w/t=2$ and a small $\gamma$, the bigon stays in plane as two circular arcs. This matches with the experimental model in Figure \ref{fig:bigonmodel}(e). A supercritical pitchfork $B_1$ connects the planar branch to a pair of buckled branches at $\gamma=45.5^{\circ}$, where the planar branch loses stability through buckling out of plane. The pair of buckled branches are observed to be stable in experiments (see Figure \ref{fig:bigonmodel}(d)). Another bifurcation $B_2$ occurs at $\gamma=91.1^{\circ}$ and connects to a pair of $S$-like configurations, which are unstable in experiments. The $S$ mode has different symmetries from the stable out-of-plane mode: it is still symmetric about the plane spanned by a bisector at one end and the chord connecting the two ends; however, instead of having a second mirror symmetry, an $S$-like mode has a $\pi$-rotational symmetry, with the line that goes through the middle point of both strips being the $C_2$ axis. Due to the symmetries, the $S$ mode is also subject to pure moments. In section \ref{se:bigonringresults}, we show that a bigon ring can be stable even with some of its bigon cells deformed into $S$ shapes. There exists other bifurcations on the horizontal axis of Figure \ref{fig:BigonAnisotropy}(a), connecting to higher modes of a bigon. These solution curves are not included here.

	Figure \ref{fig:BigonAnisotropy}(b) shows that after the bifurcations $B_1$ and $B_2$, the elastic energy of the planar branch increases much faster than the buckled solutions, i.e., the bigon manages to decrease energy cost through buckling out of plane. The stable pair connects to $B_1$ has the lowest elastic energy.
	Figure \ref{fig:BigonAnisotropy}(c) shows one of the stable buckled branches ($\alpha >0$) with different values of $w/t$. Increasing $w/t$ decreases the critical angle at the pitchfork bifurcation $B_1$, which corresponds to a small intersection angle $3.3^{\circ}$ with $w/t=8$. As $w/t \rightarrow \infty$, bifurcation $B_1$ approaches zero. This implies that with a perfectly anisotropic rod that forbids planar bending, a bigon will buckle out of plane with an infinitesimal intersection angle. In addition, beyond the bifurcation $B_1$, increasing $\gamma$ generally increases the tangent angle $\alpha$. With a fixed intersection angle $\gamma$, decreasing the anisotropy $w/t$ decreases the tangent angle $\alpha$.

	Also shown in Figure \ref{fig:BigonAnisotropy}(c) is the experimental measurements of $\alpha$ with $w/t=8$ (red; $\gamma=15^{\circ}, 30^{\circ}, 60^{\circ}, 90^{\circ}, 120^{\circ},$ and $150^{\circ}$) and $w/t=2$ (black; $\gamma=30^{\circ}, 60^{\circ}, 90^{\circ}, 120^{\circ},$ and $150^{\circ}$). Each data point reports the mean (black and red squares) and standard deviation of four tangent angles from two experimental models. Each model contributes two tangent angles that are measured from photographs of the bigons. Details of the measurements and several examples are included in Appendix \ref{app:anglemeasurement}. The agreement between the numerical predictions and the experimental results for bigons with $w/t=8$ is very good for large intersection angles ($\gamma \ge 30 ^{\circ}$); while for bigons with $w/t=2$, the measurements are slightly smaller than the numerical results, which we think could be attributed to several factors. Firstly, the printed models are observed to be very flexible; together with the gravity that always tends to flatten the structure in our experimental setup (the bigon is concave up and supported at the middle), this flexibility could decrease $\alpha$ values of $w/t=2$ models. Secondly, stretch may appear in the printed beams with $w/t=2$, which are not very thin and may not be isotropic. In general, the experimental measurements match well with  our numerical predictions.

	We also conduct two-parameter continuation with AUTO 07P to obtain the loci of the bifurcations in the plane $w/t$ versus $\gamma$. Figure \ref{fig:BigonAnisotropy}(d) shows the loci of the bifurcations $B_1$ and $B_2$ (black). On the right, the $B_1$ curve passes through the point $(180^{\circ},1)$, matching with the literature that a straight isotropic rod is about to buckle out of plane when purely bent to a circle \cite{manning2001stability,hoffman2005stability,borum2020helix}.
	The area below the $B_1$ curve represents the region where a bigon stays in plane (i.e., the planar mode is stable), while in the area above this curve, bigon buckles out of plane. Above $w/t=5$, the curve approaches $\gamma=0^{\circ}$ quickly. We conclude that for strips with a flat cross section, the bigon prefers buckling out of plane. The loci of $B_2$ is similar to the loci of $B_1$ and is slightly shifted to the right, i.e., the (unstable) second mode always appears at a larger intersection angle than the (stable) first mode.

	\begin{figure}[h!]
		\centering
		\includegraphics[width=0.6\textwidth]{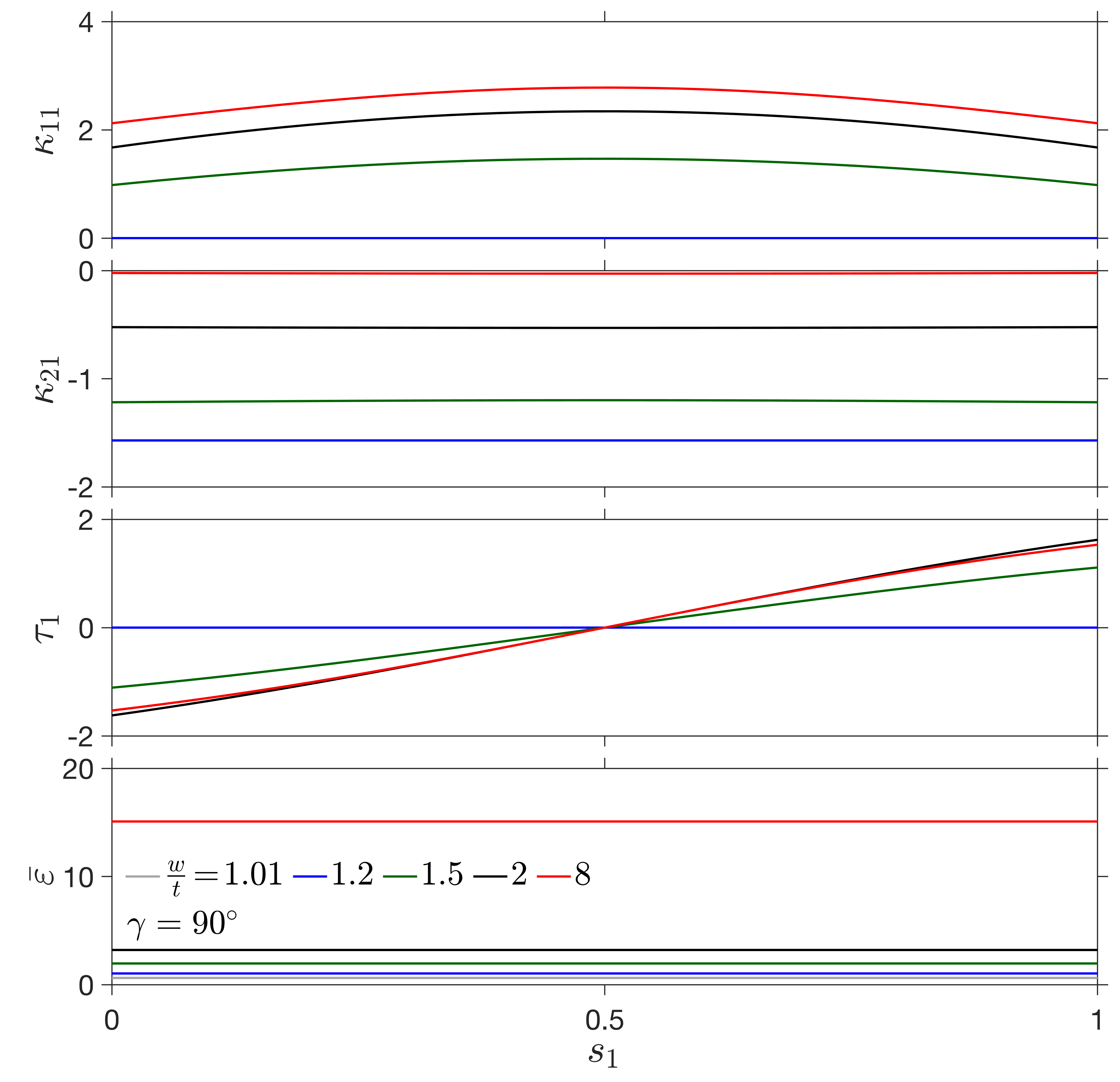}
		\caption{Curvatures, twist, and normalized energy density $\bar{\varepsilon}$ of the numerical results in Figure \ref{fig:BigonAnisotropy}(c), at $\gamma=90 ^{\circ}$. Due to symmetry, only the solutions of the first strip with $z_1 \geq 0$ are shown.}
		\label{fig:Bigontuakappaenergy}
	\end{figure}

	A pre-buckled bigon consists of two exact circular arcs with known shapes, which imply that the bifurcation points may be analytically tractable. Because of the symmetry, we only need to study a half of the bigon, corresponding to the well-known lateral buckling of a rectangular beam subject to pure moments at the two ends \cite{timoshenko2009theory}. Previous studies focused on beams with a flat cross section (i.e., large ratios of $I_2/I_1$), such that the beam only bends into a shallow circular arc before buckling out of plane. This contrasts with our numerical prediction from Kirchhoff rod, which is geometrically exact and appropriate for any $I_2/I_1$. For large $I_2/I_1$, the buckling moment is found to be $\frac{m \pi}{l} \sqrt{EI_1 GJ}$ \cite{timoshenko2009theory}, where $m$ is a positive integer that determines the number of sine waves. The buckling moment leads to an intersection angle $m \pi \sqrt{\frac{a}{b^2}}$ in our bigon. Here we plot this analytical prediction with $m=1$ and 2 in Figure \ref{fig:BigonAnisotropy}(d) (orange curves), which corresponds to the first and second buckling mode of a bigon. The comparison between the numerical results and the approximate analytical prediction shows that for the $B_1$ curve ($m=1$), the analytical result is accurate only with $w/t>>1$, and deviates quickly from the numerical prediction with $w/t<1.5$, where the beam is deformed into a deep circular arc before buckling occurs. On the other hand, the analytical prediction of the $B_2$ curve matches closely with the numerical prediction.
	
	The numerical results in Figure \ref{fig:BigonAnisotropy} restrict the intersection angle $\gamma$ to $[0^{\circ},180^{\circ}]$. For $\gamma > 180^{\circ}$, the planar branch with two circular arcs continues to exist and could be stable for $w/t <1$, which is not of interest for the construction of bigon rings. Actually, the planar solution with $\gamma> 180^{\circ}$ can be obtained by flipping inside out the two ends of a planar bigon with an intersection angle $(2 \pi - \gamma)$. The flipped state is characterized by a planar bilobate shape that becomes a creased thin strip for $w/t <<1$. Details are included in Appendix \ref{app:bilobatebigon}.  
	
	Figure \ref{fig:Bigontuakappaenergy} shows curvatures, twist, and normalized energy density $\bar{\varepsilon}$ of the numerical solutions in Figure \ref{fig:BigonAnisotropy}(c), with $\gamma = 90^{\circ}$. The length of the strip is normalized to unity. Because of the symmetry, only solutions of the first strip with $z_1 \! \geq \! 0$ is shown, corresponding to $\kappa_{11}$, $\kappa_{21}$, $\tau_1$, and $\bar{\varepsilon}=0.5 (a_{1} \kappa_{11} ^2 +b_{1} \kappa_{21} ^2 +\tau_1^2) $.
	In order to compare the energy density along the arc length with different anisotropy $w/t$, we fix the thickness $t$ and normalize the torsional rigidity of $w/t=2$ to unity, matching with the normalization in Figure \ref{fig:BigonAnisotropy}(b). 
	In general, increasing anisotropy $w/t$ makes a bigon buckle out of plane and creates nonvanishing $\kappa_{11}$ and $\tau_1$. Increasing $w/t$ also decreases the planar bending curvature $\kappa_{21}$; with $w/t=8$, $\kappa_{21}$ is almost penalized to zero everywhere.
	With $w/t=1.01$ and 1.2, $\kappa_{21}$ is constant along the arc length, matching with the conclusion from the symmetry analysis that a planar bigon is deformed uniformly into two circular arcs. With $w/t=1.5, 2$, and 8, the bigon buckles out plane, and $\kappa_{21}$ varies very slowly along the arc length. 
	In addition, the normalized energy density $\bar{\varepsilon}$ is constant along the whole arc length for all $w/t$. This matches with a known fact that for an anisotropic rod subject to end loadings, $(0.5 \bm{M} \cdot \bm{\omega} +\bm{N} \cdot \bm{d}_3)$ is conserved along the arc length \cite{van2000helical}. 
	For a bigon subject to pure moments, the second term drops out and thus $0.5 \bm{M} \cdot \omega$ is conserved, which represents exactly the elastic energy density.

	\section{Numerical results of a bigon ring} \label{se:bigonringresults}
	
	\subsection{Modeling of a bigon ring} \label{sse:modelbigonring}
	
\begin{figure}[h!]
		\centering
		\includegraphics[width=0.7\textwidth]{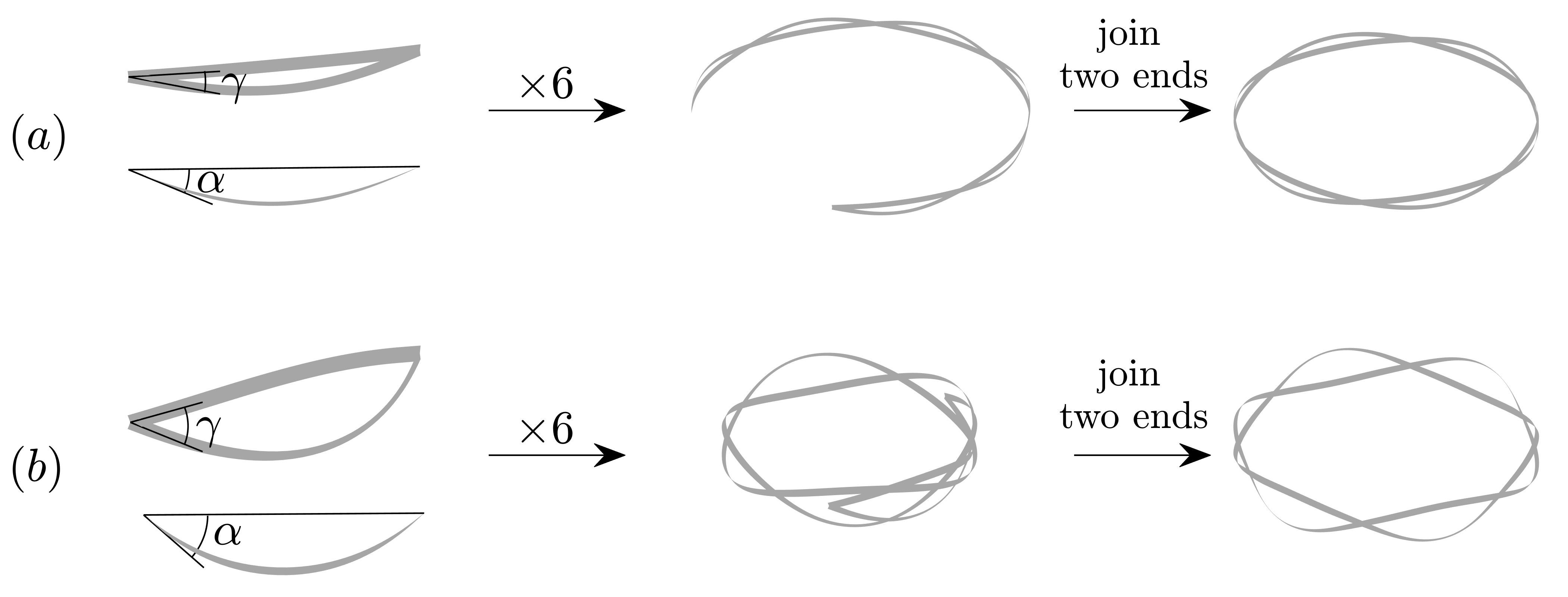}
		\caption{Construction of a 6-bigon ring with all the bigons bending inward. (a) With a small intersection angle $\gamma$, the bigon chain (middle) bends less than $2 \pi$ and needs to be closed to form a loop, leading to an overcurved bigon ring (right). (b) With a large intersection angle $\gamma$, the bigon chain (middle) bends more than $2 \pi$ and needs to be opened to make a loop, leading to an undercurved bigon ring  (right).}
		\label{fig:BigonRingClose}
	\end{figure}

\begin{figure}[h!]
		\centering
		\includegraphics[width=0.8\textwidth]{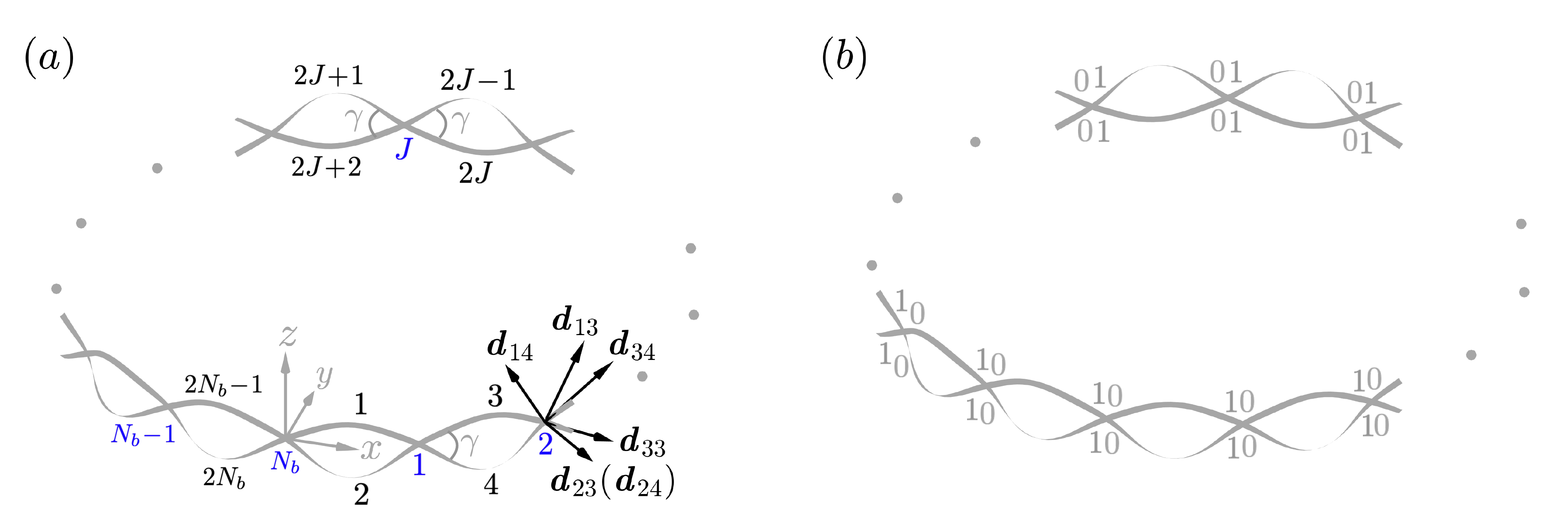}
		\caption{A bigon ring with $N_b$ bigons of the same intersection angle, $N_b$ nodes, and $2N_b$ strips. (a) Nodes are numbered (blue) in a counter clockwise direction, with the node $N_b$ fixed at the origin of a Cartesian coordinate $(x,y,z)$. All the other nodes can move and rotate freely in space. The strips are numbered (black) in a counter clockwise and alternating way. At a node, the four strips share a normal $\bm{d}_2$ and make two tangents $\bm{d}_3$. (b) An assignment of ``0" and ``1" ends to the bigon ring in (a).}
		\label{fig:bigonRingKinematics}
	\end{figure}

	To make a \emph{bigon ring}, we first connect $N_b$ bigons with the same intersection angle in series to make a \emph{bigon chain} with two free ends, shown in the middle column of Figure \ref{fig:BigonRingClose}. Then we force the two ends to meet by either opening or closing the chain, creating a ring with all bigon cells facing inward (the right column of Figure \ref{fig:BigonRingClose}). At a node, the four strips are rigidly connected, sharing a surface normal and making two continuous tangents (Figure \ref{fig:bigonRingKinematics}(a)). The degree of forcing depends on the intersection angle $\gamma$ and the number of bigons $N_b$. 
	In general, a small intersection angle with fewer bigons leads to a chain that forms an incomplete ring that needs to be closed to make a single loop (Figure \ref{fig:BigonRingClose}(a)). Likewise, a large intersection angle and a large number bigons makes a coil-like multiply-covered ring that needs to be opened to form a single loop (Figure \ref{fig:BigonRingClose}(b)). 
	The number of loops covered by the bigon chain is defined as \emph{overcurvature} and can be calculated as $N_b \alpha/180$. Here $\alpha$ is the tangent angle of a bigon, and is related to the intersection angle of a bigon $\gamma$ through Figure \ref{fig:BigonAnisotropy}(c). With $N_b \alpha /180< 1$, a bigon chain covers less than one loop and forcing the chain to close will overbend it, leading to an \emph{overcurved} bigon ring. On the contrary, with $N_b \alpha /180>1$, the bigon chain covers more than one loop and forcing the chain to close leads to an \emph{undercurved} bigon ring. 
	Figure \ref{fig:bigonmodel}(g) shows an undercurved 6-bigon ring with an intersection angle of $60^{\circ}$, which has $N_b \alpha /180  \! \approx \! 1.74$.
	
	The overcurvature in a bigon ring can be tuned by varying the intersection angle $\gamma$ - which varies the tangent angle $\alpha$ - and the number of bigons $N_b$. Another factor that influences the overcurvature of a bigon ring is the anisotropy of the component strip's cross section $w/t$, which is related to the tangent angle $\alpha$ through Figure \ref{fig:BigonAnisotropy}(c).  
	In experiments, we observed that a 6-bigon ring with a large $\gamma$ can fold into stable three loops. Similar folding behaviors are reported in a closed and naturally curved strip \cite{audoly2015buckling,manning2001stability}. Note that the rest shape of the naturally curved strip is stress-free, while the ``rest shape" of a bigon ring, i.e., the bigon chain, has internal stresses inside each bigon cell.

	We use the Kirchhoff rod model to formulate a bigon ring as a TPBVP. Figure \ref{fig:bigonRingKinematics} shows schematics of a bigon ring with $N_b$ bigons of the same intersection angle $\gamma$, a numbering of nodes ($\in [1,N_b]$, blue numbers), a numbering of strips ($\in [1,2N_b]$, black numbers), and an assignment of ``0" and ``1" ends. Other choices of ``0" and ``1" ends are possible, and the only requirement is that we must have a ``0" and ``1" end for each of the strip. The node $N_b$ is fixed at the origin of a Cartesian coordinate. In numerical continuation, we vary the intersection angle $\gamma$ from $0^{\circ}$ to $180^{\circ}$. 
	Based on the discussion in Section \ref{se: well-posed joints}, the unknowns include $28N_b$ variables from Equation \eqref{eq:normF&M}, and $3(N_b-1)$ Euler angles describing the orientations of the ($N_b-1$) free nodes. In total we have $(31N_b-3)$ unknowns. At the clamped end, we have 28 boundary conditions (i.e., $7n$ with $n=4$); at each free node, we have 31 boundary conditions (i.e., $7n+3$ with $n=4$). In total, we have $(31N_b-3)$ boundary conditions, resulting in a well-posed TPBVP. At $\gamma = 0^{\circ}$, the bigon ring degenerates to a doubly-covered loop. We use it as a start solution for numerical continuation.

	\subsection{Folding behavior of a 15-bigon ring} \label{sse:folding15bigonring}

\begin{figure}[h!]
		\centering
		\includegraphics[width=0.8\textwidth]{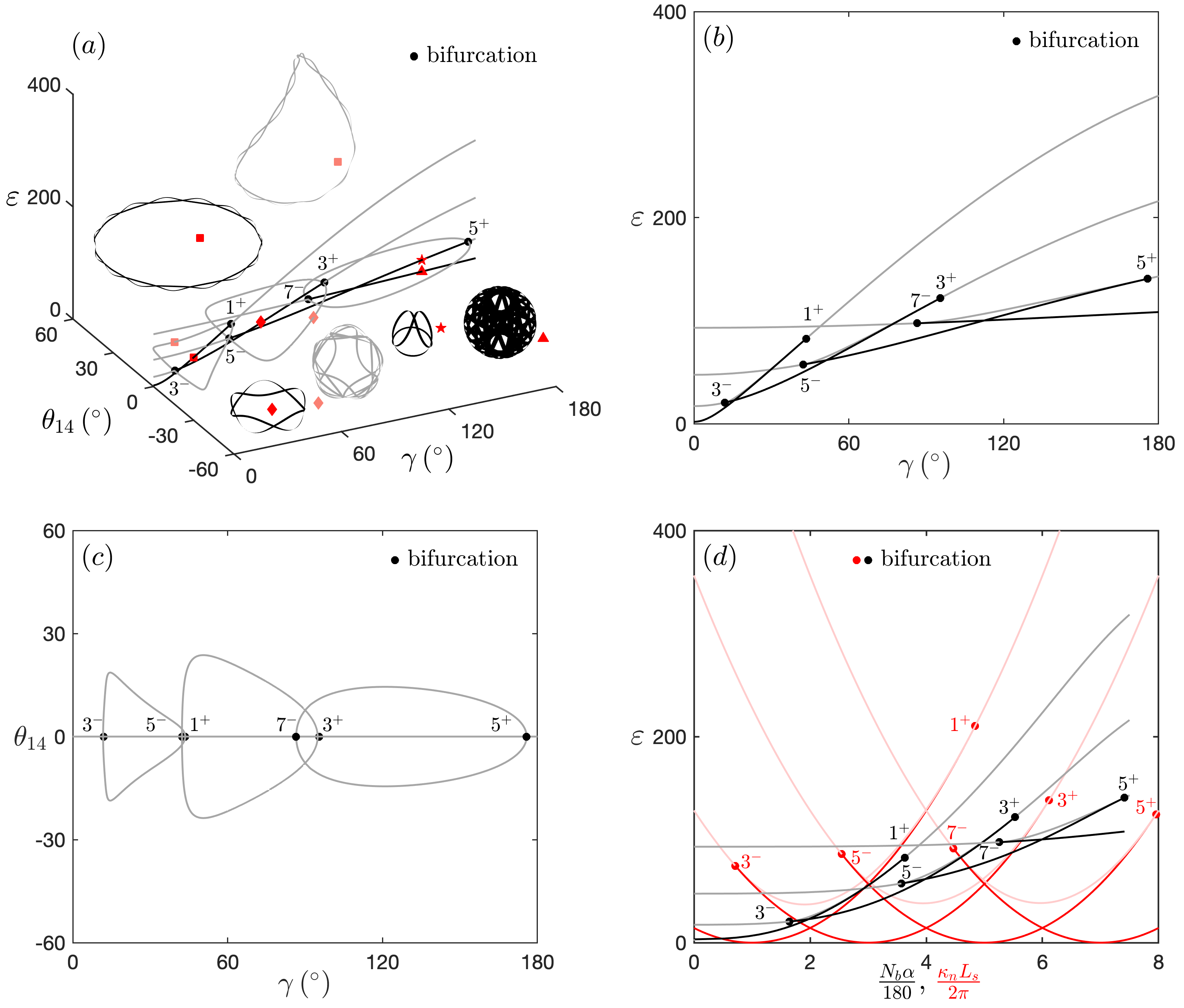}
		\caption{Comparison between the folding behavior of a 15-bigon ring and the folding behavior of a single curved strip with $w/t=8$. Solutions bounded by various bifurcations (black and red circles) are highlighted as black and red; others are colored as gray and light red. (a) Increasing the intersection angle $\gamma$ gradually folds a 15-bigon ring into three loops ($\reddiamond$), five loops ($\redstar$), and seven loops ($\redtriangle$). (b)-(c) Projections of the solutions in (a) on the $\gamma-\varepsilon$ and $\gamma-\theta_{14}$ planes. (d) The solutions in (a) presented in the $\varepsilon - N_b \alpha /180$ planes. Also shown is the folding behavior of a single curved strip with overcurvature $\tfrac{\kappa_n L_s}{2 \pi}$.}
		\label{fig:bigonRinglooping}
	\end{figure}

	Figure \ref{fig:bigonRinglooping}(a) shows the folding behavior of a 15-bigon ring with $w/t = 8$ in the $\gamma - \theta_{14} - \varepsilon$ space. $\gamma$ corresponds to the intersection angle of each constituent bigon and is implemented as a bifurcation parameter in numerical continuation. $\theta_{14}$ and $\varepsilon$ represent the second Euler angle at the $14_{\textrm{th}}$ node  (corresponding to the node $(N_b -1)$ in Figure \ref{fig:bigonRingKinematics}(a)) and the total normalized elastic energy $0.5 \sum_{i=1}^{30} \int _0 ^{1} (a_{i} \kappa_{1i} ^2 +b_{i} \kappa_{2i} ^2 +\tau_i^2) \, d s_i  $, respectively. The solution curves presented in this study are symmetric about the plane $\theta_{14} = 0^{\circ}$. A nonvanishing $\theta_{14}$ implies out-of-plane deformations, i.e., displacements in the $z$ direction. Figures \ref{fig:bigonRinglooping}(b-c) are the projections of the 3D solution curves in Figure \ref{fig:bigonRinglooping}(a) on the $\gamma - \epsilon$ and $\gamma - \theta_{14}$ plane, respectively. In this section, the 3D solution curves of bigon rings are presented together with their projections for clarity; local enlargements are also included, where necessary. The multiply-covered branches (include the single loop) bounded by various bifurcations are shown in black. The number $m^{\pm}$ represents the critical point where a bifurcation occurs on an $m-$covered loop. For example, increasing $\gamma$ folds a three-loop configuration into a five-loop configuration through bifurcation $3^{+}$, while decreasing $\gamma$ unfolds a three-loop configuration into a single loop through bifurcation $3^{-}$. This is similar to the folding behavior of a closed strip with rest curvature \cite{audoly2015buckling,manning2001stability}.

	In Figures \ref{fig:bigonRinglooping}(a-c), we start from a 15-bigon ring with all its bigon cells bending inward, and gradually increase its intersection angle $\gamma$. 
	In numerical continuation, AUTO fails to capture the pitchfork bifurcation that destabilizes the bigon ring through out-of-plane folding. A similar failure is reported when investigating the bifurcations of a Kirchhoff rod with intrinsic curvature \cite{manning2001stability}. 
	In order to track the nonplanar branch, we add a small gravity term (aligned with the negative $z$ direction) to break the pitchfork bifurcation. After getting on the nonplanar branch, we ``turn off" the gravity and continue tracing the solution curves. By increasing $\gamma$, the single loop branch of a 15-bigon ring ($\redsquare$) connects to a pair of branches with out-of-plane deformations through bifurcation point $1^{+}$ at $\gamma = 43.4^{\circ}$. Numerical continuation on the bifurcated branches ($\lightsquare$) reveals a bifurcation point $3^{-}$ at $\gamma = 12.0^{\circ}$, connecting to a planar three-loop branch ($\reddiamond$). The three-loop branch connects to a pair of branches through bifurcation point $3^{+}$ at $\gamma = 95.4^{\circ}$. 
	Numerical continuation on the bifurcated branches ($\lightdiamond$) reveals a bifurcation point $5^{-}$ at $\gamma=42.3^{\circ}$, connecting to a planar five-loop branch ($\redstar$). The five-loop branch connects to a pair of branches through bifurcation point $5^{+}$ at $\gamma = 175.8^{\circ}$. 
	Numerical continuation on the bifurcated branches reveals a bifurcation point $7^{-}$ at $\gamma=86.52^{\circ}$, connecting to a seven-loop branch ($\redtriangle$) that can be continued to $\gamma=180^{\circ}$ without additional bifurcations.

	\begin{figure}[h!]
		\centering
		\includegraphics[width=0.875\textwidth]{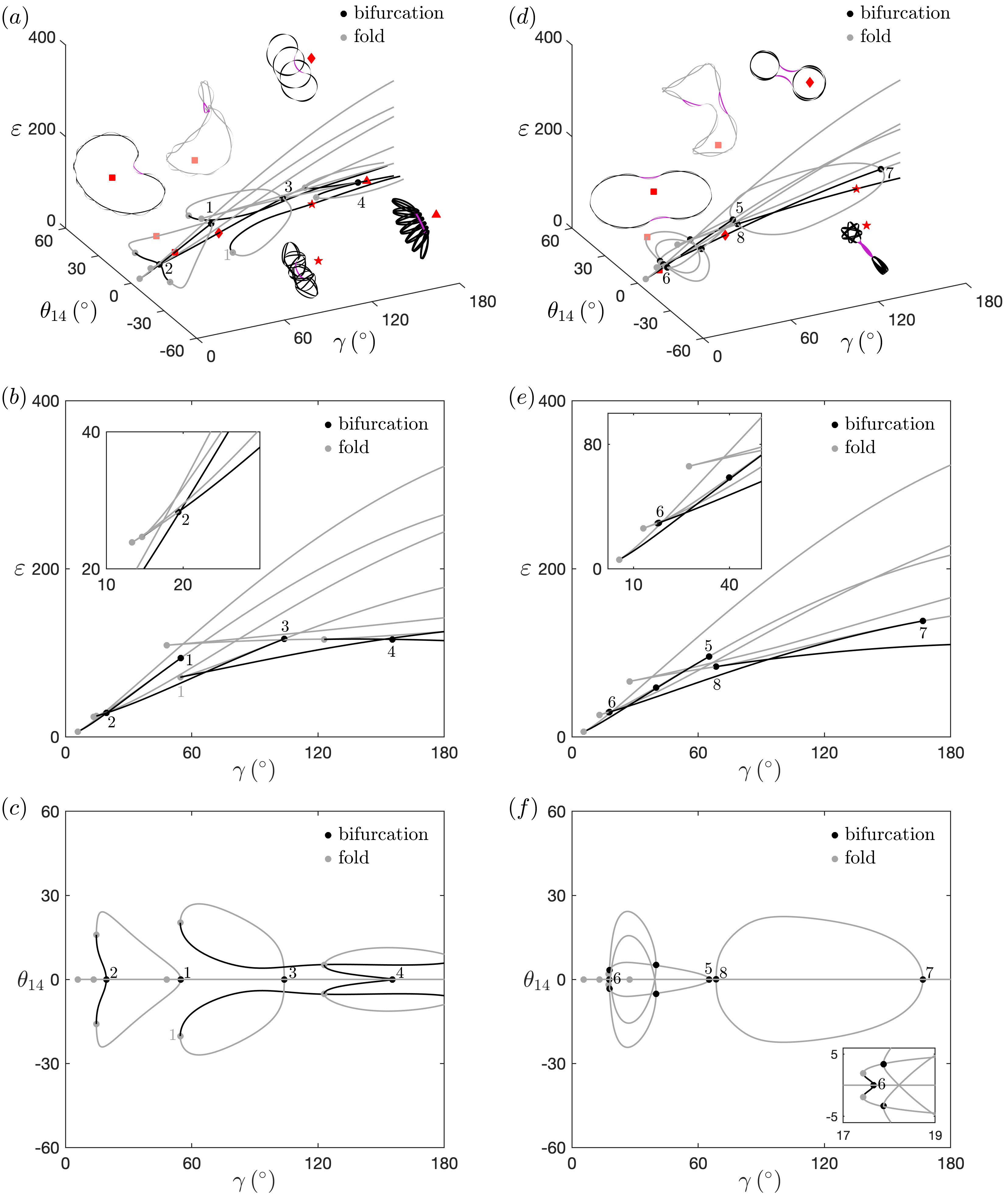}
		\caption{Additional folding behaviors of a 15-bigon ring with several bigon cells bending outward (highlighted as purple). (a) With one bigon cell bending outward ($\redsquare$), the bigon ring gradually folds into three offset loops ($\reddiamond$), a complicated shape ($\redstar$), and seven offset loops ($\redtriangle$). (b)-(c) Projections of the solutions in (a) on the $\gamma-\varepsilon$ and $\gamma-\theta_{14}$ planes. (d) With two bigon cells bending outward ($\redsquare$), the bigon ring gradually folds into two offset loops connected by the two purple cells ($\reddiamond$), and two parts connected by the two purple cells ($\redstar$). (e)-(f) Projections of the solutions in (d) on the $\gamma-\varepsilon$ and $\gamma-\theta_{14}$ planes.}
		\label{fig:bigonRingloopingmore}
	\end{figure}

	Figure \ref{fig:bigonRinglooping}(d) presents the numerical results of Figure \ref{fig:bigonRinglooping}(a) in the plane spanned by the elastic energy $\varepsilon$ and overcurvature $N_b \alpha /180$ (black and gray curves), compared with the folding behavior of a single strip with rest curvature $\kappa_n$, length $L_s$, and $w/t = 8$ (red and light red). Note that $N_b \alpha /180=0$ with $\gamma \le 3.3^{\circ}$, where a bigon stays stably in plane ($3.3^{\circ}$ corresponds the critical angle at the bifurcation in Figure \ref{fig:BigonAnisotropy}c with $w/t=8$). The elastic energy of a closed curved strip is $\varepsilon=0.5 \int _0 ^{1} (a (\kappa_{1}-\kappa_{n}) ^2 +b \kappa_{2} ^2 +\tau^2) \, d s  $; its overcurvature is $\kappa_n L_s /(2 \pi)$. In numerics, we normalized the length $L_s$ to unity. Note that the energy $\varepsilon$ scales as $L_s^{-1}$. 
	The critical overcurvatures of a single curved strip at $m^{\pm}$ obtained here through numerical continuation match exactly with the theoretical prediction in \cite{manning2001stability}. It is interesting that the critical overcurvature of destabilizing a single strip $\kappa_{n} L_s/(2 \pi)=4.84$ (red $1^{+}$) is larger than the critical overcurvature of destabilizing a 15-bigon ring $N_b \alpha/ 180=3.63$ (black $1^{+}$), even though for a bigon ring, the out-of-plane ``height" created by a nonvanishing intersection angle tends to increase its out-of-plane stiffness. This is also true for the critical 
	points $3^{+}$ and $5^{+}$. On the other hand, bifurcations $3^{-}$, $5^{-}$, $7^{-}$ of a single curved strip have lower overcurvatures than the corresponding bifurcations of a 15-bigon ring. Note that the energy of a single curved strip periodically touches zero at $\frac{\kappa_n L_s}{2 \pi}=m$ ($=$1, 3, 5 ...), where the rest configuration of the curved strip covers exactly $m$ loops. However, the 15-bigon ring does not have this feature, since its ``rest configuration" of a bigon chain is also stressed. In addition, numerical continuation of a single curved strip describes a strip that folds into an increasing number of loops (i.e., 9, 11 ...), which is not observed in the numerical continuation of a 15-bigon ring due to limitations in its geometry. It appears that the maximal number of loops that a bigon ring can fold into increases with the increase of bigon cells. Even though we propose an overcurvature parameter $N_b \alpha / 180$ that is similar to the overcurvature of a single curved strip \cite{audoly2015buckling,manning2001stability}, it appears that the folding behavior of a bigon ring is more complicated, since the geometry of the planar branch (e.g., the out-of-plane height) changes as we vary the bifurcation parameter $\gamma$. In Section \ref{se:anisotropy}, we will show that the three-loop configuration of a 6-bigon ring can also be stabilized with vanishing overcurvature.

	Another difference between a closed curved strip and a bigon ring is that a bigon ring can have various stable configurations by locally switching a bigon cell between \emph{bending inward} and \emph{bending outward}. Figures \ref{fig:bigonRingloopingmore}(a-c) and \ref{fig:bigonRingloopingmore}(d-f) report the folding behaviors of a 15-bigon ring with one and two bigon cells bending outward (highlighted as purple in the renderings), respectively. The folded branches bounded by various bifurcations and folds are drawn in black and others are colored gray. In Figures \ref{fig:bigonRingloopingmore}(a-c), the single-loop configuration ($\redsquare$) connects to a pair of branches with out-of-plane deformations through a bifurcation 1 at $\gamma = 54.8^{\circ}$. 
	Numerical continuation on the out-of-plane branch ($\lightsquare$) reveals another bifurcation 2 at $\gamma = 19.4^{\circ}$, connecting to a planar branch ($\reddiamond$) that has three offset loops. The three offset loops are connected by the purple cell. The branch with three offset loops connects to another pair of branches with out-of-plane deformations through a bifurcation 3 at $\gamma = 103.9^{\circ}$. 
	Numerical continuation on the out-of-plane branch connects to two folds (one of them is numbered as \textcolor{gray}{1}) at $\gamma = 54.5^{\circ}$, which further connects to a pair of symmetric branches (one of them is marked with $\redstar$). The configuration ($\redstar$) is folded into a complicated structure, where the purple cell is popping out of plane.
	In addition, another branch ($\redtriangle$) with seven offset loops (each loop has two bigons) connected by the purple bigon, appears through a bifurcation 4 at $\gamma = 155.4^{\circ}$.

	Figures \ref{fig:bigonRingloopingmore}(d-f) display the looping behavior of a 15-bigon ring with two cells bending outward (highlighted as purple). The two cells divide the rest of the bigon ring into two parts, containing 7 and 6 cells, respectively ($\redsquare$). 
	The single-loop configuration ($\redsquare$) connects to a pair of branches with out-of-plane deformations through a bifurcation 5 at $\gamma = 65.2^{\circ}$. Numerical continuation on the out-of-plane branch ($\lightsquare$) reveals another bifurcation 6 at $\gamma = 17.7^{\circ}$, connecting to a planar branch ($\reddiamond$) that is folded into two offset loops connected by the two purple bigon cells. The smaller loop has six bigons and the bigger loop contains seven bigons. Both of the loops are roughly doubly covered.
	The branch with two offset loops connects to a pair of branches with out-of-plane deformations through a bifurcation 7 at $\gamma = 166.7^{\circ}$. 
	Numerical continuation on the out-of-plane branch reveals another bifurcation 8 at $\gamma = 68.5^{\circ}$, which further connects to a branch ($\redstar$) that can be continued to $180^{\circ}$ without additional bifurcations. The configuration ($\redstar$) is folded to two parts, connected by the two purple cells. The star-like part has seven bigons and the tail has six bigons.  
	
	Further manipulating the orientation of individual cells likely leads to other types of folding behaviors in a 15-bigon ring. For example, we can start with a configuration that has two different cells (from Figure \ref{fig:bigonRingloopingmore}(d)) bending outward. We can also start with a configuration that has various combinations of three cells bending outward. What we present in Figures (\ref{fig:bigonRinglooping}-\ref{fig:bigonRingloopingmore}) are just a few examples of folding behaviors in a 15-bigon ring that all share a common feature, namely that increasing the intersection angle $\gamma$ will fold the bigon ring into multiply-covered loops or several parts, slowing down the increase of the elastic energy. This can be seen from Figures \ref{fig:bigonRinglooping}(b), \ref{fig:bigonRingloopingmore}(b), and \ref{fig:bigonRingloopingmore}(e), where the black lines (almost straight) keep decreasing their slopes as we increase the intersection angle. Note that some of the folded configurations have complicated shapes usually involving multiple self-contacts (such as the two $\redstar$ configurations in Figure \ref{fig:bigonRingloopingmore}) and thus could be difficult to obtain in experiments. Additionally, our numerical results show that a bigon ring has folding behavior even though the total number of bigons is not exactly divisible by three, five, seven, etc., for example, a 13-bigon ring.

	\subsection{Equilibria and bifurcations of a 6-bigon ring} \label{sse:folding6bigonring}

\begin{figure}[h!]
		\centering
		\includegraphics[width=0.8\textwidth]{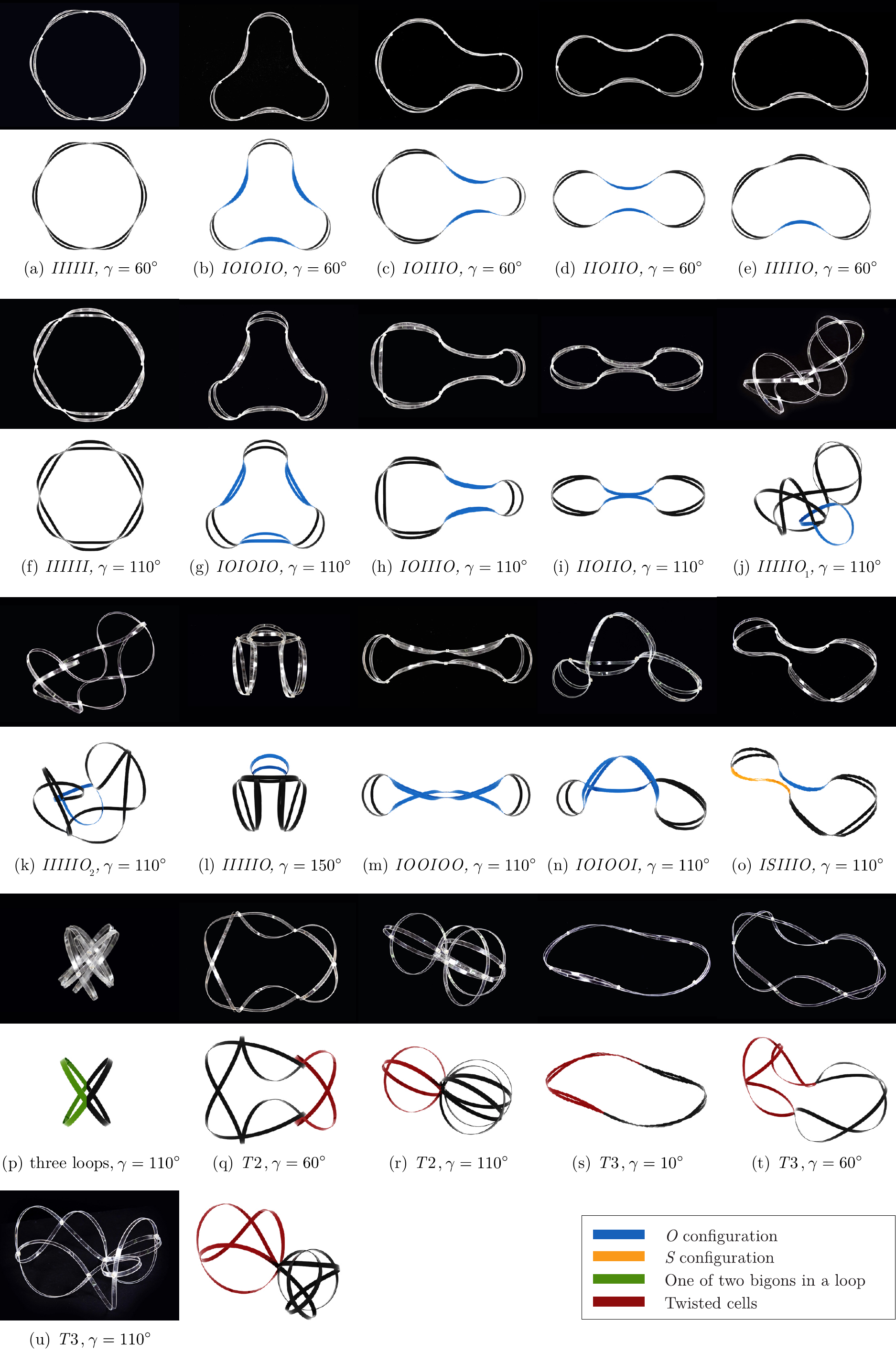}
		\caption{Comparison between experimental configurations of a 6-bigon ring and renderings of numerical results ($w/t=8$). (a)-(o) contain different combinations of $I$s, $O$s, and $S$s, which are colored as black, blue, and yellow, respectively. (p) is obtained by folding a 6-bigon ring into three loops, with each loop containing two bigons (one colored as green and the other as black). (q)-(r) are obtained by twisting two consecutive bigon cells (dark red) out of plane. (s)-(u) are obtained by twisting three consecutive bigon cells (dark red) out of plane.}
		\label{fig:ExpwithNu}
	\end{figure}

	The number of stable states in a bigon ring increases quickly with the increase of bigon cells. For a bigon ring with a large number of bigon cells, it could be hard to identify all the stable equilibria in experiments and numerics, and could require a high computational cost. In this section, we restrict our study to a 6-bigon ring and try to identify all the stable states. 
	Figure \ref{fig:ExpwithNu} shows experimental configurations of a 6-bigon ring with $l = 200$ mm, $w = 8$ mm, and $t = 1$ mm, compared with renderings of numerical results with $w/t=8$. 
	These states are generally stabilized by a nonvanishing intersection angle $\gamma$ and most of them cannot be obtained with $\gamma=0^{\circ}$, which leads to a monostable circular ring.
	In numerical continuation, we normally start with $\gamma=0^{\circ}$ (corresponding to a circle) and apply moments at the nodes to smoothly deform the circular ring into shapes with various combinations of $I$, $O$, and $S$ cells. Then we increase $\gamma$ to some finite angle, and finally remove the moments to obtain ``free-standing" configurations.

	Some of the configurations are labeled as a combination of $I$s, $O$s, and $S$s, where $I$ represents a bigon cell that bends inward, $O$ represents a bigon cell that bends outward, and $S$ represents the $S$ mode of a bigon. For example, $IIIIII$ represents a configuration with all six bigons bending inward. In our notation, the shape does not depend on which cell/direction (i.e., counter clockwise or clockwise) we start labeling with. For example, $IIIIIO$ represents the same shape with $IIOIII$, $IIIIOI$, etc., and $IIOIIO$ represents the same shape with $OIIOII$, $IOIIOI$, etc. $IIIIIO_1$ and $IIIIIO_2$ at $\gamma=110^{\circ}$ are two symmetric configurations that are similar to $IIIIIO$ at $\gamma=60^{\circ}$, except that the $O$ cell at $\gamma=110^{\circ}$ pops out of plane and makes a symmetric pair. Note that the $O$ cell goes back to be planar again at $\gamma=150^{\circ}$. The pair in Figures \ref{fig:ExpwithNu}(j-k) are rotated from $IIIIIO$ in Figure \ref{fig:ExpwithNu}(e) to reveal the out-of-plane deformation of the $O$ cell. 
	
	In order to obtain contact-free $IOIOOI$ (see Figure \ref{fig:ExpwithNu}(n)) in experimental models, we first disassembled the uppermost node, let the two adjacent $O$ cells bypass the third $O$ cell, and then rebuilt the node on the other side. Through this way, ``penetration" is allowed in experiments and the comparison between experimental configuration and numerical rendering is possible. The above operation also makes $IOIOOI$ free of contact between different strips at $\gamma=110^{\circ}$.
	In the supplementary video \emph{IOSpattern.mp4}, the $IOIOOI$ configuration is obtained directly from $IIIIII$ without disassembling, and thus contact can be seen.        
	The labeling of other configurations is based on either their shapes or how they are obtained from $IIIIII$. For example, the three-loop configuration in Figure \ref{fig:ExpwithNu}(p) is obtained by folding a 6-bigon ring into a triply-covered loop, and $T2$ in Figures \ref{fig:ExpwithNu}(q-r) and $T3$ in Figures \ref{fig:ExpwithNu}(s-u) are obtained by twisting two and three bigon cells out of plane, respectively. Each loop in Figure \ref{fig:ExpwithNu}(p) is spherical and contains two bigons, one colored as green and the other as black. More details about the spherical property of the three-loop configuration will be discussed later. 
	Note that the states in Figures \ref{fig:ExpwithNu}(q-r) could be labeled separately, since the two $T2$ states have different geometries and different symmetries, which are also true for the three $T3$ states in Figures \ref{fig:ExpwithNu}(s-u). These symmetries will be further discussed together with the numerical solutions later. In this study, we did not adopt a fine classification for these states and simply labeled them based on the number of bigon cells being twisted out of plane.
	The way to obtain various stable shapes and their dependence on intersection angle $\gamma$ are qualitatively shown in the \emph{supplementary videos}.
	
	There may exist other stable shapes that are not identified here. For example, we observed that deforming some $O$s of $IOOIOO$ (Figure \ref{fig:ExpwithNu}(m)) into $S$s seems to create additional stable configurations that are sensitive to viscous and plastic deformations, and thus cannot be obtained consistently in a time span of several days. We did not include these configurations in this study.

\begin{figure}[h!]
		\centering
		\includegraphics[width=0.95\textwidth]{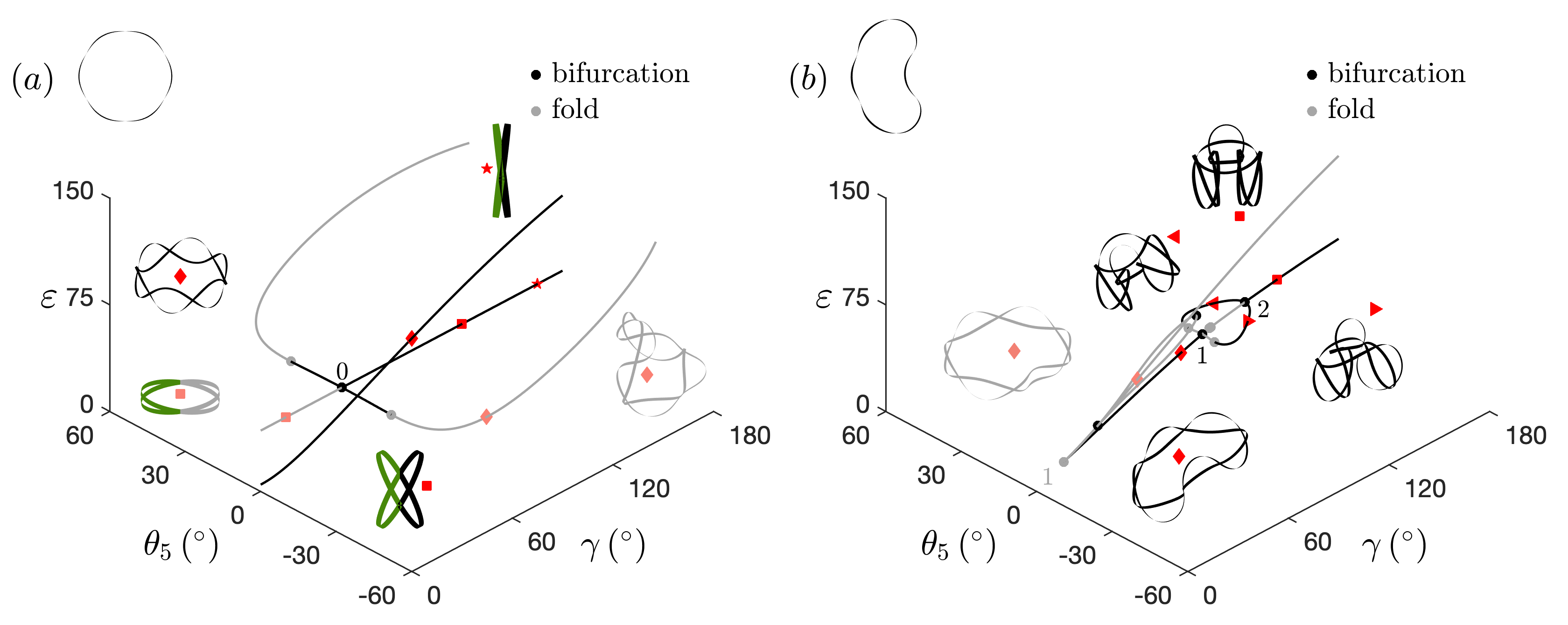}
		\caption{Numerical solutions of a 6-bigon ring with $w/t=8$. (a) $IIIIII$ ($\reddiamond$) is stable in the whole range $0^{\circ} \le \gamma \le 180^{\circ}$, and the three-loop branch ($\redsquare$) gains stability through a shallow supercritical pitchfork bifurcation 0. (b) Decreasing $\gamma$ destabilizes $IIIIIO$ ($\reddiamond$) through a fold \textcolor{gray}{1}, and increasing $\gamma$ destabilizes $IIIIIO$ through a subcritical pitchfork bifurcation 1. Further increasing $\gamma$ re-stabilizes $IIIIIO$ ($\redsquare$) through a supercritical pitchfork bifurcation 2, which connects to a pair of out-of-plane shapes ($\redtriangleleft$ and $\redtriangleright$).}
		\label{fig:6setslooping} 
	\end{figure}

\begin{figure}[h!]
	\centering
	\includegraphics[width=0.95\textwidth]{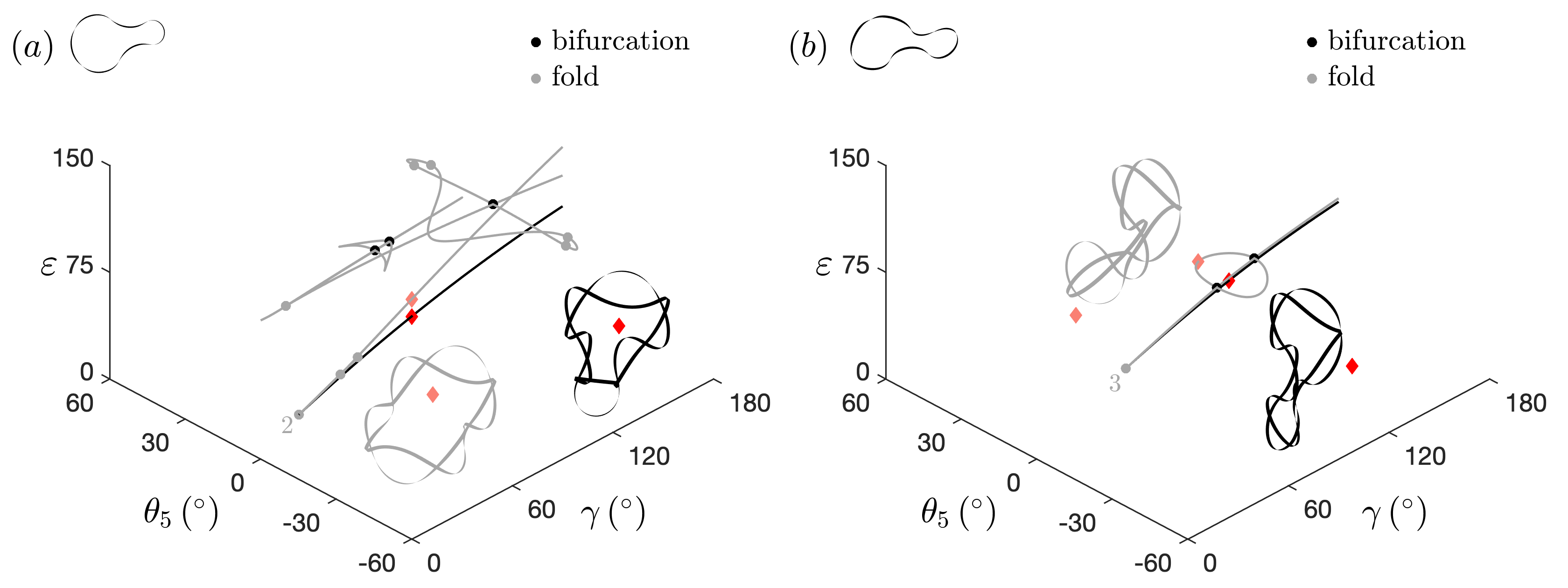}
	\caption{Numerical solutions of a 6-bigon ring with $w/t=8$. (a) $IOIIIO$ ($\reddiamond$) gains stability through fold \textcolor{gray}{2}. (b) $ISIIIO$ ($\reddiamond$) gains stability through fold \textcolor{gray}{3}.}
	\label{fig:6setstwoOs}
\end{figure}

\begin{figure}[h!]
		\centering
		\includegraphics[width=0.95\textwidth]{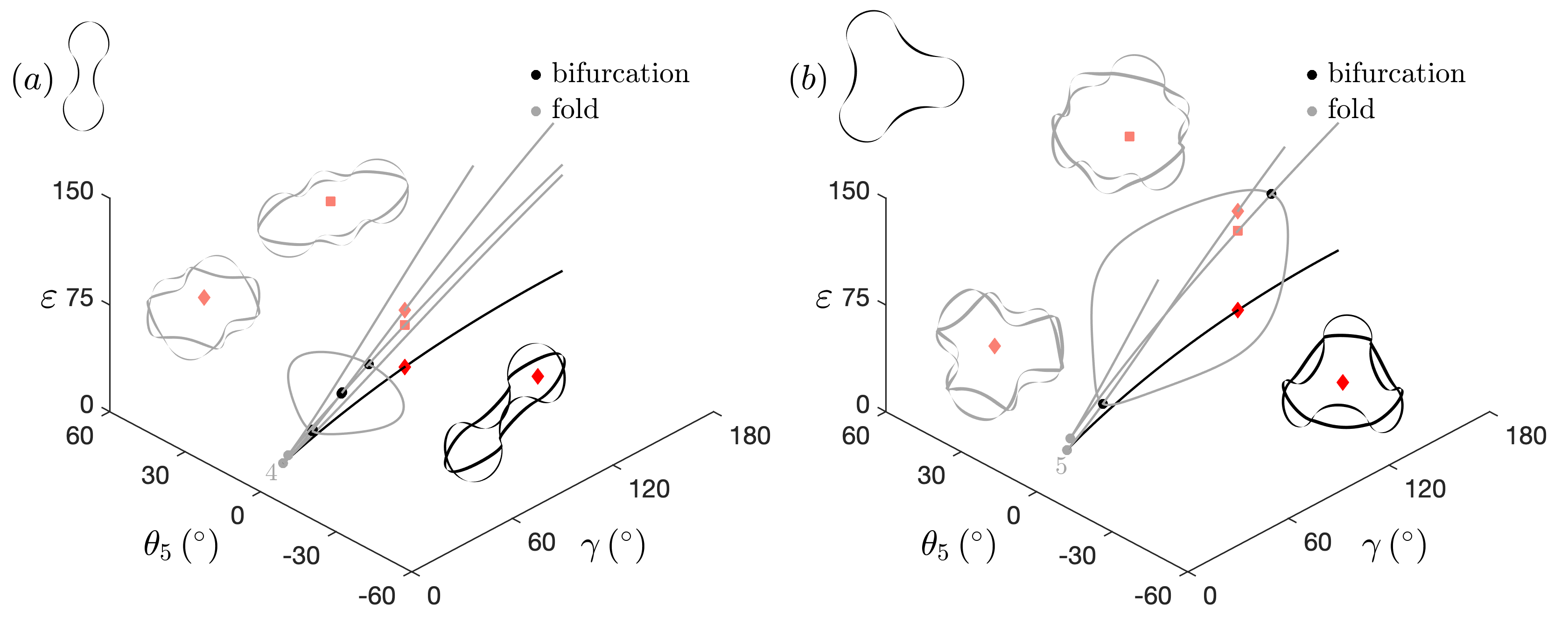}
		\caption{Numerical solutions of a 6-bigon ring with $w/t=8$. (a) $IIOIIO$ ($\reddiamond$) gains stability through fold \textcolor{gray}{4}. (b) $IOIOIO$ ($\reddiamond$) gains stability through fold \textcolor{gray}{5}.}
		\label{fig:6sets2and3Os}
	\end{figure}

\begin{figure}[h!]
		\centering
		\includegraphics[width=0.95\textwidth]{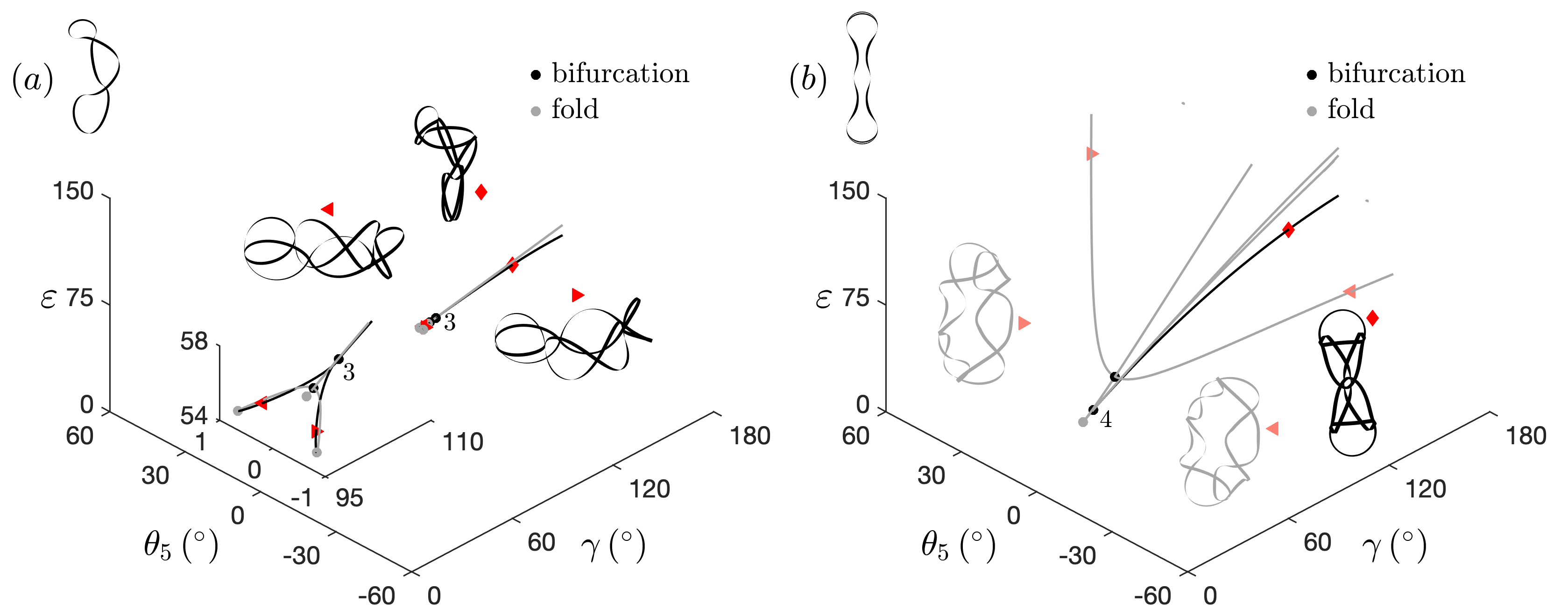}
		\caption{Numerical solutions of a 6-bigon ring with $w/t=8$. (a) $IOIOOI$ ($\reddiamond$) gains stability through a shallow supercritical pitchfork 3. (b) $IOOIOO$ ($\reddiamond$) gains stability through a subcritical pitchfork 4, which connects to a fold that further connects to a pair of unstable states ($\lighttriangleleft$ and $\lighttriangleright$).}
		\label{fig:6sets4and3Os}
	\end{figure}

\begin{figure}[h!!]
		\centering
		\includegraphics[width=0.95\textwidth]{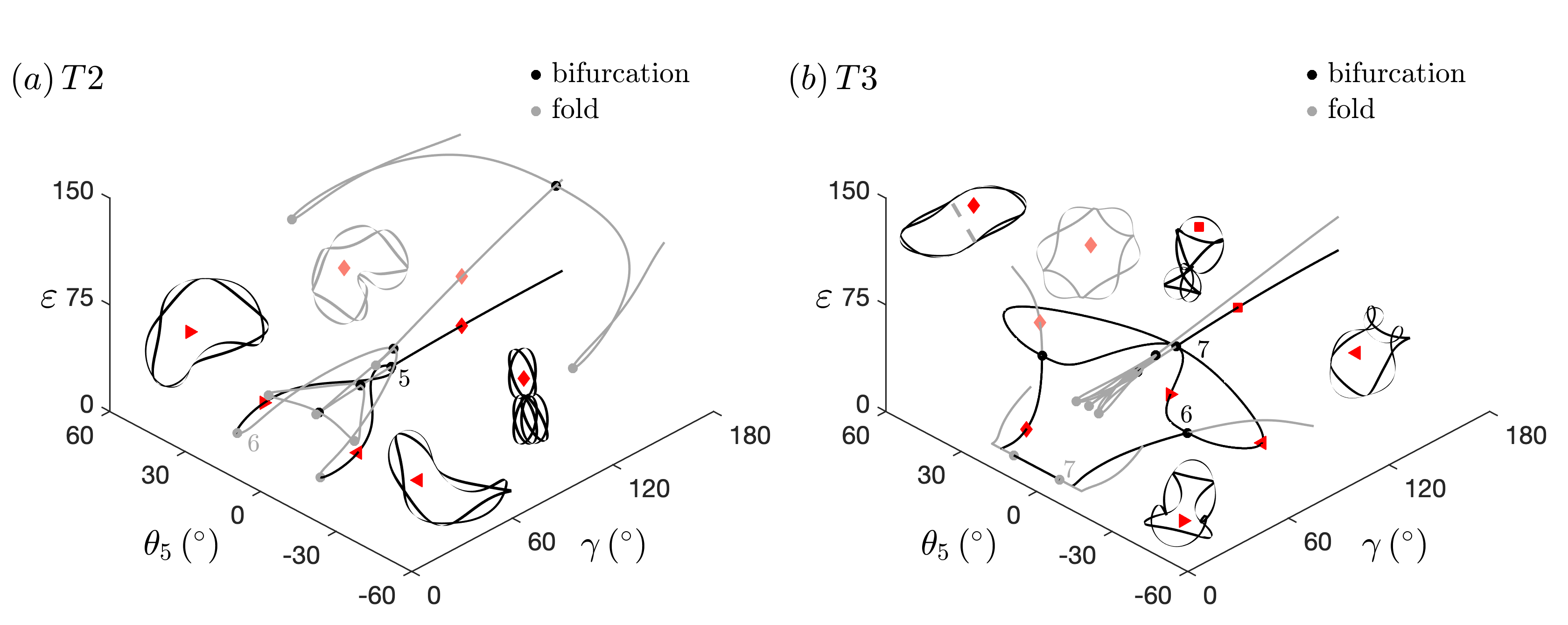}
		\caption{Numerical solutions of a 6-bigon ring with $w/t=8$. (a) $T2$: A pair of states ($\redtriangleleft$ and $\redtriangleright$) gain stability through two folds (one of them is numbered as \textcolor{gray}{6}) and merge at a supercritical pitchfork 5, which connects to a stable branch ($\reddiamond$). (b) $T3$: A pair of states gain stability through two folds (one of them is numbered as \textcolor{gray}{7}). Increasing $\gamma$ breaks the $C_2$ rotational symmetry of ($\reddiamond$) through a supercritical pitchfork bifurcation 6, creating two pairs of stable branches that are symmetric about $\theta_5=0^{\circ}$ plane.  
			One half of each pair is marked as $\redtriangleleft$ and $\redtriangleright$. Further increasing $\gamma$ makes each pair merge into a new branch ($\redsquare$) through a supercritical pitchfork 7.}
		\label{fig:6setstwisted}
	\end{figure}

	In the rest of this section, we study how the intersection angle $\gamma$ influences the stable states of a 6-bigon ring. Numerical continuation reveals interesting connections between several states. Figures (\ref{fig:6setslooping}-\ref{fig:6setstwisted}) do not include all of the solution curves. We have only presented solutions corresponding to those observed to be stable in experiments and those connected to the stable ones. It is not unexpected that we get many solutions from the anisotropic rod model. Some of the figures contain a schematic on the top left, showing the combination of $I$s, $O$s, and $S$s. Intersection angle $\gamma$ is the bifurcation parameter, $\theta_5$ measures the second Euler angle of the fifth node (see Figure \ref{fig:bigonRingKinematics} for the numbering of the node), and $\varepsilon$ represents the elastic energy. 2D projections of the solution curves in Figures (\ref{fig:6setslooping}-\ref{fig:6setstwisted}) are included in Appendix \ref{appse:2Dprojection}. The solution branches observed to be stable in experiments are highlighted as black; others are colored as gray. The stability of the solutions that directly connect to a stable branch can be inferred by bifurcation types. For example, starting from a stable configuration, a fold and a subcritical pitchfork should destabilize the state, while a supercritical pitchfork creates a pair of stable configurations. We lose track of the stability of the solutions that indirectly connect to the stable branch through several folds/bifurcations. These solutions are colored as gray. In addition, we find that the black curves in Figures (\ref{fig:6setslooping}-\ref{fig:6setstwisted}) generally have a lower elastic energy compared with the gray curves connected to them, which appears to be another implication of stability for the black solution curves.

	Figure \ref{fig:6setslooping}(a) display the solution curves of $IIIIII$. $IIIIII$ ($\reddiamond$) is stable in the whole range $0^{\circ} \le \gamma \le 180^{\circ}$, while the three-loop branch ($\redsquare$) gains stability through a shallow supercritical pitchfork bifurcation 0 at $\gamma = 48.46^{\circ}$, corresponding to an overcurvature $N_b \alpha/180 = 1.54$. 
	The supplementary video \emph{twistingandfolding.mp4} confirms that the three-loop configuration is unstable with a small intersection angle $\gamma=10^{\circ}$. 
	Unlike the three-loop branch of a 15-bigon ring that keeps folding into five and seven loops (Figure \ref{fig:bigonRinglooping}), the three-loop branch of a 6-bigon ring is stable up to $\gamma=180^{\circ}$ without additional bifurcations. In addition, we find that on the three-loop branch of a 6-bigon ring, all the strips are deformed into circular shapes with $\kappa_{1i}=\pi, \kappa_{2i}=0$, and $\tau_{i}=0$ ($i \in [1,12]$). The elastic energy on the three-loop branch is constant. The three-loop renderings in Figure \ref{fig:ExpwithNu}(p) and Figure \ref{fig:6setslooping}(a) can also be seen as two circular strips bisecting each other. Each circular strip is triply covered and contains a green and a black semicircle. By continuously increasing the intersection angle $\gamma$ from $0$ to $\pi$, the centerline of the two circles sweeps an exact spherical surface of radius $\pi$, i.e., the strips are always spherical. Two spherical renderings in Figure \ref{fig:6setslooping}(a) ($\lightsquare$ and $\redstar$) are chosen such that their intersection angle adds to $\pi$. The single strips that compose these bigon rings overlap perfectly although the two configurations differ for the way each strip is connected to the others. Actually, these configurations unfold into bigon rings with different intersection angles, and have different stability information. The spherical shape of the three-loop configuration is independent of the anisotropy of the cross section $w/t$, which will be briefly addressed in Section \ref{se:anisotropy}.

	Figure \ref{fig:6setslooping}(b) shows the solution curves of $IIIIIO$. $IIIIIO$ ($\reddiamond$) gains stability at $\gamma = 16.00^{\circ}$ through fold \textcolor{gray}{1}. Also connected to this fold is an unstable branch ($\lightdiamond$) with the $O$ cell in its higher modes. Increasing $\gamma$ destabilizes $IIIIIO$ through a shallow subcritical pitchfork 1 at $\gamma = 98.78^{\circ}$, and re-stabilizes it through a supercritical pitchfork 2 at $\gamma = 124.16^{\circ}$. The two pitchforks 1 and 2 share a pair of bifurcated stable states $IIIIIO_1$ and $IIIIIO_2$ ($\redtriangleleft$ and $\redtriangleright$), which could be obtained by pushing the $O$ cell of the unstable $IIIIIO$ out of plane. $IIIIIO_{1,2}$ are stable with $\gamma \in [98.21^{\circ}, 124.16^{\circ}]$, and the lower bound corresponds to a fold (without numbering) on the bifurcated branch that is near bifurcation 1 (see Figure \ref{fig:6setslooping}(b)). Note that in the stable range of $IIIIIO_{1,2}$, $IIIIIO$ can be seen as the energy barrier between $IIIIIO_1$ and $IIIIIO_2$, even though the elastic energy of $IIIIIO$ is just slightly higher than the energy of $IIIIIO_{1,2}$ (see the top inset in Figure \ref{fig:6setslooping2D}(c)). The mild energy barrier could make the stability of $IIIIIO_{1,2}$ sensitive to imperfections or external loads such as gravity.
		For example, in experiments we observed that with $\gamma=110^{\circ}$, gravity could favor one of the $IIIIIO_{1,2}$ states, depending on how the structure is oriented in space. 
		These numerical predictions qualitatively match with the experimental observations in Figures \ref{fig:ExpwithNu}(e) and \ref{fig:ExpwithNu}(j-l), where $IIIIIO$ is observed to be stable at $\gamma=60^{\circ}$ and $150^{\circ}$ and unstable at $\gamma=110^{\circ}$; instead, $IIIIIO_{1,2}$ is stabilized at $\gamma=110^{\circ}$. The supplementary video \emph{IOSpattern.mp4} confirms that $IIIIIO$ cannot be obtained with a small intersection angle $\gamma=10^{\circ}$ and shows the popping-out behavior of the $O$ cell at $\gamma=110^{\circ}$, leading to a pair $IIIIIO_{1,2}$. In the video, the structure is oriented in a way such that gravity influences $IIIIIO_1$ and $IIIIIO_2$ roughly in an equal manner.

	Figure \ref{fig:6setstwoOs}(a) presents the solutions of $IOIIIO$ ($\reddiamond$), which gains stability through a fold \textcolor{gray}{2} at $\gamma = 22.73^{\circ}$ and is stable up to $\gamma=180^{\circ}$. Also connected to fold \textcolor{gray}{2} is an unstable branch ($\lightdiamond$) with the two $O$ cells in a higher mode. Figure \ref{fig:6setstwoOs}(b) shows the solution curves of $ISIIIO$ ($\reddiamond$), which is obtained by deforming one of the $O$ cell in $IOIIIO$ into $S$. $ISIIIO$ gains stability through a fold \textcolor{gray}{3} at $\gamma = 53.11^{\circ}$, and is stable up to $\gamma=180^{\circ}$. The supplementary video \emph{IOSpattern.mp4} confirms that $IOIIIO$ and $ISIIIO$ are unstable with a small intersection angle $\gamma=10^{\circ}$ and are stable at $\gamma=60^{\circ}$ and $110^{\circ}$. These observations qualitatively match with our numerical predictions.

	Figure \ref{fig:6sets2and3Os}(a) shows the solution curves of $IIOIIO$ ($\reddiamond$), which gains stability through fold \textcolor{gray}{4} at $\gamma =13.35 ^{\circ}$ and is stable up to $\gamma=180^{\circ}$. Also connected to this fold is an unstable branch ($\lightdiamond$), which further connects to other solutions through bifurcations and folds. Figure \ref{fig:6sets2and3Os}(b) shows the solution curves of $IOIOIO$ ($\reddiamond$), which gains stability through fold \textcolor{gray}{5} at $\gamma = 18.07 ^{\circ}$ and is stable up to $\gamma=180^{\circ}$. Also connected to this fold is an unstable branch ($\lightdiamond$), which further connects to other solutions through bifurcations and folds. The supplementary video \emph{IOSpattern.mp4} confirms that $IIOIIO$ and $IOIOIO$ cannot be obtained with a small intersection angle $\gamma=10^{\circ}$ and are stable at $\gamma=60^{\circ}$ and $110^{\circ}$, which qualitatively match with our numerical predictions.

	Figure \ref{fig:6sets4and3Os}(a) shows the solution curves of $IOIOOI$ ($\reddiamond$), which gains stability through a supercritical pitchfork 3 at $\gamma = 104.4 ^{\circ}$ and is stable up to $\gamma=180^{\circ}$. Renderings of a pair of states ($\redtriangleleft$ and $\redtriangleright$) that connect to the pitchfork are also shown. 
	Figure \ref{fig:6sets4and3Os}(b) shows the solution curves of $IOOIOO$ ($\reddiamond$), which gains stability through a subcritical pitchfork 4 at $\gamma = 33.8^{\circ}$ and is stable up to $\gamma=180^{\circ}$. Also connected to this pitchfork is a fold that further connects to a pair of unstable branch ($\lighttriangleleft$ and $\lighttriangleright$). Supplementary video \emph{IOSpattern.mp4} confirms that $IOIOOI$ and $IOOIOO$ cannot be obtained with a small intersection angle $\gamma=10^{\circ}$. In the video, $IOOIOO$ at $\gamma=60^{\circ}$ is obtained differently and more gently than at $\gamma=110^{\circ}$, because we observed that $IOOIOO$ at $\gamma=60^{\circ}$ is less stable. In addition, $IOIOOI$ is shown to be unstable at $\gamma=60^{\circ}$ and stable at $\gamma=110^{\circ}$. These experiments qualitatively match with our numerical predictions.

	Figure \ref{fig:6setstwisted}(a) shows the solution curves of $T2$, which gains stability through a fold \textcolor{gray}{6} at $\gamma = 10.8 ^{\circ}$. In experiments, $T2$ can be obtained by twisting two consecutive bigon cells out of plane. The renderings include a symmetric pair ($\redtriangleleft$ and $\redtriangleright$), which have one mirror symmetry and merge into a single branch through a supercritical pitchfork bifurcation 5 at $\gamma = 77.9 ^{\circ}$. With $\gamma > 77.9 ^{\circ}$, the shape on the merged branch ($\reddiamond$) has two mirror symmetries and matches with the experimental configuration in Figure \ref{fig:ExpwithNu}(r). Also shown are several unstable branches and an additional rendering ($\lightdiamond$). The supplementary video \emph{twistingandfolding.mp4} shows that $T2$ is unstable at $\gamma=10^{\circ}$ and stable at $\gamma=60^{\circ}$ and $110^{\circ}$. These observations and the experimental configurations in Figures \ref{fig:ExpwithNu}(q-r) qualitatively match with our numerical predictions.

	Figure \ref{fig:6setstwisted}(b) shows the solution curves of $T3$, which gains stability through a fold \textcolor{gray}{7} at a small intersection angle $\gamma =  0.0017^{\circ}$. 
	Note that $T3$ degenerates to a doubly-covered twisted ring at $\gamma = 0^{\circ}$. In experiments, $T3$ can be obtained by twisting three consecutive bigon cells out of plane, and is observed to be stable with all three intersection angles $\gamma=10^{\circ}$, $60^{\circ}$, and $110^{\circ}$ (supplementary video \emph{twistingandfolding.mp4}).
	The rendering marked by $\reddiamond$ has two symmetries: a $C_2$ rotational symmetry about the gray dashed chord, and a mirror symmetry about the plane that is perpendicular to and bisects the chord. 
	At $\gamma =  46.6^{\circ}$, the $C_2$ rotational symmetry is broken through a supercritical pitchfork bifurcation 6, creating two pairs of stable states that contain four solutions with the same shape up to a rigid motion. Two renderings from one half of each pair ($\redtriangleleft$ and $\redtriangleright$) with $\theta_5 <0$ are shown. The solution with two symmetries become unstable with $\gamma \!>\! 46.6 ^{\circ}$, and a rendering on this unstable branch is shown ($\lightdiamond$). Further increasing $\gamma$ makes each symmetric pair (about $\theta_5=0^{\circ}$ plane) merge into a new branch through a supercritical pitchfork bifurcation 7 at $\gamma = 83.1^{\circ}$. A rendering on the merged branch is shown ($\redsquare$), which has two mirror symmetries. The numerical predictions qualitatively match with the experimental configurations in Figures \ref{fig:ExpwithNu}(s-u) and in the video \emph{twistingandfolding.mp4}.
	
	The evolution of the solution curves in Figure \ref{fig:6setstwisted} confirm our earlier statement about a possible fine classification of $T2$ and $T3$, which is based on experimental shapes. 
		For example, $T2$ in Figure \ref{fig:6setstwisted}(a) could be separated into two different states by the bifurcation point 5 ($\gamma = 77.9 ^{\circ}$), and the solutions with $\gamma < 77.9 ^{\circ}$ and $\gamma > 77.9 ^{\circ}$ correspond to the stable bifurcated branches ($\redtriangleleft$ and $\redtriangleright$) and the stable unbifurcated branch ($\reddiamond$) of a supercritical pitchfork, respectively. Compared to the solutions  on the bifurcated branches, the solutions on the unbifurcated branch reveal one additional mirror symmetry. Similarly, $T3$ in Figure \ref{fig:6setstwisted}(b) could be separated into three different states by the bifurcation points 6 and 7, and each state has different symmetries. 
		For the sake of conciseness, in this study we have labeled all of the black solutions in Figure \ref{fig:6setstwisted}(a) and Figure \ref{fig:6setstwisted}(b) as $T2$ and $T3$, respectively.

\begin{figure}[h!]
		\centering
		\includegraphics[width=0.8\textwidth]{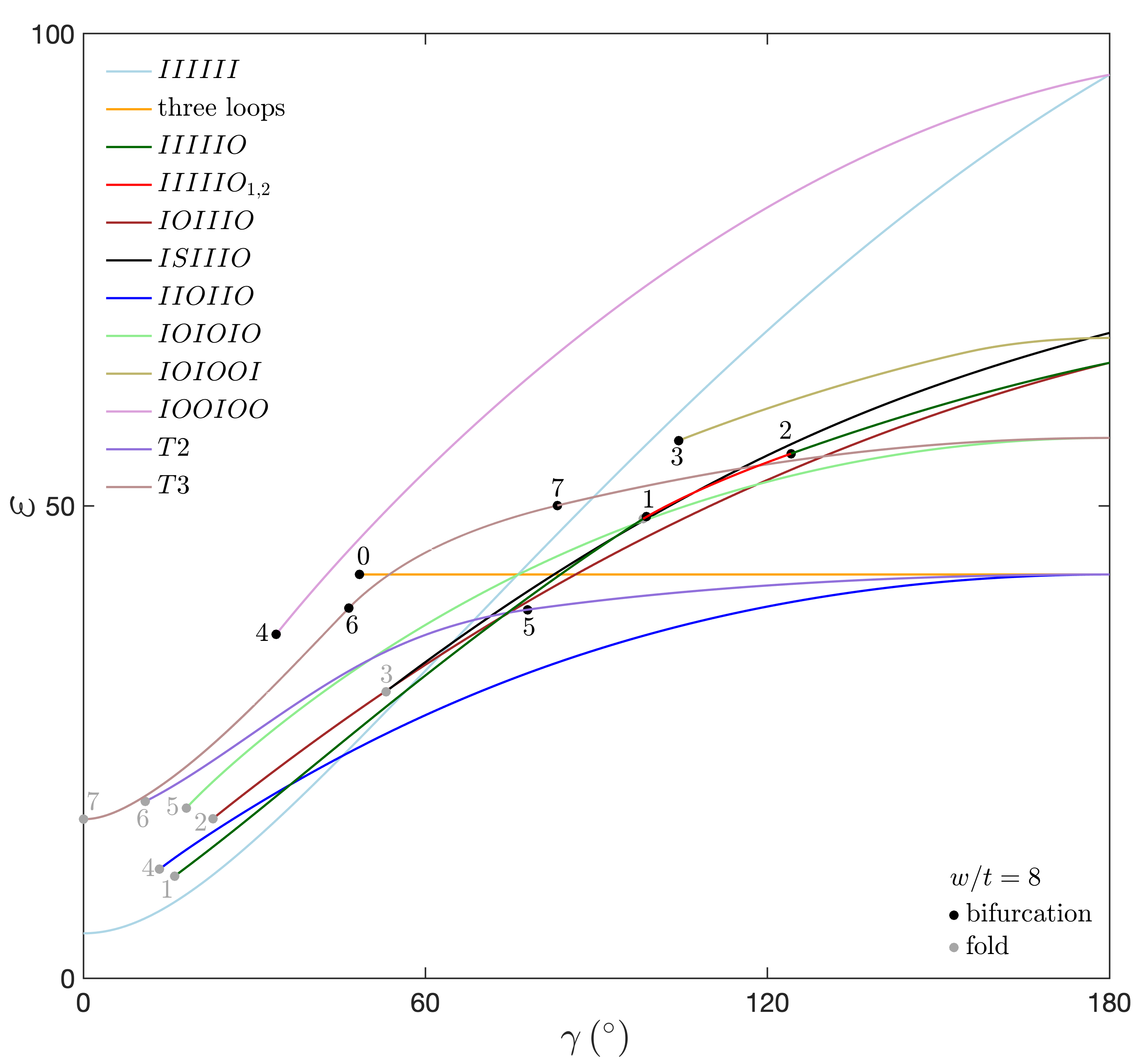}
		\caption{A summary of the states that are observed to be stable in experiments. The numbers correspond to the bifurcation/fold points in in Figures \ref{fig:6setslooping}-\ref{fig:6setstwisted}. $IIIIII: \gamma \in [0^{\circ},180^{\circ}]$. Three loops: $\gamma \in [48.46^{\circ},180^{\circ}]$. $IIIIIO: \gamma \in [16^{\circ},98.78^{\circ}] \cup [124.16^{\circ},180^{\circ}]$. $IIIIIO_{1,2}:  \gamma \in [98.21^{\circ},124.16^{\circ}]$. $IOIIIO: \gamma \in [22.73^{\circ},180^{\circ}]$. $ISIIIO: \gamma \in [53.11^{\circ},180^{\circ}]$. $IIOIIO: \gamma \in [13.35^{\circ},180^{\circ}]$. $IOIOIO:  \gamma \in [18.07^{\circ},180^{\circ}]$. $IOIOOI: \gamma \in [104.4^{\circ},180^{\circ}]$. $IOOIOO: \gamma \in [33.8^{\circ},180^{\circ}]$. $T2:  \gamma \in [10.8^{\circ},180^{\circ}]$. $T3: \gamma \in [0.0017^{\circ},180^{\circ}]$.}
		\label{fig:SixBigonRingstableAll}
	\end{figure}

	Figure \ref{fig:SixBigonRingstableAll} summarizes the stability range of the states that are stable in experiments in the $\gamma-\varepsilon$ plane. On the left end, most of the states lose stability before reaching $\gamma=0^{\circ}$, and only $IIIIII$ and $T3$ are stable with a small intersection angle. Increasing the intersection angle $\gamma$ tends to stabilize various states. On the right end, all the states can stably go to $\gamma=180^{\circ}$. $IIIIIO_{1,2}$ exists in a small interval (red lines). With such a diagram, it is clear that how many and which states are physically realistic (i.e., stable) at a specific intersection angle. With a small intersection angle ($\gamma \lessapprox 45^{\circ}$), $IIIIII$ stores the least elastic energy, which increases much faster than other states at larger intersection angles. With $45^{\circ} \lessapprox \gamma \lessapprox 160 ^{\circ}$, $IIOIIO$ contains the least elastic energy, and with $160^{\circ} \lessapprox \gamma \lessapprox 180 ^{\circ}$, $IIOIIO$, $T2$, and the three loops contain almost the same elastic energy. Similarly, $IOIOIO$ and $T3$ have almost the same energy level with $155^{\circ} \lessapprox \gamma \lessapprox 180 ^{\circ}$. After gaining stability, $IOOIOO$ has the highest elastic energy. 
	
	\newpage
	
	\section{Effects of anisotropy of the strip's cross section on the behavior of a 6-bigon ring}\label{se:anisotropy}
	
\begin{figure}[h!]
		\centering
		\includegraphics[width=0.88\textwidth]{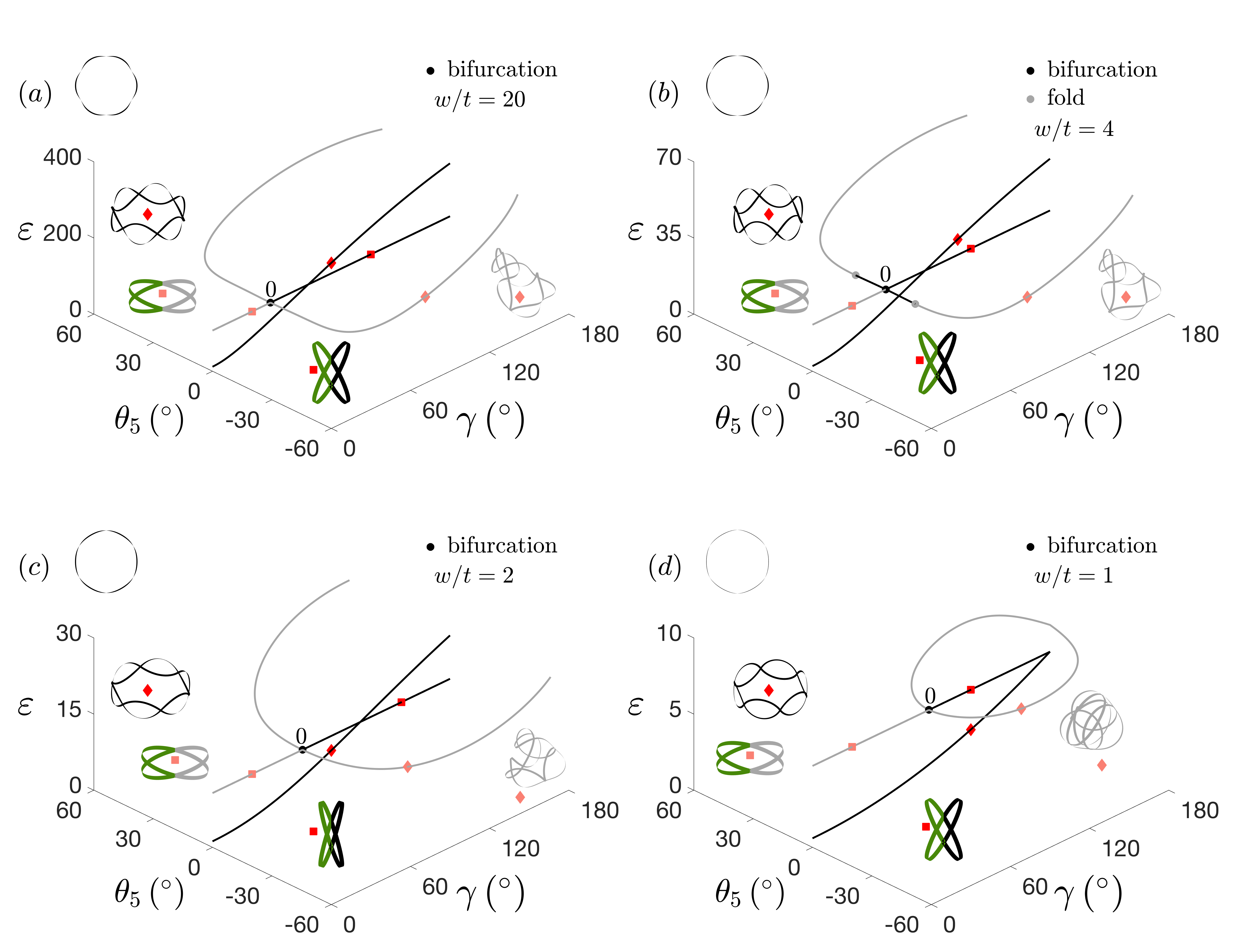}
		\caption{Numerical solutions of a 6-bigon ring with various $w/t$.
			$IIIIII$ ($\reddiamond$) is stable in the whole range $0^{\circ} \le \gamma \le 180^{\circ}$, while the three-loop branch ($\redsquare$) gains stability through a pitchfork bifurcation. Decreasing $w/t$ increases the critical intersection angle where the bifurcation occurs.} 
		\label{fig:6setsloopingAnisotropy} 
	\end{figure}

	In this section, we study the influence of the anisotropy of a strip's cross section on the mechanical behaviors of a 6-bigon ring. Figure \ref{fig:6setsloopingAnisotropy} presents several bifurcation diagrams similar to Figure \ref{fig:6setslooping}(a), with different anisotropy $w/t$. Note that the renderings in Figure \ref{fig:6setsloopingAnisotropy} have the same strip width and the varying anisotropy $w/t$ is not depicted.
	In order to make the elastic energy comparable, we fix the thickness $t$ and normalize the torsional rigidity of $w/t=8$ to unity, matching with the normalization in Figure \ref{fig:6setslooping}(a). 
	In numerical continuation, we started from a stable three-loop configuration ($\redstar$) from Figure \ref{fig:6setslooping}(a) (stability is confirmed by experimental models). We first decrease/increase $w/t$ to desired values and then vary the intersection angle $\gamma$. A small gravity force is added and then ``turned off" to track the nonplanar solutions that connect to the three-loop branch. In the continuation processes of decreasing/increasing $w/t$ and adding/removing gravity, no folds or bifurcations are encountered. We assume that the stability does not change in these steps.

	\begin{table}[h!]
		\begin{center}
			\caption{Critical intersection angle $\gamma$ and the corresponding overcurvature that stabilizes the three-loop configuration of a 6-bigon ring with various anisotropy $w/t$.}
			\label{tab:table1}
			\begin{tabular}{l|c|r} 
				\textbf{anisotropy} & \textbf{critical angle} & \textbf{critical overcurvature}\\
				$w/t$ & $\gamma$ & $N_b \alpha /180$ \\
				\hline
				20 & $43.96^{\circ}$ & 1.5190\\
				16 & $44.72^{\circ}$ & 1.5253\\
				12 & $45.98^{\circ}$ & 1.5336\\
				8 & $48.46^{\circ}$ & 1.5425\\
				4 & $55.64^{\circ}$ & 1.5122\\
				2 & $68.20^{\circ}$ & 1.1397\\
				1.75 & $ 71.36^{\circ}$ & 0.8834\\
				1.5 & $75.32^{\circ}$ & 0\\
				1 & $88.30^{\circ}$ & 0\\
			\end{tabular}
		\end{center}
	\end{table}
	
	Together with Figure \ref{fig:6setslooping}(a), we conclude that decreasing anisotropy $w/t$ generally increases the critical angle at which the three-loop configuration gains stability. Table \ref{tab:table1} summarizes critical angles and the corresponding critical overcurvatures at various anisotropy of the cross section. While the critical angle increases monotonically with the decrease of anisotropy, the critical overcurvature changes little with $w/t \ge 4$, and decreases quickly to zero with $w/t \le 4$. If we were to cut the bigon ring at a node, a vanishing overcurvature corresponds to a straight bigon chain. This is different from the folding behavior of anisotropic Kirchhoff rod, which always requires a finite amount of overcurvature to stabilize the three-loop configuration \cite{manning2001stability}. However, for a wider strip, the three-loop configuration can be stable with zero overcurvature \cite{audoly2015buckling}. It appears that the width of the strip (i.e., the out-of-plane height) helps stabilize the multiply-covered loop. For a bigon ring with rods of square cross section, increasing the intersection angle $\gamma$ ($0^{\circ} \le \gamma \le 180^{\circ} $) does not create overcurvature, but increases the out-of-plane height  and thus the out-of-plane stiffness, which appears to help stabilize the three-loop configuration. Similar behavior can be observed by folding a piece of flat paper (no matter along the short or long edges) into a stable multiply-covered roll.

\begin{figure}[h!]
		\centering
		\includegraphics[width=0.9\textwidth]{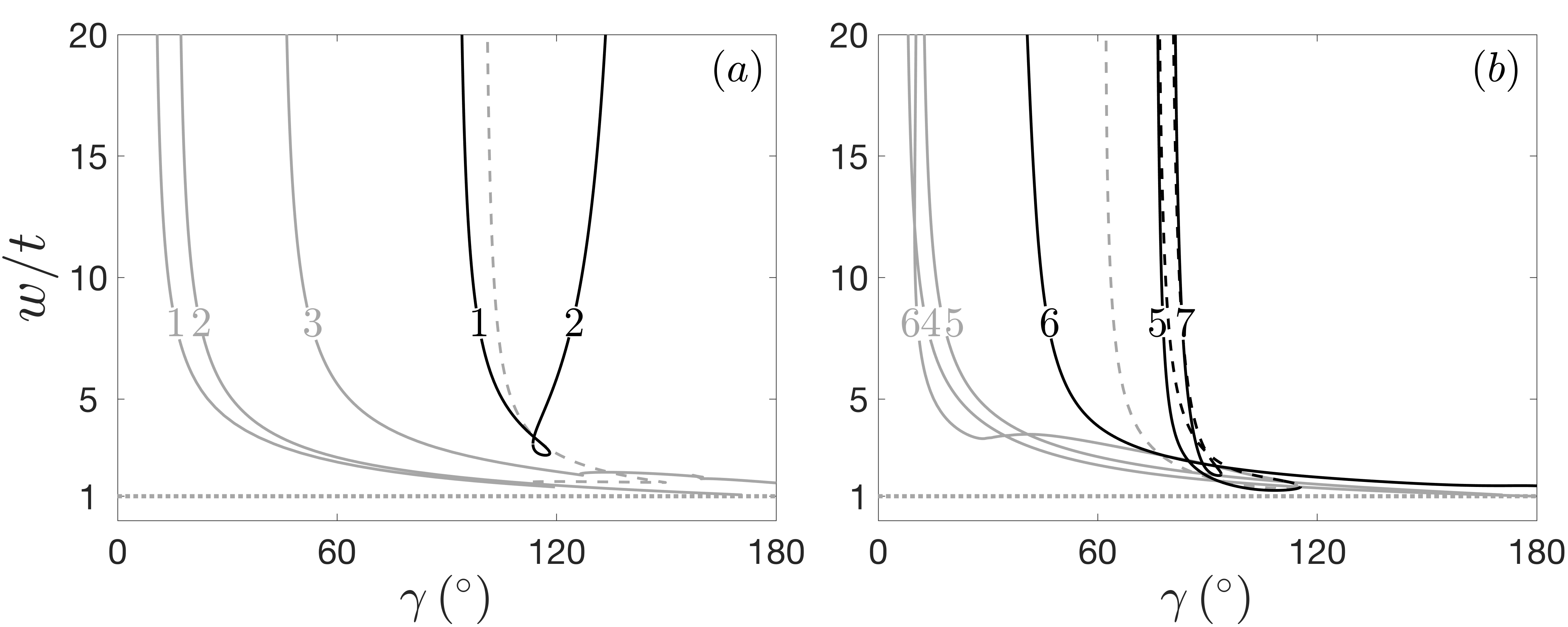}
		\caption{Loci of various bifurcations and folds in Figures \ref{fig:6setslooping}-\ref{fig:6setstwisted}, obtained by conducting two-parameter continuation in the plane spanned by anisotropy $w/t$ and intersection angle $\gamma$ (solid curves). The dashed lines connect to numbered loci curves through fold points. The dotted line represents $w/t=1$. With large $w/t$, all the loci curves approach certain intersection angles, representing a perfectly anisotropic rod. Decreasing the anisotropy $w/t$ generally decreases the stability range of each stable state.}
		\label{fig:bigonRing6setsPhase}
	\end{figure}

	
	For some of the stable states shown in Figure \ref{fig:ExpwithNu} and discussed in Section \ref{se:bigonringresults}, we conduct two-parameter continuation to track the corresponding loci of the folds/bifurcations in the space spanned by intersection angle $\gamma$ and anisotropy $w/t$. Figure \ref{fig:bigonRing6setsPhase} shows such a phase diagram. The numbers match with the folds/bifurcations in section \ref{se:bigonringresults}. All of the loci curves contain an almost vertical part, which implies that with large $w/t$, the Kirchhoff rod model approaches a perfectly anisotropic rod, where one of the bending curvatures is penalized to zero. On the other hand, the critical intersection angle $\gamma$ increases with the decrease of $w/t$, which decreases the stability range of each stable state. All the loci curves stay above $w/t=1$. This implies that before reaching a square cross section with $w/t=1$, these states either disappear or become unstable. The dashed lines connect to numbered loci curves through fold points, implying complex annihilation of folds/bifurcations in the solution space. We conclude that decreasing anisotropy $w/t$ of the strip's cross section tends to destabilize various stable states of a 6-bigon ring shown in Figures (\ref{fig:6setslooping}-\ref{fig:6setstwisted}).

	\section{Conclusion and further discussion} \label{se:conanddiscuss}
	
	By simply joining the ends of two strips through a prescribed intersection angle, we build a bistable structure \emph{bigon}, which we use to construct a novel multistable \emph{bigon ring} that connects a series of bigon cells to form a loop. We find that the intersection angle and the anisotropy of the strip's cross section are two crucial factors in determining the bistability of a bigon and also the multistability of a bigon ring. The models and insights presented in this work have potential applications in structures and materials with reversible functionalities. 
	
	A bigon can be used as a building block to construct more complicated elastic networks with target geometries. For example, connecting several bigons in series will form a bigon arm, which appears to be able to morph into a family of planar shapes by tuning the intersection angle of each bigon independently. Here, varying the intersection angle of each bigon changes its tangent angle, and thus changes the ``local curvature" of the bigon arm. 
	
	Several studies have demonstrated the potential of using strips (not necessarily straight or with uniform cross sections) as building blocks to create 3D surfaces and morphable structures. For example, using straight strips and pin joints, a planar grid can morph into freeform surfaces in which the centerlines of strips become geodesic curves of the target surfaces \cite{pillwein2020elastic}. The introduction of planar curvatures to strips can smoothen surfaces made by triaxial weaving \cite{baek2020smooth}. In another study \cite{celli2020compliant}, helicoidal ribbons (pre-twisted from ribbons with undulated edges) that have preferred bending directions are used to create shape-changing structures.
	
	In addition, we developed a numerical model to study mechanics of elastic strip networks. The numerical implementation combines several techniques from the literature \cite{healey2006straightforward,ascher1981reformulation} to formulate an elastic network as a TPBVP. Our formulation allows a convenient modeling of rigid and flexible nodes by either directly imposing the continuities of orientations in the case of a rigid node or including the constitutive laws of a flexible node. This is different from discrete elastic rods that require additional treatments for specifying rotations at coupled joints \cite{lestringant2020modeling}.In Appendix \ref{appse:hingedarm}, we apply our numerical framework to a planar bigon arm to address hinge joints. The method of formulating a MPBVP into a TPBVP is general \cite{ascher1981reformulation}, and could be applied to elastic networks consisting of general one dimensional structures, such as Cosserat rods \cite{antman1995nonlinear} and inextensible strips \cite{starostin2015equilibrium}.
	
	Together with numerical continuation, we applied the numerical model to study buckling and bifurcation behaviors of bigons and bigon rings. Numerical predictions of the folding and multistable behaviors of a bigon ring match with experimental results well. Unlike the anisotropic Kirchhoff rod that always requires a finite overcurvature to stabilize a three-loop configuration, a 6-bigon ring with vanishing overcurvature can be folded into a structure of three loops, which appears to be stabilized by the out-of-plane height/stiffness. In addition, it appears that the number of stable states in a bigon ring increases quickly with the increase of the number of bigon cells. We have investigated several examples of the folding behavior of a 15-bigon ring and the rich static equilibria of a 6-bigon ring.

	In future work, building a stability test for elastic networks composed of Cosserat rods will help identify stable states from the numerous numerical equilibria. A stability test of elastic networks that have multiple rods coupled together is developed for a tree of \emph{Elastica} \cite{o2012nonlinear} and for a parallel continuum robot of Cosserat rods with one end fixed and the free end coupled together \cite{till2017elastic}. A stability test for general elastic networks is not yet fully developed and might require coupling the ``0" and ``1'' ends together, which are not necessarily fixed in space \cite{hull2013optimal}.
	

	\section*{Acknowledgments} 
	TY and LD were supported by Princeton SEAS Project X Innovation Fund. SG and FM were supported by Council for International Teaching and Research, \emph{Global Collaborative Network: ROBELARCH at Princeton University}.
	TY thanks James Hanna, Oliver O$'$Reilly and Andy Borum for helpful discussions, John Till and Caleb Rucker for the reference \cite{hull2013optimal}, and Timothy Healey for insightful communications on the dummy parameter technique \cite{healey2006straightforward}.

	\appendix
	\section{Additional numerical example: hinged  bigon arms}\label{appse:hingedarm}
	
\begin{figure}[h!]
		\centering
		\includegraphics[width=\textwidth]{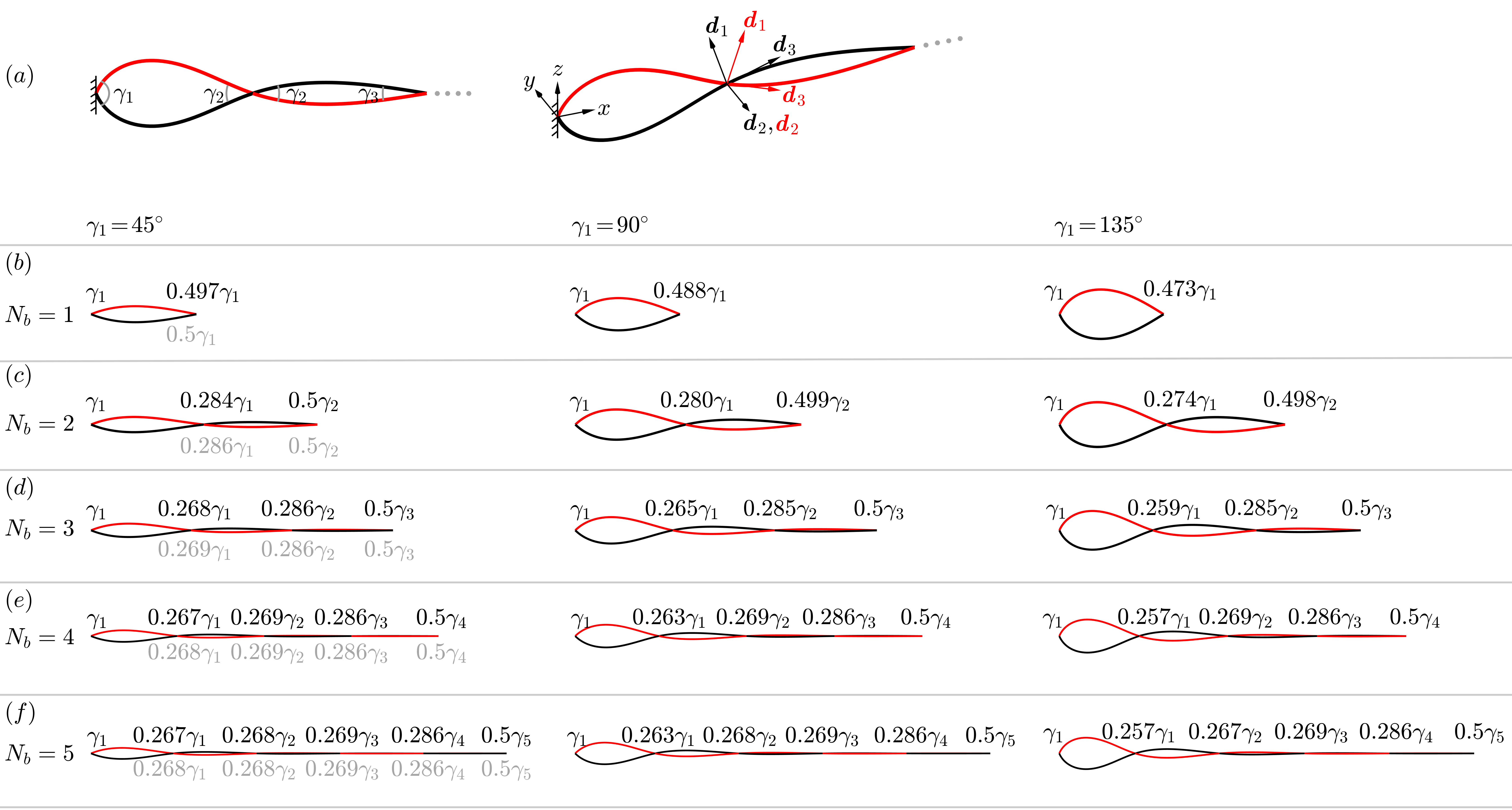}
		\caption{Numerical results (black numbers) and analytical predictions from a linear beam model (gray numbers in the first column) showing the angle propagation in a planar hinged bigon arm with different number of bigon cells $N_b$. (a) Two continuous strips (black and red) are interwoven at a series of discrete pin joints, with the left node prescribed to a fixed intersection angle $\gamma_1$; the other intersection angles $\gamma_2, \gamma_3$ etc. are unknowns. (b) $N_b=1$. (c) $N_b=2$. (d) $N_b=3$. (e) $N_b=4$. (f) $N_b=5$.}
		\label{fig:HingedBigonArmConfigSet}
	\end{figure}

	To demonstrate that our numerical framework can be applied to elastic networks with different types of joints, here we present the numerical results of a hinged bigon arm and the corresponding analytical results from a linear analysis. The problem is shown in Figure \ref{fig:HingedBigonArmConfigSet}(a). The hinged bigon arm is made by interweaving two continuous strips (black and red) at a series of discrete pin joints (without friction), with the left node clamped at the origin and prescribed to a fixed intersection angle $\gamma_1$ in the $x-z$ plane; the other pin joints are free to move and rotate in space. The arc length of the strip between adjacent joints is fixed to unity. At a pin joint, we assume that moments between the black and red strips cannot be transferred about the node normal $\bm{d}_2$ that is aligned with the pin axis. However, moments in the other directions (i.e., in the tangential plane $\bm{d}_1 - \bm{d}_3$) can be transferred. Inside each black and red strip, the moment about the node normal $\bm{d}_2$ is continuous at a pin joint.
	
	We are particularly interested in the system's capability of propagating the intersection angle, i.e., computing the sequence of intersection angles produced at other joints when the intersection angle $\gamma_1$ at the left end is assigned. Similar to a single bigon, increasing $w/t$ could make the structure buckle out of plane, resulting in deflections in the $y$ direction. Here we choose a small $w/t=1.2$ such that the hinged bigon arm stays stably in plane. Note that our formulation is fully three dimensional and could capture out-of-plane deformation. 
	
	The boundary conditions and unknowns can be identified similarly to the bigon and bigon rings. The left end is the same as the fixed end of a bigon (see Figure \ref{fig:bigonKinematics}(a)) with the same 14 unknowns and 14 boundary conditions. The difference between the right end of a hinged bigon arm and the free end of a bigon is that the intersection angle of the former is an unknown, while the intersection angle of the latter is specified. This leads to 18 unknowns at the right end of the hinged bigon arm, including 17 from the unknowns of a bigon at the free end and one corresponding to the intersection angle $\gamma_{N_{b+1}}$. 
		The boundary conditions at the right joint of a hinged bigon arm are the same with the free end of a bigon in terms of quaternions and the continuity of forces and positions (in total 14); the difference is the right end of a hinged bigon arm has four moment boundary conditions, namely, two from the continuity of moments in the tangential plane and one for each rod by imposing a vanishing moment about $\bm{d}_2$. In total, we obtain 18 boundary conditions at the right end, matching with the 18 unknowns. 
		The internal pin joints are similar to the free node in a bigon ring, except that the intersection angle is an unknown. In total, we have 32 unknowns at an internal pin joint. Compared with the free node in a bigon ring that has three moment boundary conditions, a pin joint has four, i.e., two from the continuity of moments in the tangential plane, and one for each rod by imposing the continuity of the moment about $\bm{d}_2$ at the joint. Thus, we have 32 boundary conditions at an internal pin joint. Since all the joints are well-posed, we obtain a well-posed TPBVP.

	Figures \ref{fig:HingedBigonArmConfigSet}(b-f) summarize numerical results of a hinged bigon ring with $\gamma_1=45^{\circ},90^{\circ}$ and $135^{\circ}$, and number of bigon cells $N_b=1,2,3,4,$ and 5. The numerical results can be summarized as: (1) The intersection angle decreases quickly to zero, almost following an exponential law, i.e., the next angle is slightly larger than one quarter of the previous one. (2) An interesting exception is that the rightmost intersection angle is always approximately one half of the penultimate one; (3) The exponential decay appears to be insensitive to $\gamma_1$. In addition, as long as the bigon arm stays in plane, the decay should not depend on $w/t$, since the planar bending is the only deformation. However, $w/t$ may influence the propagation of the intersection angle if the system buckles out of plane, which we did not consider in this study.

\begin{figure}[h!]
		\centering
		\includegraphics[width=0.7\textwidth]{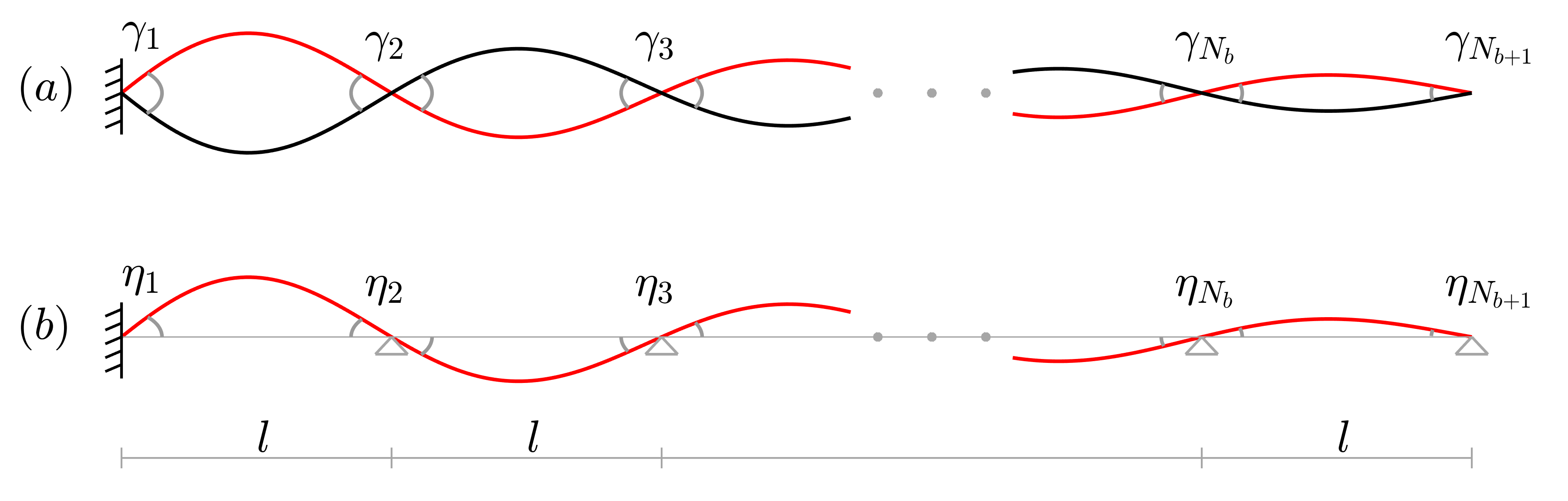}
		\caption{A plane linear continuous beam model. (a) The hinged bigon arm. (b) Because of the symmetry, half of the structure is analyzed as a continuous beam that is simply supported at the joints.}
		\label{fig:bigonarm}
	\end{figure}

	The propagation of the intersection angle is analytically tractable for a small $\gamma_1$, in which the bigon arm stays stably in the $x-z$ plane and could be analyzed by employing a plane linear beam model. Accordingly, the series of strips that compose the bigon arm can be seen as a continuous beam over multiple supports, each corresponding to a joint of the bigon arm. Because of symmetry, these supports have null transversal displacement and are free to rotate (see Figure \ref{fig:bigonarm}). Notice that the rotation angles of each beam at the supports are equal to half of the corresponding intersection angles and could be negative or positive, i.e. $\eta_i=(-1)^{i+1}\gamma_i/2$. Also, assuming small displacements and rotations, supports can be assumed to be equally spaced and their mutual distance is set to the length $l$ of the strips between adjacent joints. On the other hand, the proposed numerical formulation is geometrically exact, leading to a mutual distance that is always less than $l$. To assign the first intersection angle $\gamma_1$, the boundary condition $\eta_1=\gamma_1/2$ is prescribed at the left end and $M_{21}$ represents the corresponding bending moment there.

	By employing stiffness coefficients corresponding to the plane linear beam model, this model can be easily solved by the displacement method \cite{connor2016fundamentals}. Thus, equilibrium equations at the joints $J \in [1,N_{b+1}]$ read
		\begin{equation}\label{eq:continuousbeam}
		\begin{aligned}
		& 4 EI_2 \eta_1 / l + 2 EI_2 \eta_2 / l = M_{2\,1} \,, \quad J=1 ,
		\\
		...
		\\
		&2 EI_2 \eta_{i-1} / l + (4+4) EI_2 \eta_{i} / l  + 2 EI_2 \eta_{i+1} / l = 0 \,, \quad J=i ,
		\\
		...
		\\
		&2 EI_2 \eta_{N_b-1} / l + (4+3) EI_2 \eta_{N_b} / l = 0 \,, \quad J=N_b ,
		\\
		& \eta_{N_b+1} =-\eta_{N_b}/2 \,, \quad J=N_{b+1} ,
		\end{aligned}
		\end{equation}
		holding for $N_b > 2$. $EI_2$ is the plane bending stiffness.
		For $N_b=2$, one simply has the system of three equations corresponding to the first and the last two equations of \eqref{eq:continuousbeam}, while the case $N_b=1$ is trivially a propped cantilever beam. Notice that the last equation, which sets the angle at the right joint, corresponds to the rotation at the hinged extremity of a propped cantilever beam.
		This system of linear equations can be solved for the angles $\eta_i$ ($i=2,\,...,N_{b+1}$) and the moment $M_{2\,1}$ if $\eta_1$ is given, or it can be solved for all the angles $\eta_i$ ($i=1,\,...,N_{b+1}$) if $M_{2\,1}$ is given. These solutions will furnish the exact sequence of intersection angles for any number of bigon cells $N_b$.

	Alternatively, the above equations can be solved for $M_{2\,1}/\eta_1$ and for the ratio between successive crossing angles $r_i=\gamma_i/\gamma_{i-1}=-\eta_i/\eta_{i-1}$. This last solution allows to compute the rotational stiffness of the bigon arm extremity, which is given by $M_{21}/\gamma_1=M_{2\,1}/2\eta_1$ indeed.
		In this case the first equation of \eqref{eq:continuousbeam} gives
		\begin{equation}
		\frac{M_{2\,1}}{2\eta_1}=(2-r_2)\frac{EI_2}{l},
		\end{equation}
		while the last two equations yield
		\begin{equation}\label{eq:lasttwoangleratios}
		r_{N_b+1}=-\eta_{N_b+1}/\eta_{N_b}=0.5 \, ,
		r_{N_b}=-\eta_{N_b}/\eta_{N_b-1}=2/7.
		\end{equation}
		The remaining crossing angle rations $r_2,\,...,\,r_{N_b-1}$ are computed from the generic equilibrium equation of joint $i$, which reads
		\begin{equation}
		\eta_{i-1} + 4 \eta_{i} + \eta_{i+1} = 0 .  
		\end{equation}
		Introducing $r_i=-\eta_{i}/\eta_{i-1}$ the above equation becomes
		\begin{equation}\label{eq:sequenceofcrossingangleratios}
		1-4 r_i + r_i r_{i+1}=0
		\quad \Leftrightarrow \quad 
		r_i=\frac{1}{4-r_{i+1}},
		\end{equation}
		which can be applied in sequence from $i=N_b-1$ to $i=2$.
		This gives the series of values
		\begin{equation}\label{eq:sequenceofcrossingangleratios2}
		r_{N_b+1}=0.5\,, \quad r_{N_b} \approx 0.286\,, \quad r_{N_b-1} \approx 0.269\,, \quad r_{N_b-2} \approx 0.268\,, \quad r_{N_b-3} \approx 0.268\,, \quad ...
		\end{equation}
		where the first two are given by \eqref{eq:lasttwoangleratios}. This solution holds for any value of $N_b$. We display these approximate analytical predictions in the first column of Figure \ref{fig:HingedBigonArmConfigSet} as gray numbers, which match with our numerical predictions surprisingly well.
	
	Notice that when $N_b\rightarrow\infty$ one can reasonably assume that the ratio between two successive angles remains constant along the bigon arm, far from the boundaries. Hence, when $i << N_b$ it can be set $r = r_i \approx r_{i-1}$ and equation \eqref{eq:sequenceofcrossingangleratios}$_1$ is rewritten as
		\begin{equation}
		1 - 4 r + r^2 = 0,
		\end{equation}
		which gives the solutions $r=2\pm \sqrt{3}$.

	For the case at hand, being the constraint assigned at the first joint, we can exclude the solution with the plus sign, since it corresponds to a crossing angle that grows further from the applied action. Hence we have $r=2 - \sqrt{3} \approx 0.268$. This actually represents the asymptotic value of the sequence \eqref{eq:sequenceofcrossingangleratios2}. 
		Also, notice that the two solutions for $r$ are reciprocal one each other, i.e. $2 - \sqrt{3} = 1 / (2 + \sqrt{3})$, which fulfills left--right symmetry of the structure.
	

	\section{Measurements of the tangent angle of a bigon}\label{app:anglemeasurement}
	
\begin{figure}[h!]
		\centering
		\includegraphics[width=0.95\textwidth]{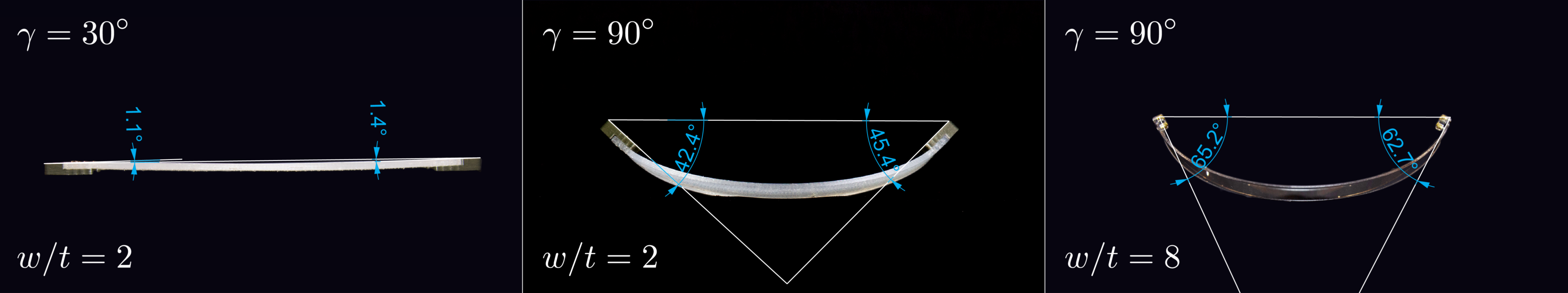}
		\caption{Examples showing the experimental measurements of the tangent angle of bigons with $w/t=2$ and 8, and different intersection angles}.
		\label{fig:experimentalpha}
	\end{figure}

	The tangent angle of a bigon is measured by taking photographs with a Canon EOS 6D DSLR camera with a 105mm zoom lens to minimize perspective distortion such that only one half of the structure is visible. The photographs are traced using CAD drafting software to extract the tangent angle. Three examples are shown in Figure \ref{fig:experimentalpha}. The bigon with $\gamma=30^{\circ}$ and $w/t=2$ is indeed flat, and the nonvanishing tangent angle is caused by gravity that slightly deforms the beams.
	
	\section{Inside-out flip of a planar bigon}\label{app:bilobatebigon}	
	
\begin{figure}[h!]
		\centering
		\includegraphics[width=0.85\textwidth]{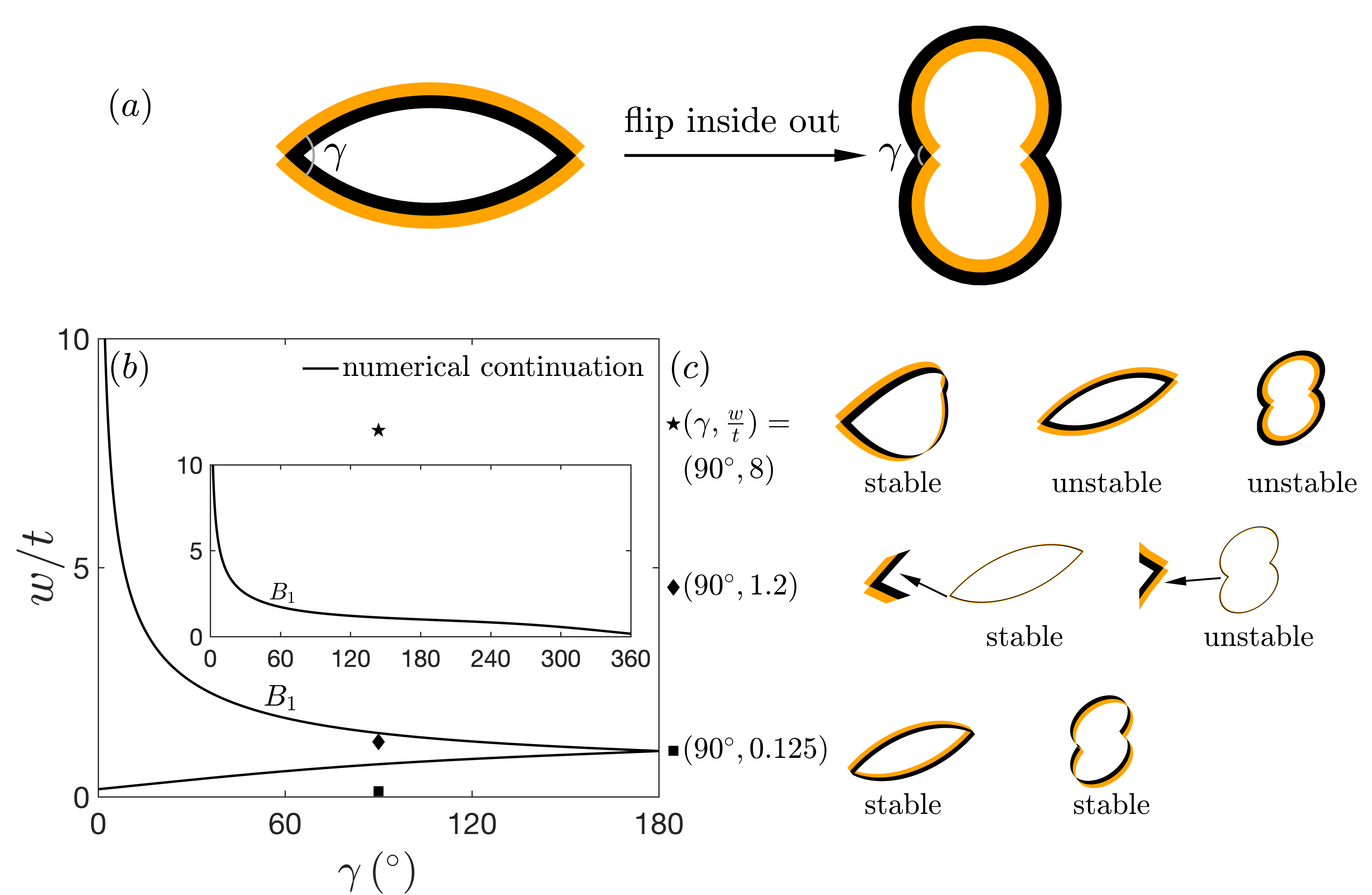}
		\caption{Numerical solutions of a bigon turned inside out. (a) The planar state of a bigon with $\gamma<180^{\circ}$ is flipped inside out to achieve a planar bilobate shape. (b) Loci of the bifurcations $B_1$. The bottom boundary is obtained by reflecting the segment $\in [180^{\circ}, 360^{\circ}]$ (see the inset) about the vertical line $\gamma=180^{\circ}$. (c) Renderings correspond to the marked locations in (b)}.
		\label{fig:BigonFlipped}
	\end{figure}

	Figure \ref{fig:BigonFlipped}(a) shows that the planar state of a bigon with an intersection angle $\gamma$ ($<180^{\circ}$) can be flipped inside out to achieve a planar bilobate shape, which is equivalent to a bigon with an intersection angle $2 \pi -\gamma$ ($>180^{\circ}$). For clarity, the internal half of the strips is colored as black and the outer half as orange.
		The bilobate shape could be stable for $w/t<1$, and its stability boundary can be achieved by reflecting the extension of the $B_1$ curve within $\gamma \in [180^{\circ}, 360^{\circ}]$ about the vertical line $\gamma=180^{\circ}$, reported in Figure \ref{fig:BigonFlipped}(b). Through this way, we have restricted the intersection angle $\gamma$ to $[0^{\circ}, 180^{\circ}]$ and also included the bilobate state in the $w/t - \gamma$ plane.
	
	The $B_1$ curve in Figure \ref{fig:BigonFlipped}(b) divides the plane into three regimes. In the top regime ($\blackstar$), only the out-of-plane state is stable and the two planar states are unstable. In the middle area ($\blackdiamond$), a bigon with $\gamma<180^{\circ}$ does not buckle out of plane and stays stably in plane; however, its flipped bilobate shape is unstable. In the bottom regime where $w/t<1$ ($\blasquare$), a bigon stays in plane and its flipped state is stable.
	
	Notice that with $w/t<<1$ (e.g., the $\blasquare$ with $w/t=0.125$), a bigon has a shape similar to a creased thin strip, whose width switches to align with $\bm{d}_2$.

		\section{2D projections of the solution curves of a 6-bigon ring}\label{appse:2Dprojection}	
		
\begin{figure}[h!]
			\centering
			\includegraphics[width=0.9\textwidth]{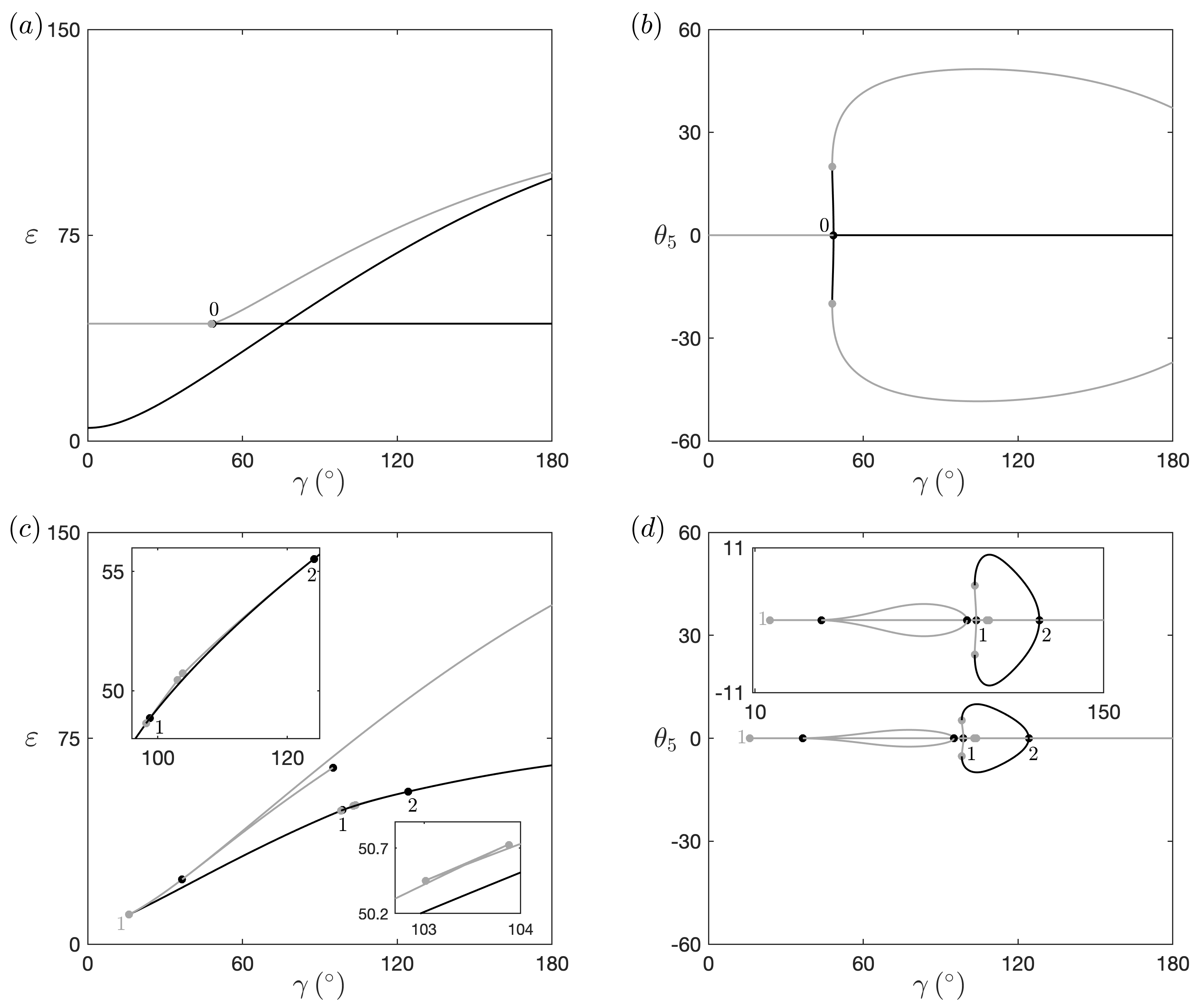}
			\caption{(a-b) Projections of the solution curves in Figure \ref{fig:6setslooping}(a) on the $\gamma-\varepsilon$ and $\gamma-\theta_{5}$ planes. (c-d) Projections of the solution curves in Figure \ref{fig:6setslooping}(b) on the $\gamma-\varepsilon$ and $\gamma-\theta_{5}$ planes.}
			\label{fig:6setslooping2D} 
		\end{figure}

\begin{figure}[h!]
		\centering
		\includegraphics[width=0.9\textwidth]{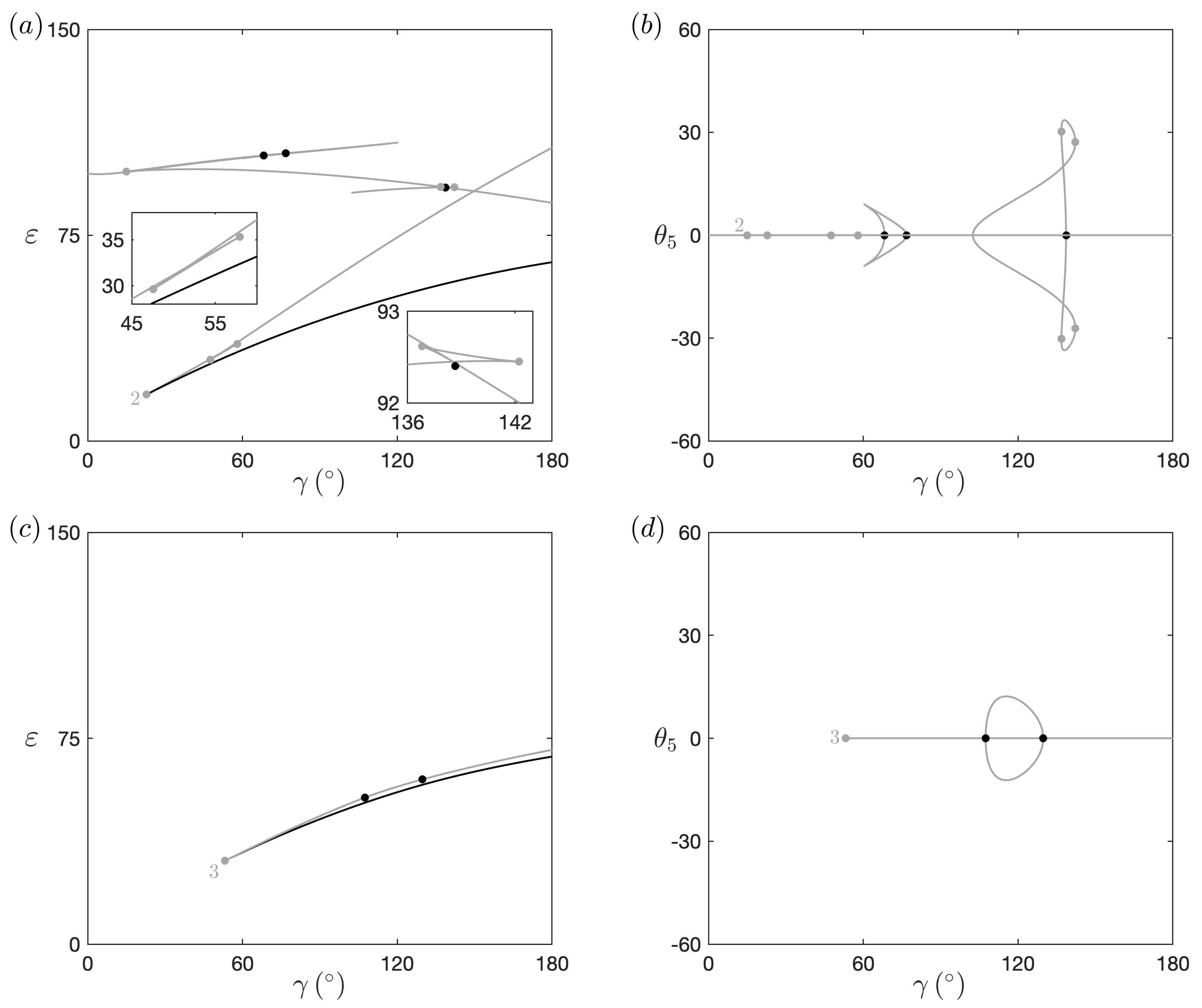}
		\caption{(a-b) Projections of the solution curves in Figure \ref{fig:6setstwoOs}(a) on the $\gamma-\varepsilon$ and $\gamma-\theta_{5}$ planes. (c-d) Projections of the solution curves in Figure \ref{fig:6setstwoOs}(b) on the $\gamma-\varepsilon$ and $\gamma-\theta_{5}$ planes.}
		\label{fig:6setstwoOs2D}
	\end{figure}

\begin{figure}[h!]
		\centering
		\includegraphics[width=0.9\textwidth]{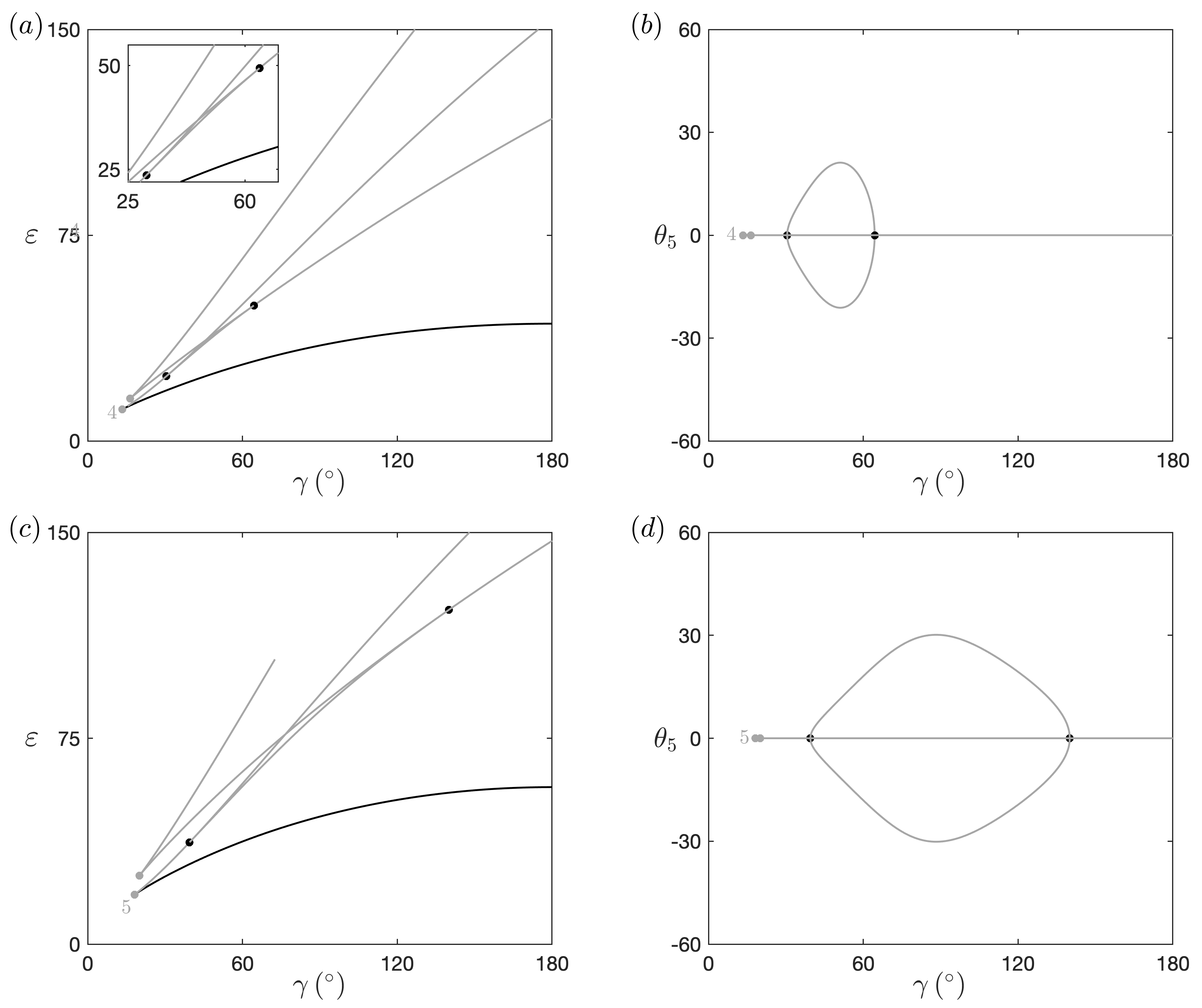}
		\caption{(a-b) Projections of the solution curves in Figure \ref{fig:6sets2and3Os}(a) on the $\gamma-\varepsilon$ and $\gamma-\theta_{5}$ planes. (c-d) Projections of the solution curves in Figure \ref{fig:6sets2and3Os}(b) on the $\gamma-\varepsilon$ and $\gamma-\theta_{5}$ planes.}
		\label{fig:6sets2and3Os2D}
	\end{figure}

\begin{figure}[h!]
	\centering
	\includegraphics[width=0.9\textwidth]{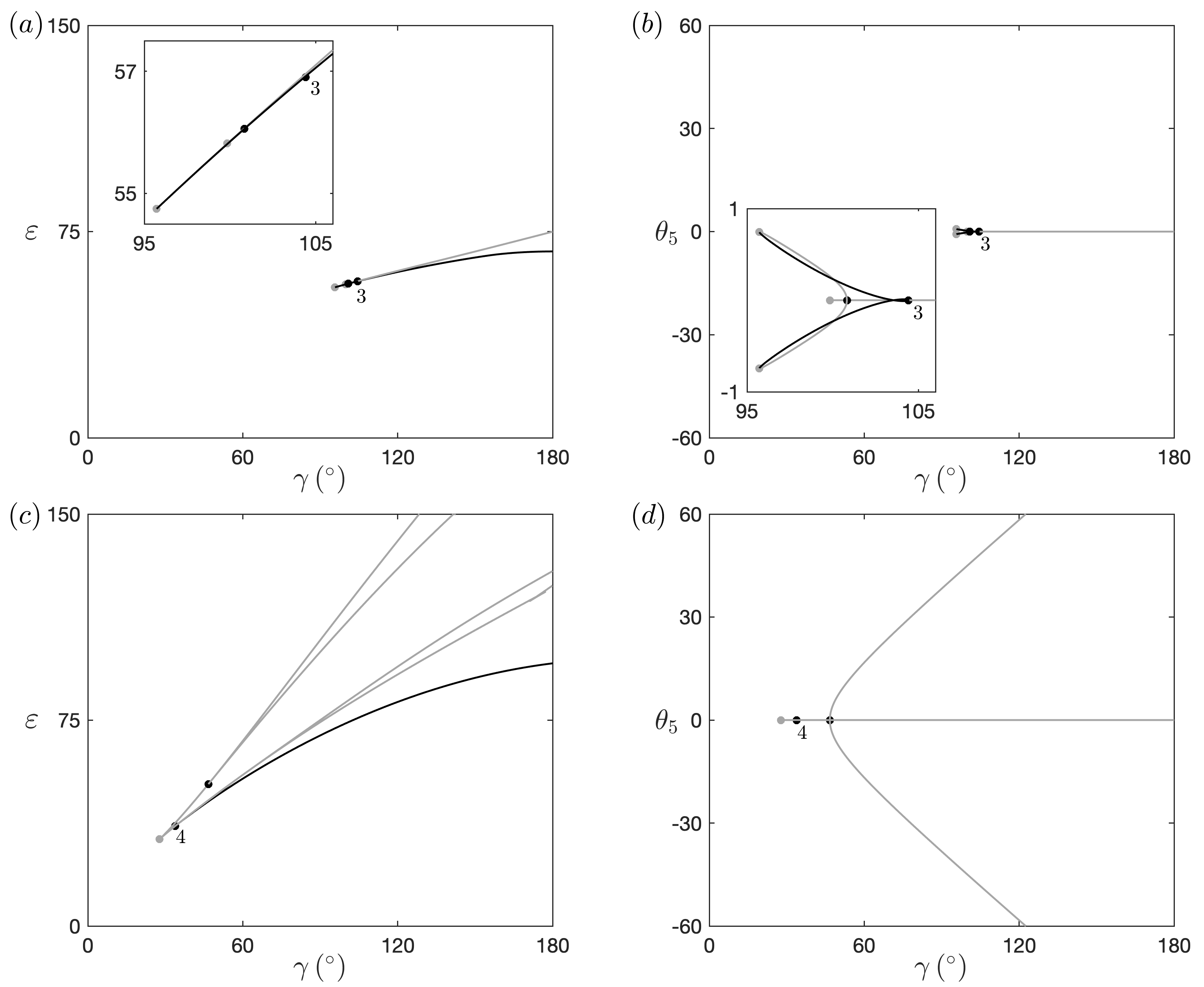}
	\caption{(a-b) Projections of the solution curves in Figure \ref{fig:6sets4and3Os}(a) on the $\gamma-\varepsilon$ and $\gamma-\theta_{5}$ planes. (c-d) Projections of the solution curves in Figure \ref{fig:6sets4and3Os}(b) on the $\gamma-\varepsilon$ and $\gamma-\theta_{5}$ planes.}
	\label{fig:6sets4and3Os2D}
\end{figure}

\begin{figure}[h!!]
	\centering
	\includegraphics[width=0.9\textwidth]{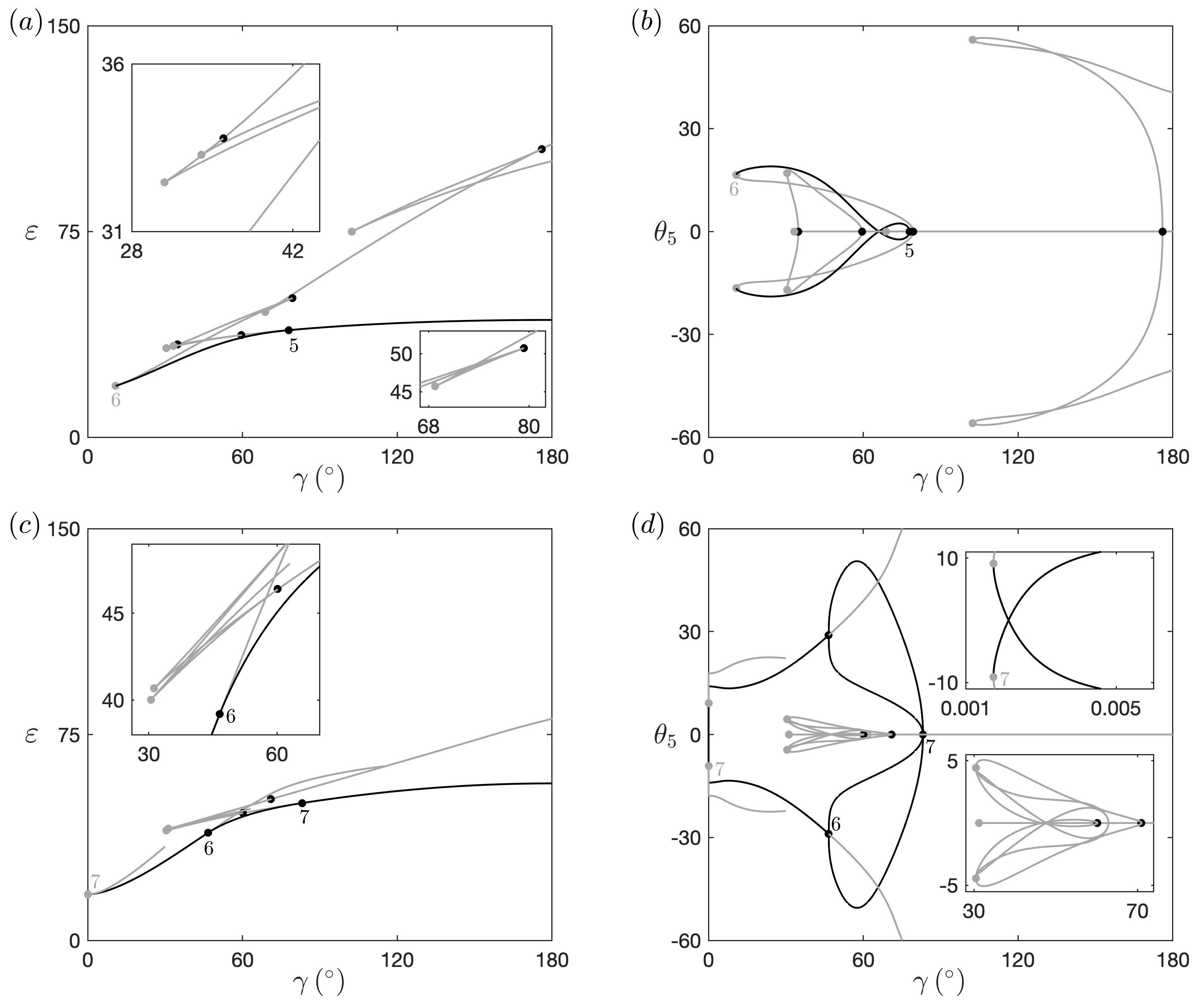}
	\caption{(a-b) Projections of the solution curves in Figure \ref{fig:6setstwisted}(a) on the $\gamma-\varepsilon$ and $\gamma-\theta_{5}$ planes. (c-d) Projections of the solution curves in Figure \ref{fig:6setstwisted}(b) on the $\gamma-\varepsilon$ and $\gamma-\theta_{5}$ planes.}
	\label{fig:6setstwisted2D}
\end{figure}

		2D projections of the solution curves of the 6-bigon ring in Figures (\ref{fig:6setslooping}-\ref{fig:6setstwisted}) are documented in Figures (\ref{fig:6setslooping2D}-\ref{fig:6setstwisted2D}). 
		
	\clearpage
	
	\bibliographystyle{unsrt}
	\bibliography{Bigon}
\end{document}